\documentclass[%
 reprint,
amsmath,amssymb,
aps,
]{revtex4-2}

\usepackage{graphicx}
\usepackage{dcolumn}
\usepackage{bm}
\usepackage{orcidlink} 
\usepackage{float}

\begin{document}

\title{\boldmath Matter effect in presence of a sterile neutrino and resolution of the octant degeneracy using a liquid argon detector}


\author{Animesh~Chatterjee$^{1,2,3}$ \orcidlink{0000-0002-2935-0958}}
\email{animesh.chatterjee@cern.ch}
\author{Srubabati~Goswami$^1$ \orcidlink{0000-0002-5614-4092}}
\email{sruba@prl.res.in}
\author{Supriya~Pan$^{1,4}$ \orcidlink{0000-0003-3556-8619}}
\email{supriyapan@prl.res.in}

\affiliation{$^1$ Physical Research Laboratory, Ahmedabad, Gujarat, 380009, India}
\affiliation{$^2$ Department of Physics and Astronomy, University of Pittsburgh, Pittsburgh, PA 15260}
\affiliation{$^3$ European Organization for Nuclear Research (CERN), 1211 Geneva 23, Switzerland}
\affiliation{$^4$ Indian Institute of Technology, Gandhinagar, Gujarat, 382355, India}

\begin{abstract}Results from the experiments like LSND, and MiniBooNE hint towards the possible presence of an extra eV scale sterile neutrino. The addition of such a neutrino will significantly impact the standard three flavour neutrino oscillations; in particular, it can give rise to additional degeneracies due to new sterile parameters. In our work, we investigate how the sensitivity to determine the octant of the neutrino mixing angle $\theta_{23}$ is affected by introducing a sterile neutrino to the standard three generation framework. We compute the oscillation probabilities analytically in the presence of a sterile neutrino, using the approximation that $\Delta_{21}$, the smallest mass squared difference, is zero. We use these probabilities to understand the degeneracies analytically at different baselines. We present our results of the sensitivity to octant of $\theta_{23}$ for beam neutrinos using a liquid argon time projection chamber (LArTPC). We also obtain octant sensitivity using atmospheric neutrinos using the same LArTPC detector. For the latter, we present our results assuming (i) no charge identification capability, and (ii) partial charge identification capability using the charge tagging ability of muon capture in Argon. In addition, we include the charge tagging capability of muon capture in argon which allows one to differentiate between muon neutrino and antineutrino events. The combined sensitivity of beam and atmospheric neutrinos in a similar experimental setup is also delineated. We observe that by combining simulated data from the beam and atmospheric neutrinos (including charge-id for muons), the sensitivity to the octant of $\theta_{23}$ for true values of $\theta_{23}=41^\circ(49^\circ)$ exceeds $4\sigma(3\sigma)$ for more than $50\%$ values of true $\delta_{13}$.
\end{abstract}

\maketitle
\flushbottom
\section{Introduction}
\label{sec:intro}
    Coming a long way since the first observation of neutrino oscillations, the standard three flavour neutrino oscillation paradigm is well established now, with most of the parameters being measured with considerable precision\cite{Esteban:2016qun, deSalas:2020pgw, Capozzi_2020}. The parameters describing the standard three flavour oscillations are the three mixing angles $\theta_{12},\theta_{13},\theta_{23}$ corresponding to mixing between the mass eigenstates with mass eigenvalues $m_1,m_2,m_3$, the Dirac CP phase $\delta_{13}(\delta_{CP})$, the two mass squared differences $\Delta_{21}=m_2^2-m_1^2$ driving the solar neutrino transitions and $\Delta_{31}=m_3^2-m_1^2$ governing the atmospheric neutrino oscillations. The oscillation probabilities are also dependent on the experimental parameters like the baseline distance $L$ traversed by the neutrinos from the source to the detector and the energy of the neutrinos $E_\nu$. Currently, the unknowns in the standard oscillation sector are the mass ordering among the three neutrino states, the octant of the atmospheric mixing angle $\theta_{23}$, and the value of the CP violating phase $\delta_{CP}$. The mass ordering refers to whether the sign of the atmospheric mass squared difference $\Delta_{31}$ is positive (Normal Ordering/NH) or negative (Inverted Ordering/IH). The octant of $\theta_{23}$ signifies if the value of the angle lies above (Higher Octant/HO) or below (Lower Octant/LO) $45^\circ$. One of the most impeding factors in the precise determination of these three parameters is the occurrence of degeneracies, i.e., various sets of different values of unknown parameters giving rise to equal probability, making an unambiguous determination of these parameters difficult. In view of the current unknowns in the three flavour framework, the existing degeneracies can be understood through a generalized \textit{hierarchy-octant-$\delta_{CP}$} degeneracy\cite{Ghosh:2015ena}. Measuring these parameters with considerable precision and alleviating the existing degeneracies are the focus of the ongoing and upcoming experiments such as T2K\cite{T2K:2019bcf}, NO$\nu$A\cite{NOvA:2007rmc}, T2HK, DUNE \cite{DUNE:2020ypp}, ESS$\nu$SB\cite{Alekou:2022emd}, etc. Planned atmospheric neutrino experiments like HyperKamiokande\cite{Hyper-KamiokandeProto-:2015xww}, KM3NeT\cite{Brunner:2017zcc}, PINGU\cite{IceCube:2016xxt}, INO\cite{ICAL:2015stm}, etc can also throw light on these parameters. Synergy and complementarity between atmospheric and beam experiments have been explored in the context of three generation framework in \cite{ Ghosh:2014rna, Barger:2013rha, Ghosh:2013yon, Chatterjee:2013qus, Ghosh:2014dba, Ghosh:2012px, Chakraborty:2019jlv, Fukasawa:2016yue, Ghosh:2015tan}.

    Although neutrino oscillation is unequivocally the leading solution for the flavour conversion to explain solar, atmospheric, reactor, and accelerator observations, the possibility of other beyond standard model (BSM) effects at a sub-leading level cannot be precluded. Several new physics effects have been discussed in the literature, including sterile neutrinos, long rang forces, neutrino decay, non-standard interactions, etc. Among these BSM scenarios, the existence of a light sterile neutrino is one of the most promising new physics hypotheses to explain the anomalies observed by LSND and MiniBooNE experiments.

    The idea of the light sterile neutrino was initiated in view of the results from the LSND experiment, which reported a signature of $\bar{\nu}_\mu \rightarrow \bar{\nu}_e$ oscillation at $3.8\sigma$\cite{Hill:1995gf}. Later MiniBooNE experiment with the same $L/E$ ratio also confirmed this result at $4.8\sigma$ significance\cite{Aguilar-Arevalo:2020nvw}. If we want to interpret these results through effective two flavour oscillations then corresponding to $ L/E\sim 1$ GeV/km in these experiments, there should be a new mass-squared difference $ \Delta_s \sim 1$ eV$^2$. This new mass-squared difference doesn't fit in the standard three flavour oscillation scheme, requiring one to incorporate at least one extra neutrino with a mass of eV scale. The result of the invisible decay width of the $ Z $ boson at CERN suggests that there can only be three different neutrinos below the mass range of half of the Z boson\cite{Abrams:1989yk}. Hence, this extra neutrino has to be inert with no standard model gauge interactions.

    Additional support in favour of an additional light sterile neutrino came from the observation of electron neutrino $\nu_e$ deficit in gallium-based radio-chemical experiments SAGE and GALLEX (\textit{Gallium anomaly})\cite{Giunti:2010zu, Acero:2008zz} which has been reinforced with the recent results from BEST experiment\cite{https://doi.org/10.48550/arxiv.2109.11482} at $5\sigma$. There was also the \textit{reactor antineutrino anomaly} in which several reactor neutrino experiments showed a deficit in the measured flux with an improved calculation of the inverse beta decay cross-section \cite{Huber_2011, Mueller_2011}. These could also be explained in terms of a sterile neutrino with a mass of the order of eV. However, the results from reactor experiments such as DANSS\cite{Svirida:2019kbq,https://doi.org/10.48550/arxiv.2112.13413}, NEOS\cite{PhysRevLett.118.121802}, STEREO\cite{PhysRevD.102.052002}, and PROSPECT\cite{PhysRevD.103.032001} excluded most of the \textit{reactor antineutrino anomaly} region\cite{Minotti:2022yae} at more than $90\%$ C.L. So far, the \textit{3+1 framework} including a light sterile neutrino with a mass of 1 eV first introduced in \cite{Goswami:1995yq}, offers the most economical scenario to explain these anomalies. However, the \textit{3+1 framework} suffers from a tension between the $\nu_\mu$ disappearance and appearance data. This tension\cite{Dentler:2018sju} originates from the non-observation of any similar supportive signal in the accelerator based disappearance experiments in $P_{\mu \mu}$ channel like CDHSW, MINOS\cite{Abazajian:2012ys, MINOS:2017cae}, Super-Kamiokande\cite{T2K:2019efw}, IceCube DeepCore\cite{Vannerom:2022cpf}, MicroBooNE\cite{Arguelles:2021meu}, NO$\nu$A\cite{Hausner:2022fli}. Reactor based electron disappearance searches in the experiments Bugey3\cite{Hu:2020uvx} and DayaBay\cite{MINOS:2020iqj} also didn't provide any evidence in support of sterile neutrino. The global fit performed in \cite{Gariazzo:2017fdh}, allowed three narrow regions around $\Delta_{41}\approx 1-2$ eV$^2$ with $0.00048<\sin^2{2\theta_{\mu e}}<0.002 $. However, after adding Bugey3, DayaBay, and MINOS+ data, the goodness of fit decreases drastically\cite{MINOS:2020iqj}. The most recent results from the MicroBooNE experiment did not report any evidence of electron neutrino disappearance in their three years of data\cite{https://doi.org/10.48550/arxiv.2110.14054,MicroBooNE:2022wdf}. However, it was shown in \cite{https://doi.org/10.48550/arxiv.2111.10359} that MicroBooNE data can not exclude the electron neutrino excess observed in  MiniBooNE in a model independent way. The joint analysis of results from MiniBooNE, and MicroBooNE experiments preferred the 3+1 scenario over no oscillation\cite{https://doi.org/10.48550/arxiv.2201.01724}.
    
    One of the sternest challenges for the existence of sterile neutrino comes from cosmology\cite{B_ser_2020}. The inclusion of an extra sterile neutrino increases the effective no of neutrinos relevant for the Big Bang Nucleosynthesis. It was proposed in \cite{Chu:2018gxk} that secret interaction between sterile neutrinos can remedy this situation but it was later disfavoured by cosmic microwave background analysis \cite{Goswami:2021eqy}. Recently, a joint analysis of short baseline and cosmological data showed that a sterile neutrino with a mass around 1 eV can be allowed for interaction with a new light pseudo scalar. To summarize, the existence of sterile neutrinos is still an open question while more experimental efforts are underway to resolve this.
    
    The upcoming TRISTAN detector at the KATRIN\cite{KATRIN:2022spi}, SBN\cite{MicroBooNE:2015bmn} at Fermilab, JSNS$^2$ detector\cite{JSNS2:2021hyk} at J-PARC are following up the results of LSND, MiniBooNE. The results from these experiments are expected to help in reaching a definitive conclusion about the existence of an eV scale sterile neutrino. If these experiments confirm the presence of an eV scale neutrino, then some new physics will be required to explain the tension between the disappearance and the appearance data. Some ideas in this direction can be found \cite{Babu:2022non, Hardin:2022muu}.

	If we consider the sterile neutrino hypothesis to be true, then the standard framework of neutrino oscillations is going to see some important modifications. The addition of a light sterile neutrino comes with three extra active-sterile sector mixing angles and two additional CP phases. These will compound the effect of the parameter degeneracies that already exist in the standard three flavour framework. In particular, it was shown in \cite{Agarwalla:2016xlg} that for the 3+1 oscillation framework, the octant degeneracy is more pronounced due to the effect of an additional interference term in the $\nu_\mu\rightarrow\nu_e$ vacuum oscillation probability relevant at long baseline setups in the context of the DUNE detector. It is well known that the addition of neutrino and anti-neutrino can evade the octant-$\delta_{13}$ degeneracy for three flavour case\cite{KumarAgarwalla:2013fko, Ghosh:2015ena}. However, in presence of a sterile neutrino, the octant-$\delta_{14}$ degeneracy can't be resolved even after the addition of neutrino plus anti-neutrino\cite{Ghosh:2017atj}. Implications of additional octant degeneracies associated with the new phases in the 3+1 framework have also been studied in the context of the NO$\nu$A\cite{Choubey_2019, Ghosh:2017atj} experiment. Other studies in the context of long baseline experiments in presence of a sterile neutrino can be found, for instance, in \cite{Dutta:2016glq, Singha:2022btw, Berryman:2015nua, Gandhi:2015xza, Agarwalla:2016xxa, Reyimuaji:2019wbn, Denton:2022pxt, Choubey:2017cba}.\\

	The primary focus of our paper is to study the octant sensitivity if an additional light sterile neutrino is present. We perform a comprehensive study of the octant sensitivity usinga LArTPC detector. LArTPC, first proposed in \cite{Rubbia:1977zz} constitutes one of the most important classes of scintillator detectors at present because of its superior capabilities, which provide several advantages in the precise reconstruction of neutrino events. Some of the current and future detectors using this technology are MicroBooNE, SBND, DUNE, etc. Earlier studies performed for three neutrino generations and atmospheric neutrinos in a liquid argon (LAr) detector can be found, for instance, in \cite{Chatterjee:2013qus, Barger:2013rha, Gandhi:2008zs}. In this paper, we extend our scope to investigate if the effect of additional degeneracies arising from an extra light sterile neutrino can be reduced in the presence of a large matter effect encountered at higher baselines. This has been studied for the combined analysis of beam neutrinos at a baseline of 1300 km and atmospheric neutrinos, which provide larger baselines as well as higher energies in this experimental setup, along with a separate study for each. Additionally, we present the results, including the charge tagging capability of muon capture in liquid argon, allowing one to differentiate between $\mu^+$ and $\mu^-$ events in the context of atmospheric neutrinos.
	
	In order to have a proper understanding of the octant degeneracy seen from numerical analysis, the study of the analytic expressions of neutrino oscillation probabilities is important. We obtain analytic expressions of the neutrino oscillation/survival probabilities assuming the solar mass squared difference $\Delta_{21}$ to be negligible as compared to the mass squared differences $\Delta_{31}$, and $\Delta_{41}=m_4^2-m_1^2$ driving the atmospheric and sterile neutrino oscillations respectively. We use the analytic expressions to understand the octant degeneracy at some representative baselines, e.g., 1300 km and 7000 km. There are other analytical calculations of oscillation probabilities in the presence of sterile neutrino in matter using the rotation methods\cite{PhysRevD.101.056005}, an exact analytical method\cite{Li_2018}. We discuss the region of validity and the error of the analytic expressions compared to the exact numerical probabilities.
	
	Studies related to sterile neutrinos in the context of atmospheric neutrino observations at India-based Neutrino Observatory (INO) experiment have been performed in \cite{Behera:2016kwr, Gandhi:2011jg}. More recently, an analysis in \cite{Thakore:2018lgn} considers sterile neutrinos in atmospheric baselines for a wide $\Delta_{41}$ mass squared range $10^{-5}:100$ eV$^2$ in the context of the INO experiment. This paper obtained bound on the active-sterile mixing angles as well as the sensitivity to the neutrino mass ordering in the 3+1 oscillation framework. Our study in this paper focuses on the impact of resonant matter effect on the probabilities at very long baselines and its influence on the sensitivity to determine the octant. To the best of our knowledge, this kind of study of the degeneracies in presence of a light sterile neutrino under the influence of resonance matter effect at very long baselines has not been looked into in past. We also explore this aspect in the context of atmospheric and beam neutrinos in a long baseline experimental setup of 1300 km both separately as well as together using a LArTPC detector and examine the complementarities between these two. Such studies in the context of a generic LAr detector have been performed for the three generation case earlier in \cite{Chatterjee:2013qus}.
	
	The plan of the paper is as follows. To start with, we establish the analytic framework for neutrino oscillations in presence of sterile neutrino in \autoref{sec:3+1}. The subsequent \autoref{sec:Prob} details the calculation of the probabilities. Next, \autoref{sec:degeneracy} contains the discussion on octant degeneracy for different baselines and energies as well as the dependence on the CP phases $\delta_{13}$, and $\delta_{14}$. In \autoref{sec:analysis}, we describe the experimental details for the LArTPC detector and outline the procedure of $\chi^2$ analysis adopted. We discuss the results in \autoref{sec:discussion}. Finally, we conclude in \autoref{sec:conclusion}.

\section{3+1 Framework}
\label{sec:3+1}
	The minimal scheme postulated to explain the results of LSND, and MiniBooNE is the 3+1 framework in which one light sterile neutrino is added to the three active neutrinos in the SM. In the \textit{3+1 oscillation framework}, the mixing matrix $ U $ depends on three additional mixing angles $ \theta_{14},\theta_{24},\theta_{34} $ corresponding to mixing between the light sterile neutrino $ \nu_s$ and the active sector neutrinos, two new CP phases $ \delta_{14},\delta_{34} $ along with the standard oscillation parameters $\theta_{12},\theta_{13},\theta_{23},\delta_{13}$ and can be expressed as,
    \begin{align}\label{eq:U-st}
     	U=\tilde{R}_{34}(\theta_{34},\delta_{34})R_{24}(\theta_{24})\tilde{R}_{14}(\theta_{14},\delta_{14})*\nonumber\\R_{23}(\theta_{23})\tilde{R}_{13}(\theta_{13},\delta_{13})R_{12}(\theta_{12})
    \end{align}
	where $ \tilde{R}_{ij}=U^\delta_{ij}(\delta_{ij})R_{ij}(\theta_{ij})U^{\dagger\delta}_{ij}(\delta_{ij}) $, $R_{ij}(\theta_{ij})$'s are the rotation matrices in i-j plane and $U^\delta_{ij}$ are diagonal unitary matrices with the CP phases $\delta_{ij}$'s. In the presence of an additional light sterile neutrino, there is a new mass squared difference $\Delta_{41}$. The 3+1 picture considered here is $m_4>>m_3>>m_2>>m_1$ which corresponds to $m_4$ being the heaviest mass state. The case with $m_4$ as the lowest state is disfavoured from cosmology. The mass ordering for three generation is considered to be NH.

    Recent studies about the best-fit values and allowed ranges of the parameters associated with eV scale sterile neutrino can be found in \cite{Gariazzo:2017fdh, Dentler:2018sju, Acero:2022wqg}. In particular, the global analysis of data performed in \cite{Gariazzo:2017fdh} illustrates the following  3$\sigma$ bounds and best-fits in sterile mixing angles for $\Delta_{41}=1.3$ eV$^2$,
    \begin{table}[H]
        \centering
        \begin{tabular}{|c|c|c|}
         \hline Parameters & $3\sigma$ range & Best Fit\\  \hline $\sin^2 2\theta_{14}$ & 0.04 - 0.09 & 0.079\\  $\theta_{14}$ & $5.76^\circ-8.73^\circ$ & $8.15^\circ$\\\hline $\sin^2 \theta_{24}$ & $6.7 \times 10^{-3}-  0.022 $ & 0.015\\ $\theta_{24}$ & $4.68^\circ - 8.6^\circ$ & $7.08^\circ$\\ \hline
        \end{tabular}
        \caption{$3\sigma$ Levels and Best fit values extracted from \cite{Gariazzo:2017fdh}}
        \label{tab:3sigma-mixing}
    \end{table}
    
    However, the analysis performed in \cite{MINOS:2020iqj} including the MINOS+ data disfavoured the allowed regions in $\theta_{24}$ from above with a new bound at 90\% C.L. $\sin^2 \theta_{24} \leq 0.006$, i.e., $\theta_{24} \leq 4.5^\circ$. Also, the analysis of DayaBay and Bugey3 gives at 90\% C.L. $\sin^2 2\theta_{14} \leq 0.046$. i.e., $\theta_{14} \leq 6.2^\circ$.	
 
\section{Oscillation Probability}
\label{sec:Prob}
    The effective matter interaction Hamiltonian in flavour basis is given as follows,
	\begin{equation}\label{eq:H-innt}
     \begin{split}
		H_{int}& =diag(V_{CC},0,0,-V_{NC}) \\
      &= diag(\sqrt{2}G_F N_e,0,0,\sqrt{2}G_F N_n/2)
      \end{split}
	\end{equation} where $V_{CC} =\sqrt{2}G_F N_e $ is the charge current interaction potential, $V_{NC} =-\sqrt{2}G_F N_n/2 $ is the neutral current interaction potential, $G_F$ is the Fermi coupling constant, $ N_e$, and $N_n $ correspond to electron density and neutron density, respectively, of the medium in which neutrinos travel. In order to obtain the probabilities in the matter, one has to solve the neutrino propagation equation with the total Hamiltonian given as follows.
    \begin{equation}\label{eq:H-tot-st}
	    H=\frac{1}{2E_\nu}U\begin{bmatrix}
		            0&0&0&0\\0&\Delta_{21}&0&0\\0&0&\Delta_{31}&0\\0&0&0&\Delta_{41}
		            \end{bmatrix}U^\dagger + \frac{1}{2E_\nu}\begin{bmatrix}
		            A&0&0&0\\0&0&0&0\\0&0&0&0\\0&0&0&\frac{A}{2}
		            \end{bmatrix}
	\end{equation}
	where the propagation medium has been considered to be the earth matter with neutron density being equal to electron density, i.e, $N_e=N_n$ and the matter potential term is $A=2\sqrt{2}G_F N_e E_\nu$ with neutrino energy $E_\nu$ and the mass squared differences are given as $\Delta_{ij}=m_j^2-m_i^2$ where $m_i$'s are mass eigenvalues.
    This would require diagonalization of the total Hamiltonian to go to the matter mass basis. However, this poses difficulty even in the three flavour case and one has to resort to approximate methods. A comprehensive review of the various approximations used in the three flavour case has been discussed in \cite{Parke:2019jyu}. There are two well-known methods: (I) OMSD approx., (II) $\alpha-s_{13}$ approx. These two methods can also be used for the 3+1 framework. The $\alpha-s_{13}$ method for the sterile case has been done in ref.\cite{Chattopadhyay:2022ftv}.

    In the context of this work, we have considered the two mass scale dominance(TMSD) approximation with $\Delta_{21}$ set as zero, similar to the well known one mass scale dominance(OMSD) approximation\cite{Banuls:2001zn} in three flavour case. TMSD approximation allows us to obtain compact analytic expressions for the probabilities in the matter, which can facilitate the understanding of the underlying physics in the 3+1 framework. Although we have used the TMSD approx. for analytical calculations, in the numerical analysis, we consider the current best fit value of $\Delta_{21}$.
\subsection{TMSD approximation}
    
    In the TMSD approximation, we choose $ \Delta_{21}=0$ since from the experimental data $\Delta_{21}<<\Delta_{31}<<\Delta_{41}$. As a consequence, the contribution of the solar angle $ \theta_{12} $ drops out of mixing matrix $U$ \eqref{eq:U-st} as $R_{12}$ commutes with the mass matrix $M$ in this approximation. The $ \Delta_{21}=0$ approximation holds well for $\frac{\Delta_{21}L}{E_\nu}<<1$\cite{Banuls:2001zn}. In our study, we further assume $\theta_{34}=0$ which is allowed within current bounds \cite{Capozzi_2020, Esteban:2016qun}. Thus we have only two additional non-zero mixing angles $ \theta_{14},\theta_{24} $ and a non-zero phase $ \delta_{14}$. This leads to the effective vacuum mixing matrix,
    \begin{widetext}
        \begin{equation}\label{eq:U_mix_24}
	\begin{split}
	    \tilde{U}&=R_{24}(\theta_{24})\tilde{R}_{14}(\theta_{14},\delta_{14})R_{23}(\theta_{23})U_{\delta13}R_{13}(\theta_{13}) \\
	    &=\begin{bmatrix}
			c_{13}c_{14}&0&c_{14}s_{13}&e^{-\iota\delta_{14}}s_{14}\\
			-e^{\iota\delta_{13}}c_{24}s_{13}s_{23}-e^{\iota\delta_{14}}c_{13}s_{14}s_{24}&c_{23}c_{24}&e^{\iota\delta_{13}}c_{13}c_{24}s_{23}-e^{\iota\delta_{14}}s_{13}s_{14}s_{24}&c_{14}s_{24}\\
			-e^{\iota\delta_{13}}c_{23}s_{13}&-s_{23}&e^{\iota\delta_{13}}c_{13}c_{23}&0\\
			-e^{\iota\delta_{14}}c_{13}c_{24}s_{14}+e^{\iota\delta_{13}}s_{13}s_{23}s_{24}&-c_{23}s_{24}&-e^{\iota\delta_{14}}c_{24}s_{13}s_{14}-e^{\iota\delta_{13}}c_{13}s_{23}s_{24}&c_{14}c_{24}
		\end{bmatrix}
	\end{split}
	\end{equation}
    \end{widetext}
	where we have used notations $s_{ij}=\sin\theta_{ij},c_{ij}=\cos\theta_{ij}$. Since the allowed values of the vacuum mixing angles $\theta_{13}, \theta_{14}, \text{and }\theta_{24}$ are of a similar order, these small parameters can be expressed in terms of $\mathcal{O}(\lambda^n)$ with $\lambda\sim 0.15$ as follows;
    \begin{align}\label{eq:small-param-lamda}
        \sin\theta_{13}\simeq \mathcal{O}(\lambda), \sin\theta_{14}\simeq \mathcal{O}(\lambda), \sin\theta_{24}\simeq \mathcal{O}(\lambda),\nonumber\\ \Delta_{21}\simeq\mathcal{O}(\lambda^5), \Delta_{31}\simeq\mathcal{O}(\lambda^3), A\simeq\mathcal{O}(\lambda^3)
    \end{align}
    We can split the total Hamiltonian $H$ into two parts as
	\begin{equation}\label{eq:H-tot-split}
	    H=\frac{1}{2E_\nu}(H_0+H_p)
	\end{equation}
	 where $H_p$, the perturbed Hamiltonian, is proportional to the order of $\Delta_{31}, A [\mathcal{O}(\lambda^3)]$ whereas the unperturbed Hamiltonian $H_0$ is proportional to $\Delta_{41}$. These can be written as follows,
	\begin{eqnarray}
		H_0&=&\Delta_{41}\begin{bmatrix}
			s^2_{14}&e^{-\iota\delta_{14}}c_{14}s_{14}s_{24}&0&e^{-\iota\delta_{14}}c_{24}c_{14}s_{14}\\
			e^{\iota\delta_{14}}c_{14}s_{24}s_{14}&c^2_{14}s^2_{24}&0&c^2_{14}c_{24}s_{24}\\
			0&0&0&0\\
			e^{\iota\delta_{14}}c_{24}c_{14}s_{14}&c^2_{14}c_{24}s_{24}&0&c^2_{14}c^2_{24}
		\end{bmatrix}\mathrm{,}\label{eq_H0_3}\\
	    H_p&=&\tilde{U}\begin{bmatrix}
	        0&0&0&0\\0&0&0&0\\0&0&\Delta_{31}&0\\0&0&0&0
	    \end{bmatrix}\tilde{U}^\dagger + 
	    \begin{bmatrix}
		A&0&0&0\\0&0&0&0\\0&0&0&0\\0&0&0&\frac{A}{2}
		\end{bmatrix}\label{eq_Hp}
	\end{eqnarray}
    The unperturbed and perturbed Hamiltonian can be expressed in terms of the small parameter $\lambda$ in the following manner,
	\begin{equation}\label{eq:H0-Hp-lambda}
	    H_0\sim \begin{bmatrix}
	        \lambda^2 & \lambda^2 & 0 & \lambda\\
	        \lambda^2 & \lambda^2 & 0 & \lambda\\
	        0 & 0 & 0 & 0\\
	        \lambda & \lambda & 0 & 1
	    \end{bmatrix}\mathrm{,}
	    H_p\sim\begin{bmatrix}
	    \lambda^5 & \lambda^4 & \lambda^4 & -\lambda^5 \\
	    \lambda^4 &\lambda^3  & \lambda^3 &-\lambda^4 \\
	    \lambda^4 & \lambda^3 & \lambda^3 & -\lambda^4 \\
	    -\lambda^5 & -\lambda^4 & -\lambda^4 & \lambda^5
	    \end{bmatrix}
	\end{equation}
	The unperturbed Hamiltonian has the smallest terms proportional to $\mathcal{O}(\lambda^2)$, which is at least an order less than the largest term in $H_p$, the perturbed Hamiltonian. \textbf{The terms proportional to $\Delta_{21}$ are of higher order than $\lambda^5$ and thus are neglected}. The eigenvalues of $H_0$ are $ \lambda_{01}=0 $, $ \lambda_{02}=0 $, $ \lambda_{03}=0 $, $ \lambda_{04}=\Delta_{41} $. This implies the need for degenerate perturbation theory to determine the modified energy eigenvalues in the presence of the matter potential. The modified energy eigenvalues evaluated using degenerated perturbation theory in ascending order of energy are as follows,
	\begin{equation}\label{E_m_24}
		\begin{split}
			E_{1m}&=\frac{1}{2E_\nu}[\Delta_{31}\sin^2(\theta_{13}-\theta_{13m}) -A^{'} \sin^2\theta_{24}\cos 2\theta_{13m} \\&+ A^{'}\cos^2\theta_{13m}(1+\cos^2\theta_{14}+\cos^2\theta_{14}\sin^2\theta_{24})\\&-A\sin2\theta_{24}\sin\theta_{14}\sin\theta_{23}\sin2\theta_{13m}\cos\delta/2]\mathrm{,}\\
			E_{2m}&=0\mathrm{,}\\
			E_{3m}&=\frac{1}{2E_\nu}[\Delta_{31}\cos^2(\theta_{13}-\theta_{13m}) + A^{'} \sin^2\theta_{24}\cos 2\theta_{13m} \\&+ A^{'}\sin^2\theta_{13m}(1+\cos^2\theta_{14}+\cos^2\theta_{14}\sin^2\theta_{24})\\&+ A\sin2\theta_{24}\sin\theta_{14}\sin\theta_{23}\sin2\theta_{13m}\cos\delta/2]\mathrm{,}\\
			E_{4m}&=\frac{1}{2E_\nu}[\Delta_{41} + A^{'}(1+\sin^2\theta_{14}-\cos^2\theta_{14}\sin^2\theta_{24})]
		\end{split}
	\end{equation} 
	where $A^{'}=A/2=\sqrt{2}G_F N_e$, the modified angle $ \theta_{13m} $ in the matter is related to the original angles, and the new phase $\delta=(\delta_{13}-\delta_{14})$ as,
	\begin{eqnarray}
	    \sin 2\theta_{13m} &= \frac{\Delta_{31}\sin 2\theta_{13}+A^{'} \cos\delta \sin\theta_{14}\sin\theta_{23}\sin 2\theta_{24}}{f} \mathrm{,}\label{sin_2th13m_d}\\
		\cos 2\theta_{13m} &= [\Delta_{31}\cos 2\theta_{13}- A^{'}(1+\cos^2\theta_{14}+\nonumber \\&\cos^2\theta_{14}\sin^2\theta_{24}-2\sin^2\theta_{24})]/f\label{cos_2th13m_d}
	\end{eqnarray}
	where $ f $ is defined as,
	
	\begin{equation}\label{f_24}
	\begin{split}
	    f&=\left([\Delta_{31}\sin 2\theta_{13}+A^{'} s_{14}s_{23}\sin 2\theta_{24}\cos\delta]^2 \right.\\&\left.+ [\Delta_{31}\cos 2\theta_{13}-A^{'}(1+c^2_{14}+c^2_{14}s^2_{24}-2 s^2_{24})]^2 \right)^{1/2}
	\end{split}
	\end{equation}
	It is noteworthy that for the 3+1 framework, the modified angle $\theta_{13m}$ depends on cp phases, unlike in the three generation framework. Now if we put $\sin 2\theta_{13m}=1$, i.e., $\cos 2\theta_{13m}=0$, we will get maximum $\theta_{13m}$, i.e., resonance in this sector for the matter. The corresponding resonance energy is given by,
	\begin{equation}\label{eq:e-res}
	    E_{res}=\frac{\Delta_{31}\cos 2\theta_{13}}{\sqrt{2}G_F N_e(1+\cos^2\theta_{14}+\cos^2\theta_{14}\sin^2\theta_{24}-2\sin^2\theta_{24})}
	\end{equation}
   The resonance energy for 1300 km and 7000 km are $\sim 11$ GeV, and $8$ GeV respectively corresponding to $\theta_{14}=\theta_{24}=7^\circ, \theta_{13}=8.57^\circ, \Delta_{31}=2.515\times10^{-3}\mathrm{eV}^2$. It only changes minimally from the three generation case. The modified active-sterile mixing angles $ \theta_{14m},\theta_{24m} $ are related to the vacuum angles as,
	
	\begin{gather}
		\sin\theta_{14m} = \sin\theta_{14}[1+\frac{A^{'}}{\Delta_{41}}\cos^2\theta_{14}(1+s^2_{24})]\mathrm{,}\\
		\cos\theta_{14m} = \cos\theta_{14}[1-\frac{A^{'}}{\Delta_{41}}\sin^2\theta_{14}(1+s^2_{24})]\mathrm{,}\label{th14m}\\
		\sin\theta_{24m} = \sin\theta_{24}[1-\frac{A^{'}}{\Delta_{41}}\cos^2\theta_{14}\cos^2\theta_{24}]\mathrm{,}\\
	\cos\theta_{24m} = \cos\theta_{24}[1+\frac{A^{'}}{\Delta_{41}}\cos^2\theta_{14}\sin^2\theta_{24}]\label{th24m}
	\end{gather}
	The mixing matrix in matter obtained from the modified eigenvectors using degenerate perturbation theory is as follows,
    \begin{widetext}
        \scriptsize 
	\begin{equation}\label{eq:U_mix_24-mod}
	\begin{split}
	     &\tilde{U}_m=R_{24}^m(\theta_{24m})\tilde{R}_{14}^m(\theta_{14m},\delta_{14})R_{23}(\theta_{23})U_{\delta13}R_{13}^m(\theta_{13m})R_{12}^m(\theta_{12m})\\
	     &= \begin{bmatrix}
			c_{13m}c_{14m}&(U_m)_{12}&c_{14m}s_{13m}&e^{-\iota\delta_{14}}s_{14m}\\
			-e^{\iota\delta_{13}}c_{24m}s_{13m}s_{23}-e^{\iota\delta_{14}}c_{13m}s_{14m}s_{24m}&c_{23}c_{24m}&e^{\iota\delta_{13}}c_{13m}c_{24m}s_{23}-e^{\iota\delta_{14}}s_{13m}s_{14m}s_{24m}&c_{14m}s_{24m}\\
			-e^{\iota\delta_{13}}c_{23}s_{13m}&-s_{23}&e^{\iota\delta_{13}}c_{13m}c_{23}&0\\
			-e^{\iota\delta_{14}}c_{13m}c_{24m}s_{14m}+e^{\iota\delta_{13}}s_{13m}s_{23}s_{24m}&-c_{23}s_{24m}&-e^{\iota\delta_{14}}c_{24m}s_{13m}s_{14m}-e^{\iota\delta_{13}}c_{13m}s_{23}s_{24m}&c_{14m}c_{24m}
		\end{bmatrix}
	\end{split}
	\end{equation}
    \end{widetext}
    \normalsize
	where the original vacuum angles are replaced by modified angles as given by \eqref{sin_2th13m_d}, \eqref{cos_2th13m_d}, \eqref{th14m}, \eqref{th24m} and null value of the element $(\tilde{U})_{12}$ in vacuum mixing matrix $\tilde{U}$\eqref{eq:U_mix_24} is modified as $(U_m)_{12}=\frac{A}{\Delta_{41}}e^{-\iota\delta_{14}}c_{14}c_{23}c_{24}s_{14}s_{24} \sim \mathcal{O}(\lambda^5)$. This is due to the fact that the matter effect introduces correction of mixing angle $\theta_{12}$ which was absent before due to the approximation $\Delta_{21}=0$. The other terms related to $\theta_{12}$ don't show up as they are $>\mathcal{O}(\lambda^5)$. Now we can calculate the oscillation(survival) probabilities using the elements of $\tilde{U}_m $ in place of $ U $ and $ \Delta_{ij}^m=2E_\nu(E_{im}-E_{jm}) $ replacing $ \Delta_{ij} $ in \eqref{eq:Prob-N} assuming constant matter density,
	\begin{align}\label{eq:Prob-N}
 			P_{\alpha\beta} &=  \delta_{\alpha \beta} - 4\sum_{i>j}^{N}Re(U_{\alpha i}^\ast U_{\beta i} U_{\alpha j} U_{\beta j}^\ast)\sin^2\frac{1.27\Delta_{ij}L}{E_\nu}\nonumber
 			\\& + 2\sum_{i>j}^{N}Im(U_{\alpha i}^\ast U_{\beta i} U_{\alpha j} U_{\beta j}^\ast)\sin 2 \frac{1.27\Delta_{ij}L}{E_\nu}
 	\end{align}
 	On the other hand, the exact numerical probability at constant matter density can be evaluated as,
 	\begin{equation}\label{eq:Prob-exact}
 	    P_{\alpha\beta}^\mathrm{num}=|[e^{-\iota H L}]_{\alpha \beta}|^2 ,
 	\end{equation}
  where $H$ is the total Hamiltonian without any approximation given by \eqref{eq:H-tot-st}.
\subsubsection{$P_{\mu e}$ Channel}
\begin{figure}[H]
		\centering
		\includegraphics[width=0.47\linewidth]{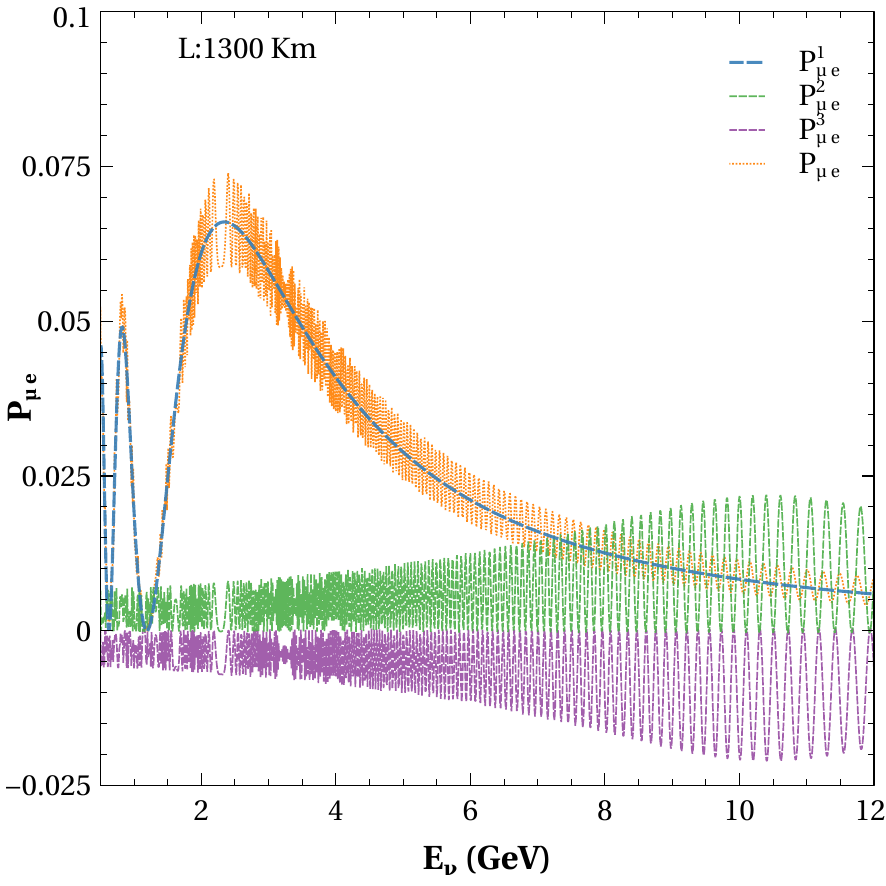}
		\includegraphics[width=0.47\linewidth]{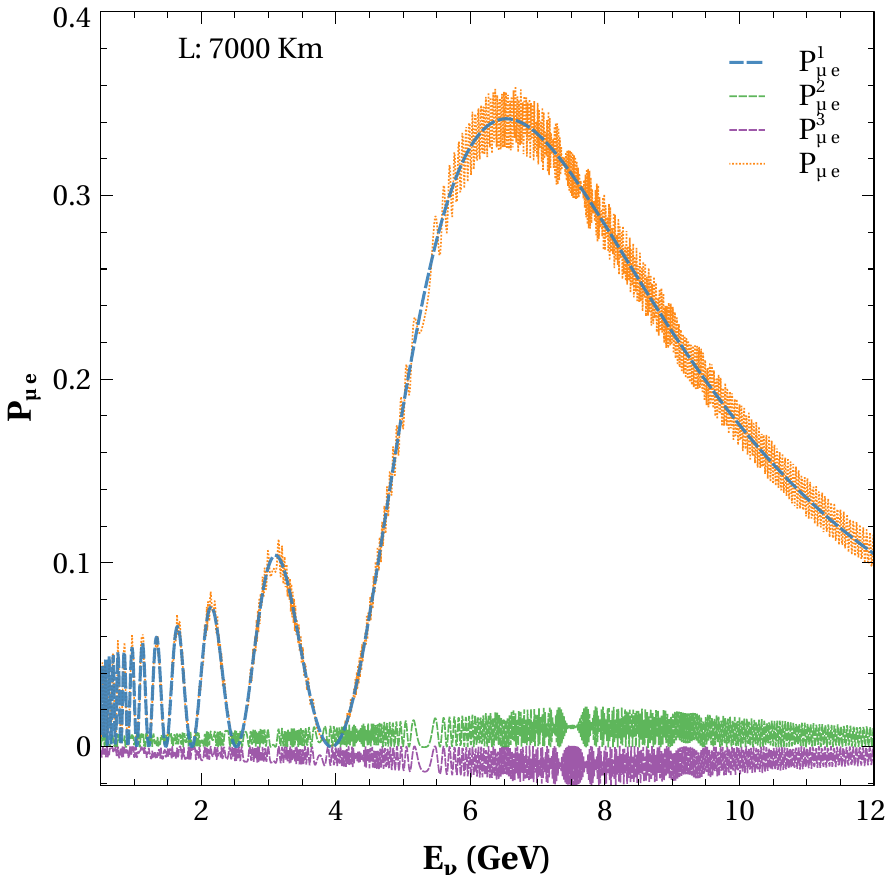}
		\includegraphics[width=0.47\linewidth]{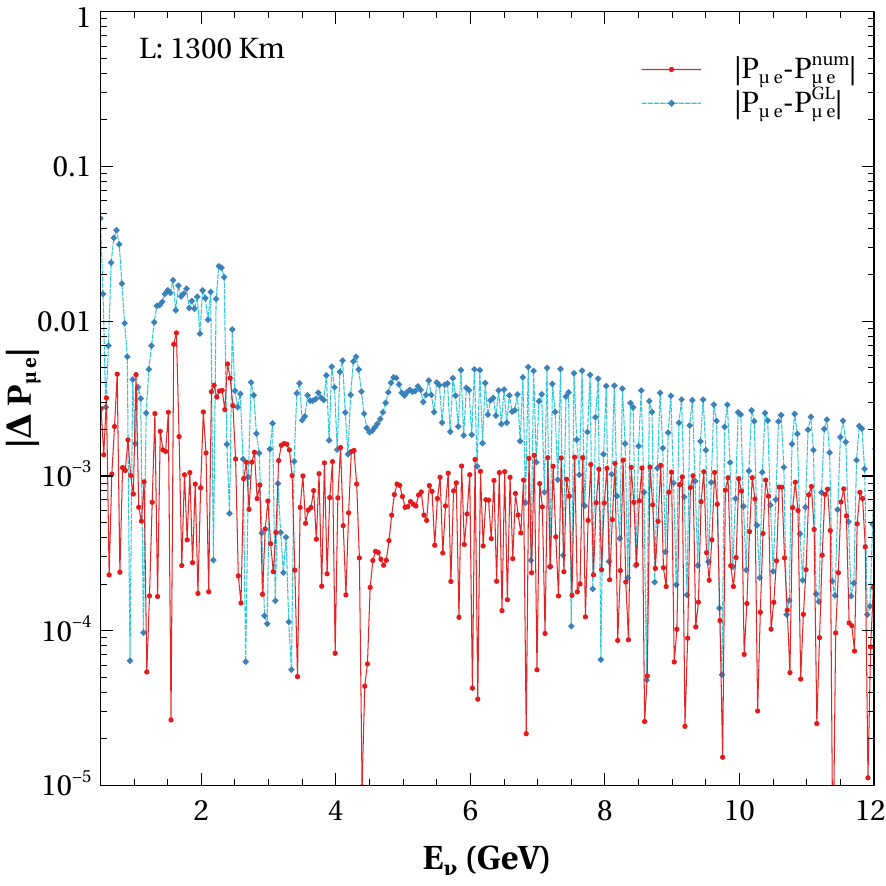}
		\includegraphics[width=0.47\linewidth]{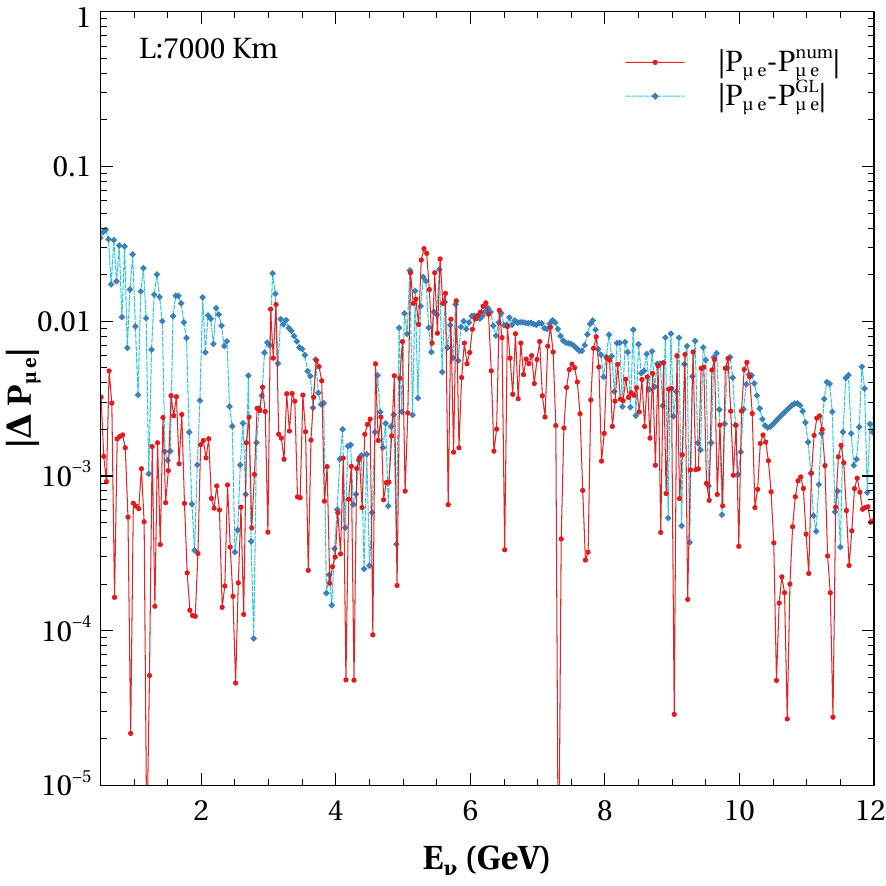}
		\caption{The total analytic probability $P_{\mu e}$ (orange) along with its other dominant terms in the top panel and the absolute differences $|P_{\mu e}-P_{\mu e}^\mathrm{num}|$ (red) and $|P_{\mu e}-P_{\mu e}^\mathrm{GL}|$ (cyan) in the bottom panel at 1300 km(left), and 7000 km(right) baselines.}
		\label{fig:pme_terms}
	\end{figure}

	The appearance channel, i.e., $ \nu_\mu \rightarrow \nu_e $ oscillation probability is given by,
	\begin{equation}\label{eq:pme_st}
	    P_{\mu e}=P_{\mu e}^1 + P_{\mu e}^2 + P_{\mu e}^3 + \mathcal{O}(\lambda^6)
	\end{equation}
	where the different significant terms of the probability $P_{\mu e}$ are as follows,
    \begin{widetext}
        \begin{align}
	    \begin{split}
			P_{\mu e}^1&=4\cos^2\theta_{13m}\cos^2\theta_{14m}\sin^2\theta_{13m}(\cos^2\theta_{24m}\sin^2\theta_{23}-\sin^2\theta_{14m}\sin^2\theta_{24m})\sin^2\frac{1.27\Delta_{31}^m L}{E}\\
			&+2\cos^3\theta_{13m}\cos^2\theta_{14m}\sin\theta_{13m}\sin\theta_{14m}\sin 2\theta_{24m}\sin\theta_{23}\sin{\frac{1.27\Delta_{31}^m L}{E}}\sin(\frac{1.27\Delta_{31}^m L}{E}+\delta)\\
			&-2\cos\theta_{13m}\cos^2\theta_{14m}\sin^3\theta_{13m}\sin\theta_{14m}\sin 2\theta_{24m}\sin\theta_{23}\sin{\frac{1.27\Delta_{31}^m L}{E}}\sin(\frac{1.27\Delta_{31}^m L}{E}-\delta)\mathrm{,}
		\end{split}\label{Pme24_1}\\
	    \begin{split}
	        P_{\mu e}^2 &=\cos^2\theta_{14m}\sin 2\theta_{13m}\sin\theta_{14m}\sin\theta_{23}\sin 2\theta_{24m}\sin \frac{1.27\Delta_{41}^m L}{E} \sin(\frac{1.27\Delta_{41}^m L}{E} - \delta)\\ &+\sin^2 2\theta_{14m}\sin^2\theta_{24m}\cos^2\theta_{13m}\sin^2 \frac{1.27\Delta_{41}^m L}{E}\mathrm{,}
	    \end{split}\label{Pme24_2}\\
	    \begin{split}
	        P_{\mu e}^3 &= -\cos^2\theta_{14m}\sin 2\theta_{13m}\sin\theta_{14m}\sin\theta_{23}\sin 2\theta_{24m}\sin \frac{1.27\Delta_{43}^m L}{E} \sin(\frac{1.27\Delta_{43}^m L}{E}- \delta)\\
	        &+\sin^2 2\theta_{14m}\sin^2\theta_{24m}\sin^2\theta_{13m}\sin^2 \frac{1.27\Delta_{43}^m L}{E}
	    \end{split}\label{Pme24_3}
	\end{align}
    \end{widetext}
    The total analytic probability $P_{\mu e}$(orange) and the dominant terms contributing to it are plotted at 1300 km and 7000 km baselines as a function of neutrino energy $E_\nu$ in the top panel of \autoref{fig:pme_terms}. For the plots, and calculations of $P_{\mu e}, P_{\mu \mu}$ in this section, we have considered $\theta_{12}=33.44^\circ,\theta_{13}=8.57^\circ,\theta_{23}=49^\circ,\theta_{14}=\theta_{24}=7^\circ,\delta_{13}=195^\circ,\delta_{14}=30^\circ,\Delta_{31}=2.515\times10^{-3}$eV$^2$, and $\Delta_{41}=1$eV$^2$.
    The analytic expression of $P_{\mu e}$ consists of three significant terms, although there are other higher order terms [$\mathcal{O}(\lambda^6)$] that are neglected. The first term in \eqref{Pme24_1}(blue curve) which is proportional to the modified mass squared difference $\Delta_{31}^m$, is the most dominant one and provides the average curve of the total probability as seen in \autoref{fig:pme_terms}. The fast oscillations seen \autoref{fig:pme_terms} are a manifestation of the terms in \eqref{Pme24_2} (green curve), \eqref{Pme24_3} (violet curve) which are proportional to the modified mass squared differences related to the sterile neutrino mass states $\Delta_{41}^m, \Delta_{43}^m$ respectively. The fast oscillations are not reflected in experiments, as we can only get the average probability. Also, these terms are relatively much smaller than the $P_{\mu e}^1$ around probability maxima, so in the next section, while discussing the degeneracies, we will only use the term $P_{\mu e}^1$. Putting $\theta_{14},\theta_{24}$ angles to zero in equations \eqref{Pme24_1}, \eqref{Pme24_2}, \eqref{Pme24_3} gives the standard three flavour oscillation probability from the very first term of the \eqref{Pme24_1} as the other terms go to zero due to presence of $\sin\theta_{14m},\sin\theta_{24m}$.\\

	We have shown the comparison of the absolute differences $|\Delta P|$ of the analytic probability $P_{\mu e}$ \eqref{eq:pme_st} with the exact probability $P_{\mu e}^\mathrm{num}$\eqref{eq:Prob-exact} (red) as well as with the probability $P_{\mu e}^\mathrm{GL}$ (cyan) obtained using GLoBES\cite{Huber:2004ka} as a function of neutrino energy at the bottom panel in \autoref{fig:pme_terms}. We can see the value of $|\Delta P|$ is around $10^{-3}$ for most of the energies. $|\Delta P|$ values are smaller around the resonance energy of $11$ GeV for 1300 km and $8$ GeV for 7000 km. Also, we can see the energies at which the value of probability is smaller we get smaller values of $|\Delta P|$. Overall we can conclude that the analytic probability $P_{\mu e}$ using TMSD approximation is in good agreement with both numerical and exact ones, better with the exact one $P_{\mu e}^\mathrm{num}$ for all energies ($>0.5$ GeV), especially around the resonance. This is similar to probabilities derived using OMSD approximation matching well with numerical ones in the standard three flavour case in the region with significant matter effect\cite{Gandhi:2007td, Choubey:2005zy}.
 
\subsubsection{$P_{\mu\mu}$ Channel}	
\begin{figure}[H]
		\centering
		\includegraphics[width=0.48\linewidth]{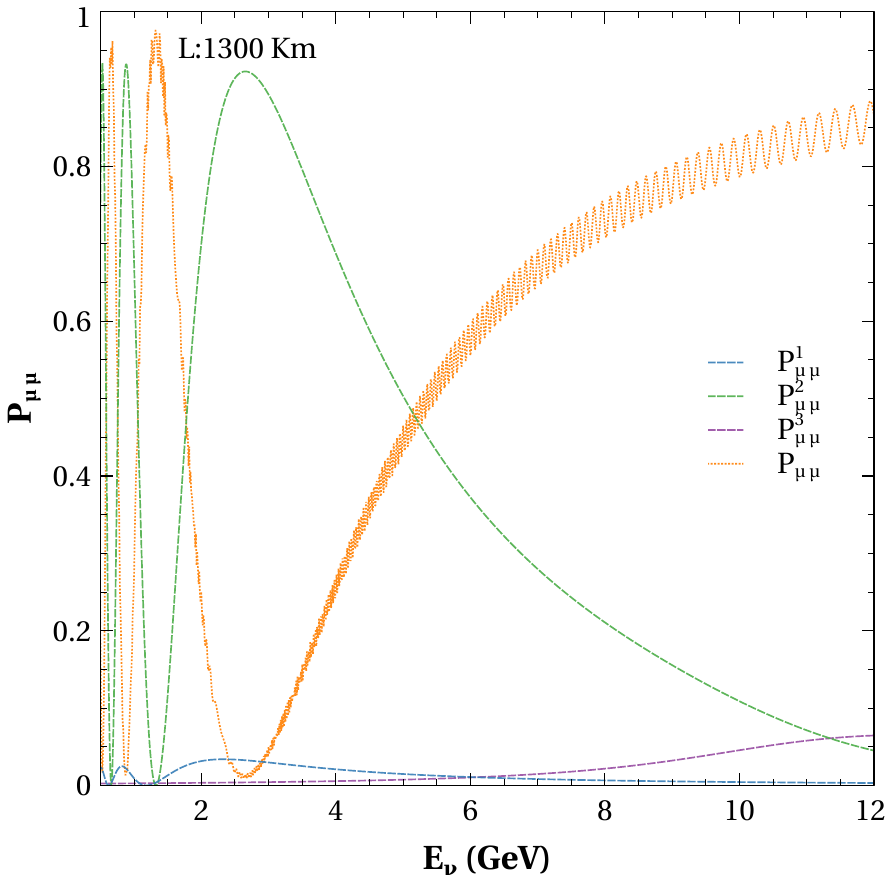}
		\includegraphics[width=0.48\linewidth]{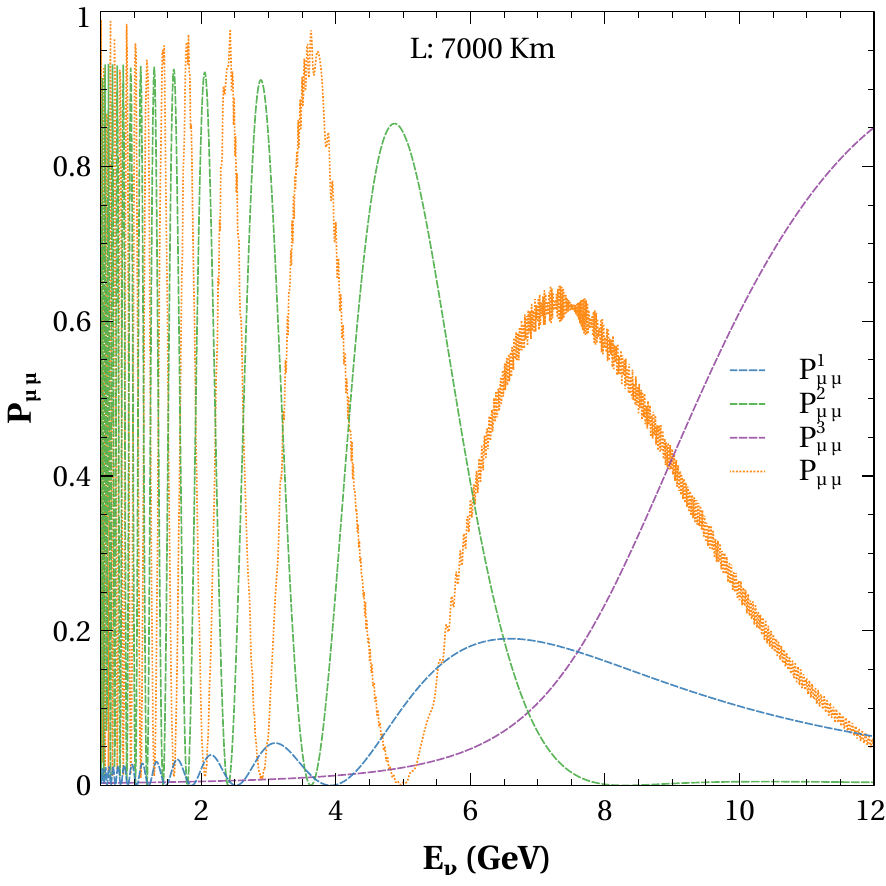}
		\includegraphics[width=0.48\linewidth]{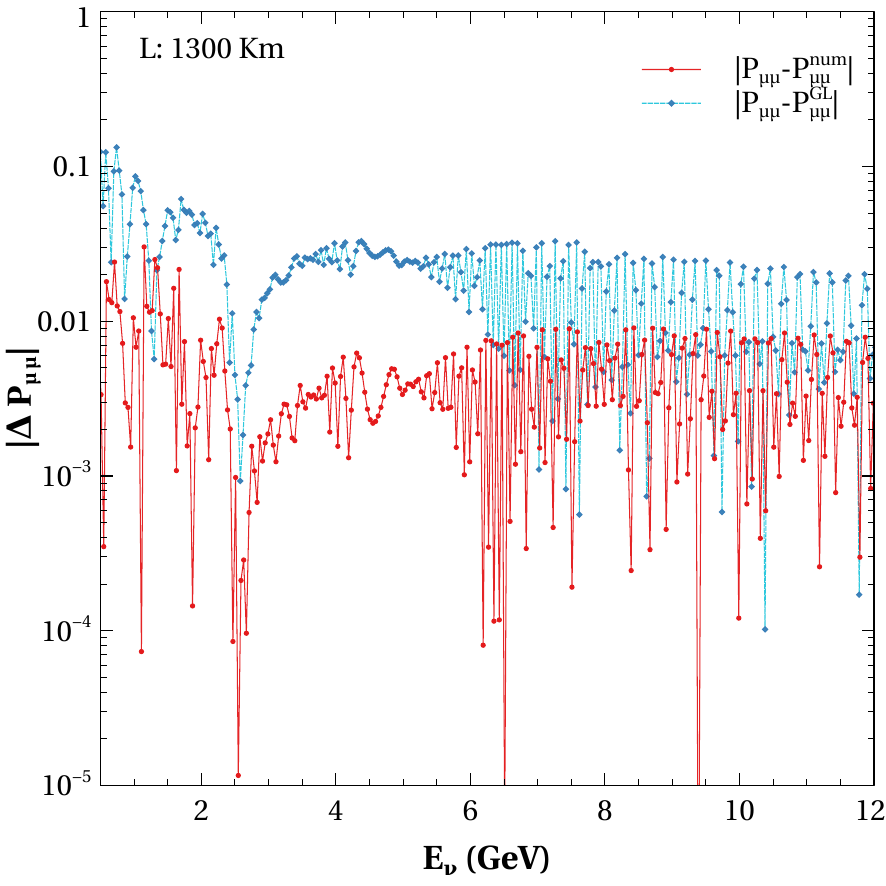}
		\includegraphics[width=0.48\linewidth]{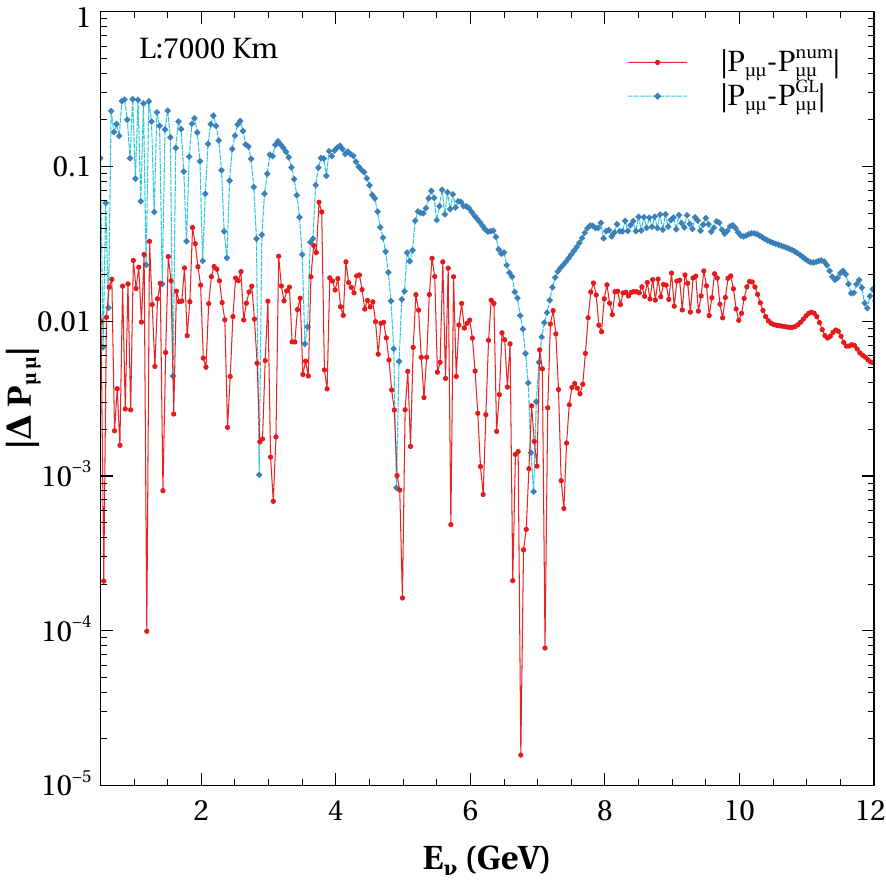}
		\caption{The total analytic probability $P_{\mu \mu}$ (orange) along with its other dominant terms in the top panel and the absolute differences $|P_{\mu\mu}-P_{\mu\mu}^\mathrm{num}|$ (red) and $|P_{\mu\mu}-P_{\mu\mu}^\mathrm{GL}|$ (cyan) in the bottom panel at 1300 km(left), and 7000 km(right).}
		\label{fig:pmm_terms}
	\end{figure}

    The disappearance channel, i.e., $ \nu_\mu \rightarrow \nu_\mu $ survival probability is given by,
    \begin{equation}\label{eq:Pmm-st}
        P_{\mu\mu}=1-P_{\mu\mu}^1 - P_{\mu\mu}^2 - P_{\mu\mu}^3 + \mathcal{O}(\lambda^6)
    \end{equation}
    where the significant terms of the probability are as follows,
    \begin{widetext}
            \begin{align}
	    \begin{split}
	    	P_{\mu\mu}^1&= \cos^4\theta_{24m}\sin^2 2\theta_{13m}\sin^4\theta_{23}\sin^2 \frac{1.27\Delta_{31}^m L}{E}+\sin^4\theta_{24m}\sin^4\theta_{14m}\sin^2 2\theta_{13m}\sin^2 \frac{1.27\Delta_{31}^m L}{E}\\& +\sin 2\theta_{24m}\sin\theta_{14m}\sin 4\theta_{13m}\sin\theta_{23} \cos\delta(\cos^2\theta_{24m}\sin^2\theta_{23}-\sin^2\theta_{24m}\sin^2\theta_{13m}) \sin^2 \frac{1.27\Delta_{31}^m L}{E}\\
	    	&+4\cos^2\theta_{24m} \sin^2\theta_{24m} \sin^2\theta_{14m}\sin^2\theta_{23}(1-\frac{\sin^2 2\theta_{13m}}{2}- \sin^2 2\theta_{13m}\cos^2\delta) \sin^2 \frac{1.27\Delta_{31}^m L}{E}\mathrm{,}
    	\end{split}\label{Pmm24_1}\\
		\begin{split}
			P_{\mu\mu}^2&=\cos^4\theta_{24m} \cos^2\theta_{13m}\sin^2 2\theta_{23}\sin^2 \frac{1.27\Delta_{32}^m L}{E}\\&+ 4 \cos^2\theta_{24m}\sin^2\theta_{24m}\sin^2\theta_{14m}\sin^2\theta_{13m}\cos^2\theta_{23}\sin^2 \frac{1.27\Delta_{32}^m L}{E}\\
			&-4 \cos^3\theta_{24m} \sin\theta_{24m} \sin\theta_{14m} \sin 2\theta_{13m} \cos^2\theta_{23} \sin\theta_{23} \cos\delta \sin^2 \frac{1.27\Delta_{32}^m L}{E}\mathrm{,}
		\end{split}\label{Pmm24_2}\\
		\begin{split}
			P_{\mu\mu}^3&=\cos^4\theta_{24m}\sin^2\theta_{13m}\sin^2 2\theta_{23}\sin^2 \frac{1.27\Delta_{21}^m L}{E} \\&+ 4 \cos^2\theta_{24m}\sin^2\theta_{24m}\sin^2\theta_{14m}\cos^2\theta_{13m}\cos^2\theta_{23} \sin^2 \frac{1.27\Delta_{21}^m L}{E}\\
			&+ 4 \cos^3\theta_{24m} \sin\theta_{24m} \sin\theta_{14m} \sin 2\theta_{13m}\cos^2\theta_{23} \sin\theta_{23} \cos\delta \sin^2 \frac{1.27\Delta_{21}^m L}{E}
		\end{split}\label{Pmm24_3}
	\end{align}
    \end{widetext}
    We show the total analytic probability $P_{\mu \mu}$ (orange) and the different terms contributing significantly to it at 1300 km and 7000 km baselines in the top panel of \autoref{fig:pmm_terms} as a function of neutrino energy.
    The analytic expression of $P_{\mu\mu}$ consists of three significant terms \eqref{Pmm24_1}, \eqref{Pmm24_2}, and \eqref{Pmm24_3}, although there are three other fast oscillating terms that are neglected. Here, the fast oscillating terms are proportional to the sterile mass squared differences $\Delta_{41}^m,\Delta_{42}^m,\Delta_{43}^m$ and are of higher orders [$\mathcal{O}(\lambda^6)$]. The first term in \eqref{Pmm24_1} (blue curve), which is proportional to the modified mass squared difference $\Delta_{31}^m$, has a dependence on octant of $\theta_{23}$ in the leading order due to the presence of $\sin^4 \theta_{23} $. $P_{\mu\mu}^1$ grows with energy initially and decreases after resonance energy. The second, and third terms in \eqref{Pmm24_2} (green curve), \eqref{Pmm24_3} (violet curve) which are proportional to the modified mass squared differences $\Delta_{32}^m, \Delta_{21}^m$ respectively, show no octant dependence in the leading order due to the presence of $\sin^2 2\theta_{23}$. The second term is the most dominant one before resonance energy but almost becomes zero after resonance energy, whereas the third term only grows after the resonance energy. In the case of 7000 km at oscillation maxima of 7.5 GeV, $P_{\mu \mu}^1,P_{\mu \mu}^2,P_{\mu \mu}^3$ all have significant contributions. Putting the $\theta_{14},\theta_{24}$ angles to zero, we will get back the three flavour oscillation probability from the first term of the equations \eqref{Pmm24_1}, \eqref{Pmm24_2}, and \eqref{Pmm24_3}.\\

    It has also been shown in the bottom panel of \autoref{fig:pmm_terms}, the absolute differences $|\Delta P|$ of the analytical probability $P_{\mu \mu}$\eqref{eq:Pmm-st} with the exact probability $P_{\mu\mu}^\mathrm{num}$\eqref{eq:Prob-exact} (red) and the probability $P_{\mu\mu}^\mathrm{GL}$ (cyan) obtained using GLoBES at 1300 km and 7000 km baselines. We observe that value of $|\Delta P|$ is mostly around $10^{-3}$. The $|\Delta P|$ values are seen to be lower around resonance energies. We can also see the $|\Delta P|$ value going down at the minima or at the regions where the value of probability is less. The $|\Delta P_{\mu\mu}|$ for 7000 km is increasing after resonance energy as the dominant term in those energies is $P_{\mu\mu}^3$ that is proportional to $\Delta_{21m}$ and hence is affected by the $\Delta_{21}=0$ approximation\footnote{In the appendix we have shown that with non-zero $\Delta_{21}$ in the Cayley Hamilton method we get better fit at these regions as well as at very low energies.}. Hence, we can conclude that the analytical probability $P_{\mu\mu}$ using TMSD approximation is in agreement with exact and numerical probabilities to a good extent, matching better with the exact one $P_{\mu\mu}^\mathrm{num}$.
	
\section{Octant Degeneracy}
\label{sec:degeneracy}
    The degeneracy in the determination of the octant of $\theta_{23}$ can arise from both the survival/oscillation probabilities as follows:
    \begin{itemize}
        \item When the probability is a function of $sin^{2}{2\theta_{23}}$, it is not possible to differentiate between the probabilities arising due to $\theta_{23}$ and $\frac{\pi}{2} - \theta_{23}$. This is called intrinsic octant degeneracy\cite{Fogli:1993ck}.
        
        \item When the probability is a function of $sin^{2}{\theta_{23}}$ or $cos^{2}{\theta_{23}}$, the degeneracy of the octant arises due to the uncertainties in the Dirac CP phase $\delta_{CP}$.
        \begin{equation}
            P(\theta_{23}^{\rm{right}},\delta_{13})=P(\theta_{23}^{\rm{wrong}},\delta_{13}^{'})
        \end{equation}
        
        \item Addition of a light sterile neutrino brings an extra phase $\delta_{14}$ which will also affect the determination of octant just like in the above case through additional degeneracies.
        \begin{equation}
            P(\theta_{23}^{\rm{right}},\delta_{14})=P(\theta_{23}^{\rm{wrong}},\delta_{14}^{'})
        \end{equation}
        
        \item Considering known hierarchy and two unknown phases, there will be a new 8 fold \textit{octant-$\delta_{13}$-$\delta_{14}$} degeneracy.
        \begin{equation}
            P(\theta_{23}^{\rm{right}},\delta_{13},\delta_{14})=P(\theta_{23}^{\rm{wrong}},\delta_{13}^{'},\delta_{14}^{'})
        \end{equation}
    \end{itemize}
	
	We consider the normal hierarchy ($\Delta_{31}=2.515\times 10^{-3}$ eV$^2$) for our octant degeneracy study. Therefore, we have a 8-fold \textit{octant-$\delta_{13}$-$\delta_{14}$} degeneracy in presence of a sterile neutrino as depicted in \autoref{tab:8-deg}. For unknown hierarchy, this will become a 16-fold degeneracy.\\
	\begin{table}[h!]
	    \centering
	    \begin{tabular}{c|c}
            \hline Solution with right octant & Solution with wrong octant\\
	         \hline RO-R$\delta_{13}$-R$\delta_{14}$ &  WO-R$\delta_{13}$-R$\delta_{14}$\\
	         RO-R$\delta_{13}$-W$\delta_{14}$ &  WO-R$\delta_{13}$-W$\delta_{14}$\\
	         RO-W$\delta_{13}$-R$\delta_{14}$ &  WO-W$\delta_{13}$-R$\delta_{14}$\\
	         RO-W$\delta_{13}$-W$\delta_{14}$ &  WO-W$\delta_{13}$-W$\delta_{14}$\\\hline
	    \end{tabular}
	    \caption{New degeneracies in presence of unknown octant and phases with fixed hierarchy.}
	    \label{tab:8-deg}
	\end{table}
	
	In order to understand the degeneracy analytically, we follow the method outlined in \cite{Agarwalla:2016xlg} and use the TMSD probabilities derived in the earlier section. The current $3\sigma$ range of $\theta_{23}$ is $ [39.7^\circ,50.9^\circ] $ \cite{Esteban:2016qun} for normal hierarchy. We can express $\theta_{23}$ w.r.t. $\pi/4$ as,
	\begin{equation}\label{eq:th23_eta}
		\theta_{23}=\frac{\pi}{4}\pm \eta
	\end{equation}
	where the deviation in value of $ \theta_{23} $ from current global analysis fit is given by $ \eta\sim 0.1 $ with the plus and minus sign in \eqref{eq:th23_eta} indicating higher octant(HO), and lower octant (LO) of $ \theta_{23} $ respectively. The octant sensitivity will be there if there is a difference between probabilities of the two opposite octants even when the phases $ \delta_{13},\delta_{14} $ vary in the range $[-\pi,\pi] $. The octant sensitivity from the appearance channel probability $P_{\mu e}$ is defined as,
	\begin{equation}\label{eq:delP_oct}
		\Delta P_{oct,1} \equiv P_{\mu e}^{1HO}(\delta_{13}^{HO},\delta_{14}^{HO})- P_{\mu e}^{1LO}(\delta_{13}^{LO},\delta_{14}^{LO})> 0
	\end{equation}
	As $ \eta $ is small, we can have the following expansion
	\begin{equation}\label{eq:sin-cos-eta}
		\sin^2\theta_{23}\simeq\frac{1}{2}\pm\eta,\sin\theta_{23}\simeq\frac{1}{\sqrt{2}}(1\pm\eta),\cos\theta_{23}\simeq\frac{1}{\sqrt{2}}(1\mp\eta)
	\end{equation}
	Putting $P_{\mu e}^1$ from \eqref{Pme24_1} in \eqref{eq:delP_oct} and using the above expressions of \eqref{eq:sin-cos-eta}, we get three contributions to $\Delta P_{oct,1}$ corresponding to the three terms in $P_{\mu e}^1$,
	\begin{equation}\label{eq:Del_P0-P1-P2}
		\begin{split}
			\Delta P_0 &=8\eta\cos^2\theta_{13m}\cos^2\theta_{14m}\cos^2\theta_{24m}\sin^2\theta_{13m}\sin^2D_{31}^m\mathrm{,}\\
			\Delta P_1 &=X_1[\sin(D_{31}^m+\delta^{HO})-\sin(D_{31}^m+\delta^{LO})]\\&+ \eta X_1[\sin(D_{31}^m+\delta^{HO})+\sin(D_{31}^m+\delta^{LO})]\mathrm{,}\\
			\Delta P_2 &=-Y_1[\sin(D_{31}^m-\delta^{HO})-\sin(D_{31}^m-\delta^{LO})]\\&- \eta Y_1[\sin(D_{31}^m-\delta^{HO})+\sin(D_{31}^m-\delta^{LO})]
		\end{split}
	\end{equation}
        The contribution of the fast oscillation terms $P_{\mu e}^2,P_{\mu e}^3$ to the octant sensitivity is,
        \begin{align}\label{eq:Del_P-fast}
            \Delta P_{fast}&=\sum_{k=1,3} Z_{k}[\sin(D_{4k}^m-\delta^{HO})-\sin(D_{4k}^m-\delta^{LO})]\nonumber \\&+ \eta Z_{k}[\sin(D_{4k}^m-\delta^{HO})+\sin(D_{4k}^m-\delta^{LO})]
        \end{align}
	Where $D_{ij}^m=\frac{1.27 \Delta_{ij}^m}{E}$. Now we can rewrite \eqref{eq:delP_oct} for octant sensitivity as,
	\begin{equation}\label{del_P}
		\Delta P_{oct,1} = \Delta P_0 + \Delta P_1 + \Delta P_2+ \Delta P_{fast}
	\end{equation}
	Among the terms of $\Delta P_{oct,1}$ \eqref{eq:Del_P0-P1-P2}, $\Delta P_0$ has no dependence on phase and is positive whereas the values of $ \Delta P_1, \Delta P_2, \Delta P_{fast} $ can be both positive and negative as they contain phases. Thus degeneracy can occur when $ \Delta P_1+\Delta P_2+\Delta P_{fast} $ is negative and is of the same order as $ \Delta P_0 $, making $ \Delta P $ zero. $ X_1,Y_1$ the positive definite amplitudes of $\Delta P_1$, $\Delta P_2$ respectively as well as the amplitudes $Z_1,Z_3$ of $\Delta P_{fast}$ are as follows,
	\begin{equation}\label{eq:X1-Y1}
		\begin{split}
            X_1&=\sqrt{2}\cos^3\theta_{13m}\cos^2\theta_{14m}\sin\theta_{13m}\sin\theta_{14m}\sin 2\theta_{24m}\sin D_{31}^m\mathrm{,}\\
            Y_1&=\sqrt{2}\cos\theta_{13m}\cos^2\theta_{14m}\sin^3\theta_{13m}\sin\theta_{14m}\sin 2\theta_{24m}\sin D_{31}^m\mathrm{,}\\
            Z_1&=\cos^2\theta_{14m}\sin 2\theta_{13m}\sin\theta_{14m}\sin 2\theta_{24m}\sin D_{41}^m/\sqrt{2}\mathrm{,}\\
            Z_{3}&=-\cos^2\theta_{14m}\sin 2\theta_{13m}\sin\theta_{14m}\sin 2\theta_{24m}\sin D_{43}^m/\sqrt{2}\\
		\end{split}
	\end{equation}
	Now, if we inspect the possibility of the octant degeneracy through the probabilities at a baseline of 1300 km. We use the following values of the oscillation parameters: $\theta_{12}=33.47^\circ, \theta_{13}=8.54^\circ, \theta_{14}=7^\circ, \theta_{24}=7^\circ, \Delta_{31}=2.515\times 10^{-3}\rm{eV^2}, \Delta_{41}=1\rm{eV^2}$. For 1300 km at oscillation maxima of $2.5$ GeV, the values of various terms of $\Delta P_{oct,1}$ are,
	\begin{align}\label{del_p-me-1300-cal}
	  \Delta P_0=0.0279,X_1=0.0073,Y_1=0.0003,\nonumber\\
        Z_1=-0.0056 , Z_3=0.0064 
	\end{align}
	Therefore, $\Delta P_2$ is negligible compared to $\Delta P_0$, $\Delta P_{fast}$, and $\Delta P_1$ due to presence of extra $\sin^2\theta_{13m}$ in $Y_1$\eqref{eq:X1-Y1}. It can be seen from \eqref{eq:Del_P0-P1-P2} the square bracketed terms multiplying $X_1, Z_1, Z_3$ can vary from $-2:+2$ and therefore, for certain phase choices, cancellation can occur resulting in loss of octant sensitivity in 1300 km baseline when fast oscillations considered. However, in the absence of fast oscillations, there is octant sensitivity.
	
	Next, we use the analytic expressions in \eqref{del_P} to understand the octant sensitivity at 7000 km. In the case of 7000 km at oscillation maxima of $E=6.5$ GeV, the values of the different terms contributing to $ \Delta P $ are,
	\begin{align}\label{del_p-me-7k-cal}
	    \Delta P_0=0.1453,X_1=0.0133,Y_1=0.0040,\nonumber\\
            Z_1=0.0001, Z_3=-0.0164 
	\end{align}
	It shows that $ \Delta P_0, X_1, Z_3$ are the dominant contributions, and any combination of phases can not make $ \Delta P_{oct,1}=0 $ as the value of $ P_0 $ is one order greater than $ X_1 $. It shows probabilities $ (P_{\mu e}) $ corresponding to two different octants will always be well separated from each other, i.e., the octant-$ \delta_{13} $-$ \delta_{14}$ degeneracy will be removed. This suggests unlike in 1300 km here, even with the variation of phases in both octants, we can have significant octant sensitivity at higher baselines. This is mainly because $\Delta P_0$ has much higher values than others at the higher baselines due to higher matter effects. Note that if the values of $\theta_{14},\theta_{24}$ are decreased, the dominant contribution, $\Delta P_0$ becomes larger whereas other contributions $X_1 , Y_1, Z_1, Z_3$ get smaller. Therefore, octant sensitivity will be higher for smaller values of sterile mixing angles.
 
    The octant sensitivity from the disappearance channel probability $P_{\mu\mu}$ is defined as,
	\begin{equation}\label{eq:delP_oct-2}
		\Delta P_{oct,2} \equiv P_{\mu \mu}^{1HO}(\delta_{13}^{HO},\delta_{14}^{HO})- P_{\mu \mu}^{1LO}(\delta_{13}^{LO},\delta_{14}^{LO})> 0
	\end{equation}
    As we have seen earlier, the largest octant sensitive term in $P_{\mu\mu}$ comes from \eqref{Pmm24_1}, we put that in the above \eqref{eq:delP_oct-2} to get the difference in opposite octant probabilities as,
	\begin{equation}
        \begin{split}
          \Delta P_{oct,2}&=\cos^2 \theta_{24m}\sin 2\theta_{24m} \sin\theta_{14m} \sin 4\theta_{13m}*\\&(\cos\delta^{HO}-\cos\delta^{LO})\frac{1+3\eta}{2\sqrt{2}}\sin^2 D^m_{31}\\
          &+\cos^2 \theta_{24m} 2\eta \cos^2 \theta_{24m} \sin^2 2\theta_{13m} \sin^2 D^m_{31}
        \end{split}
	\end{equation}
    It can be noted from the above expression that, the first term has phase dependence while the second term is independent of the phases.
\subsection{Degeneracy in $\cos\theta_\nu-E_\nu$ Plane}
	\begin{figure}[H]
		\centering
		\includegraphics[width=\linewidth]{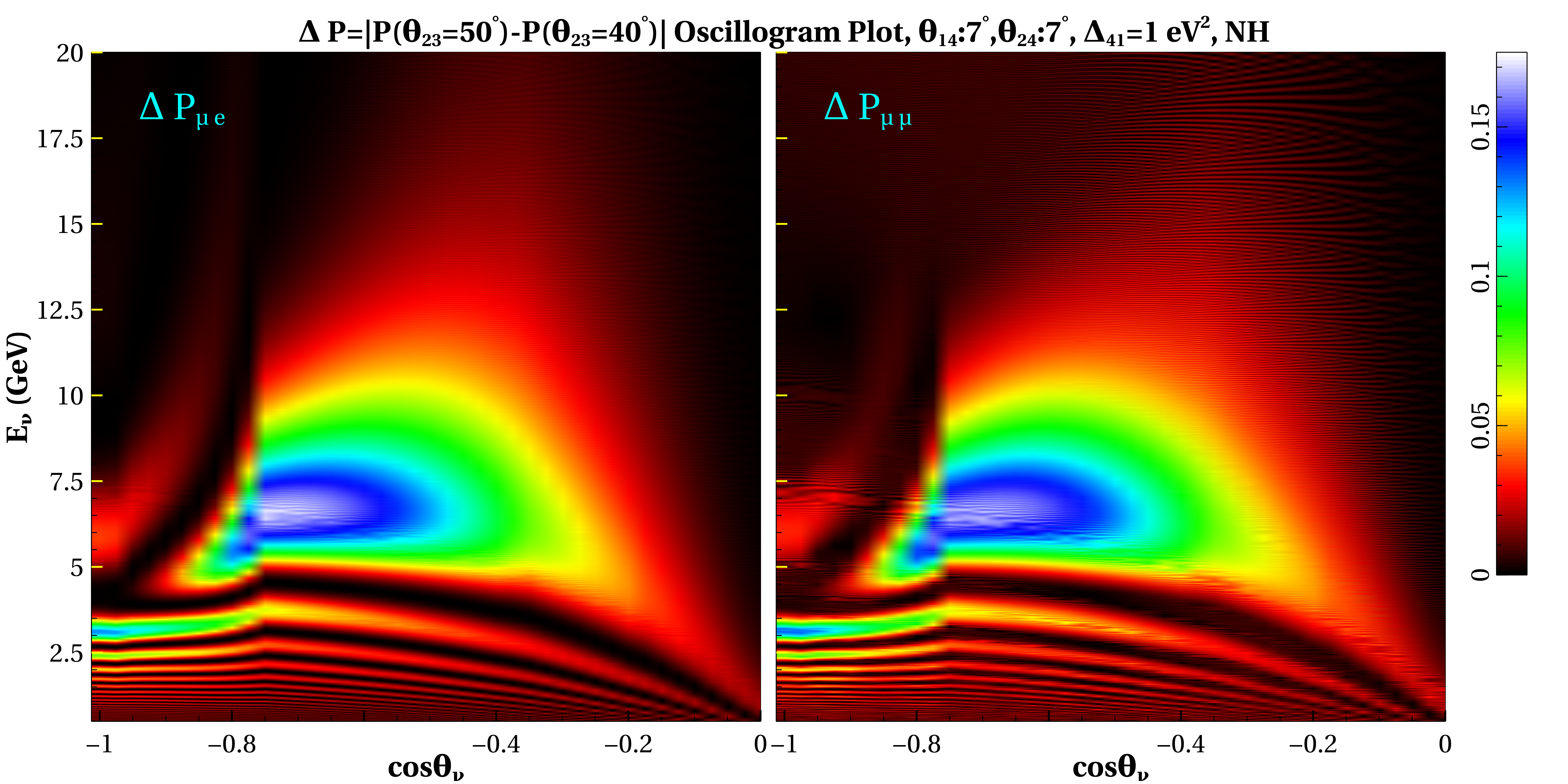}
		\caption{ $\Delta P_{\mu e}$(left), $\Delta P_{\mu\mu} $(right), i.e, the absolute differences in probabilities for $\theta_{23}$ values from opposite octant with fixed value of $\delta_{13},\delta_{14}$ in $\cos\theta_\nu -E_\nu$ plane.}
		\label{fig:pme_pmm_oscillogram}
	\end{figure}
	To probe the octant sensitivity spanning over all the baselines and energies, we present the oscillogram plots of the differences in probabilities corresponding to the value of $\theta_{23}=40^\circ$(LO) and $\theta_{23}=50^\circ$(HO) in $\cos\theta_\nu -E_\nu$ plane for normal hierarchy in \autoref{fig:pme_pmm_oscillogram}. The phases are kept fixed at same $\delta_{13}=195^\circ,\delta_{14}=30^\circ$ for both the octants. From the figure, it can be seen that the maximum difference is obtained at the energy range of $5:10$ GeV for $\cos\theta_\nu$ in the range of $-0.5:-0.8$ which roughly translates to baselines around 5000-10000 km. This figure serves as a reference to show at which baselines and energies the octant sensitivity can be maximum and motivates us to add the contribution from atmospheric neutrinos to obtain better octant sensitivity in our analysis. 
	
\subsection{Degeneracy with variation of $\delta_{13}, \delta_{14}$ at fixed baseline}
\begin{figure}[H]
        \centering
        \includegraphics[width=0.96\linewidth]{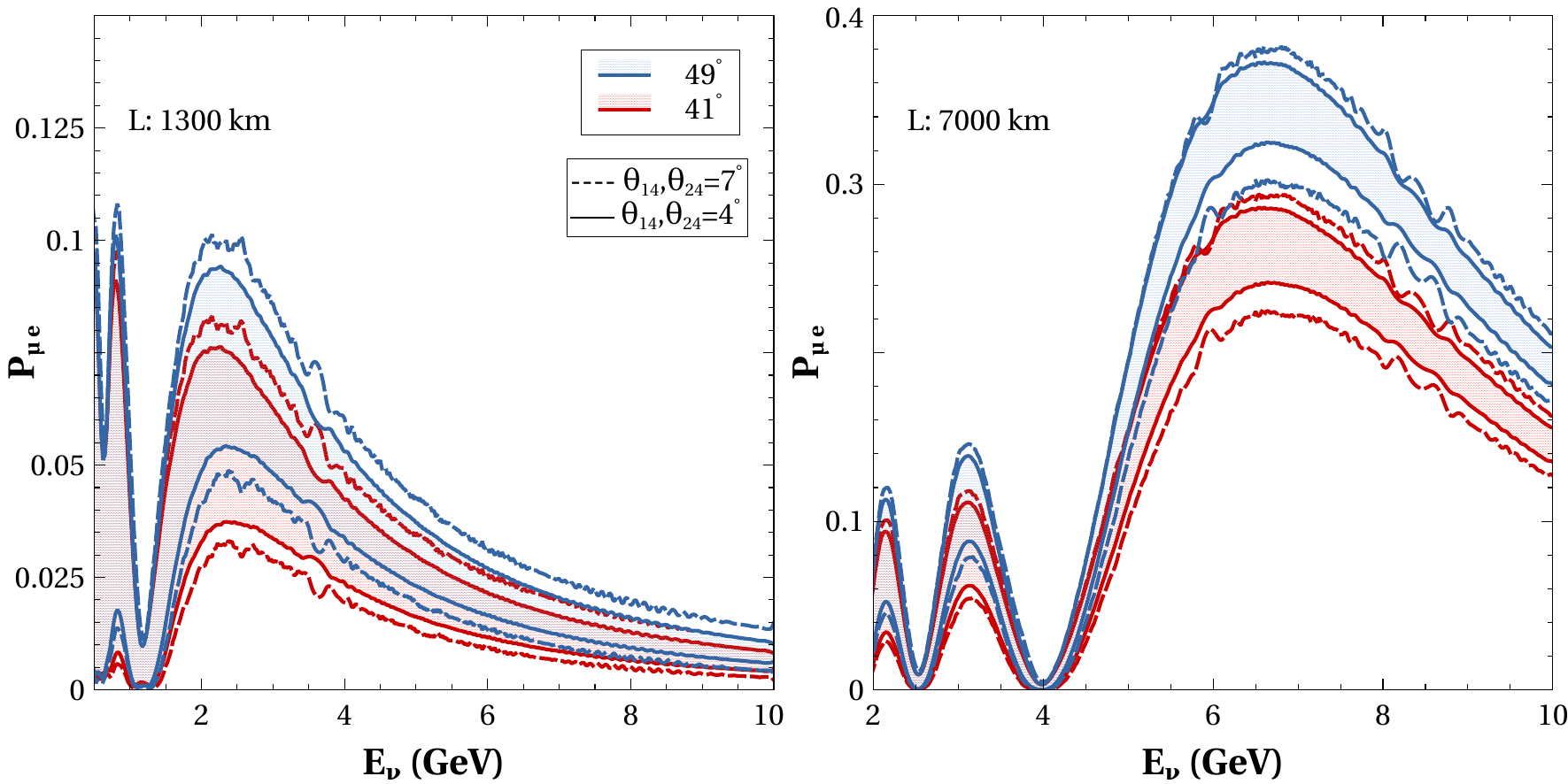}
        \caption{$P_{\mu e}$ as a function of energy at 1300 km (left), and 7000 km (right). Blue and red bands are due to variation of $\delta_{13},\delta_{14}$ for $\theta_{23}=49^\circ$, $41^\circ$ using $\theta_{14}=\theta_{24}=4^\circ$. The regions between blue, and red dotted curves are for $49^\circ$, $41^\circ$ respectively, considering  $\theta_{14}=\theta_{24}=7^\circ$.}
        \label{fig:pme_eng-oct}
    \end{figure}

\begin{figure}[H]
		\centering
		\includegraphics[width=0.96\linewidth]{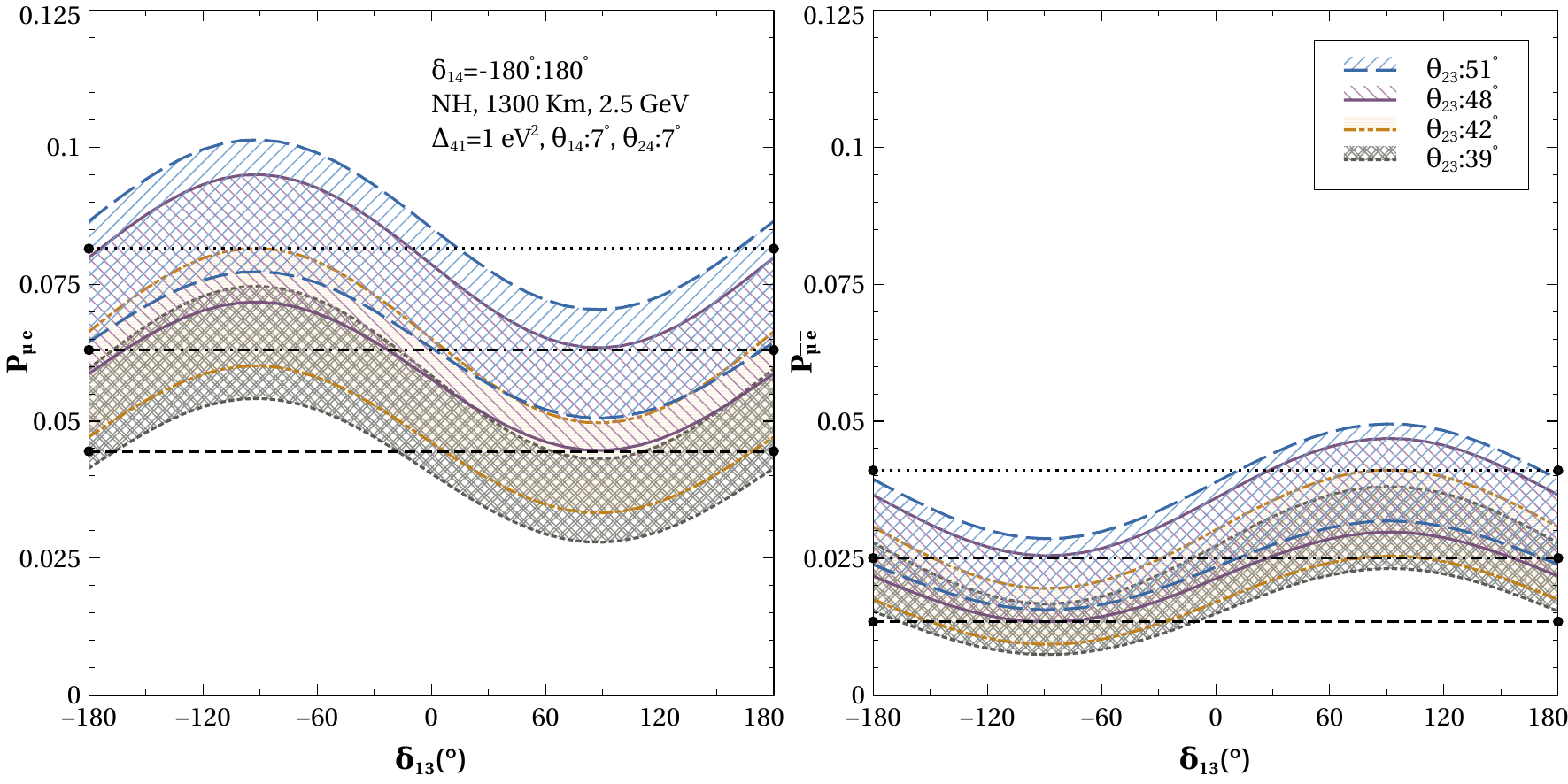}
		\includegraphics[width=0.96\linewidth]{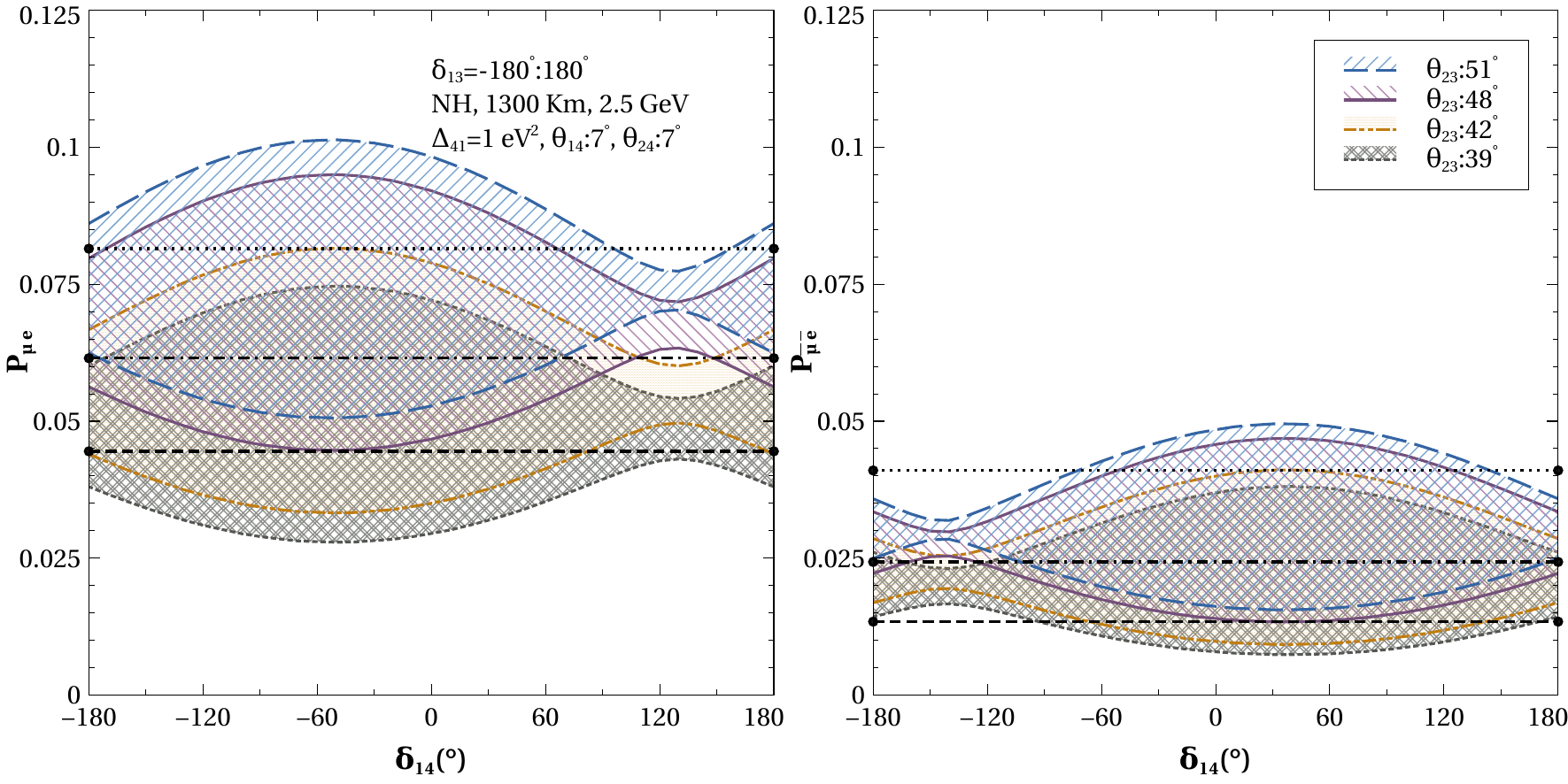}
		\caption{$ P_{\mu e}$ (left), and $P_{\bar{\mu}\bar{e}}$ (right) as a function of $\delta_{13}$ (top), $\delta_{14}$ (bottom) for variation of the respective another phase at neutrino energy 2.5 GeV at 1300 km baseline for NH.}
		\label{fig:pme-pmeb_2.5_oct_1300}
	\end{figure}

\begin{figure}[H]
		\centering
		\includegraphics[width=0.96\linewidth]{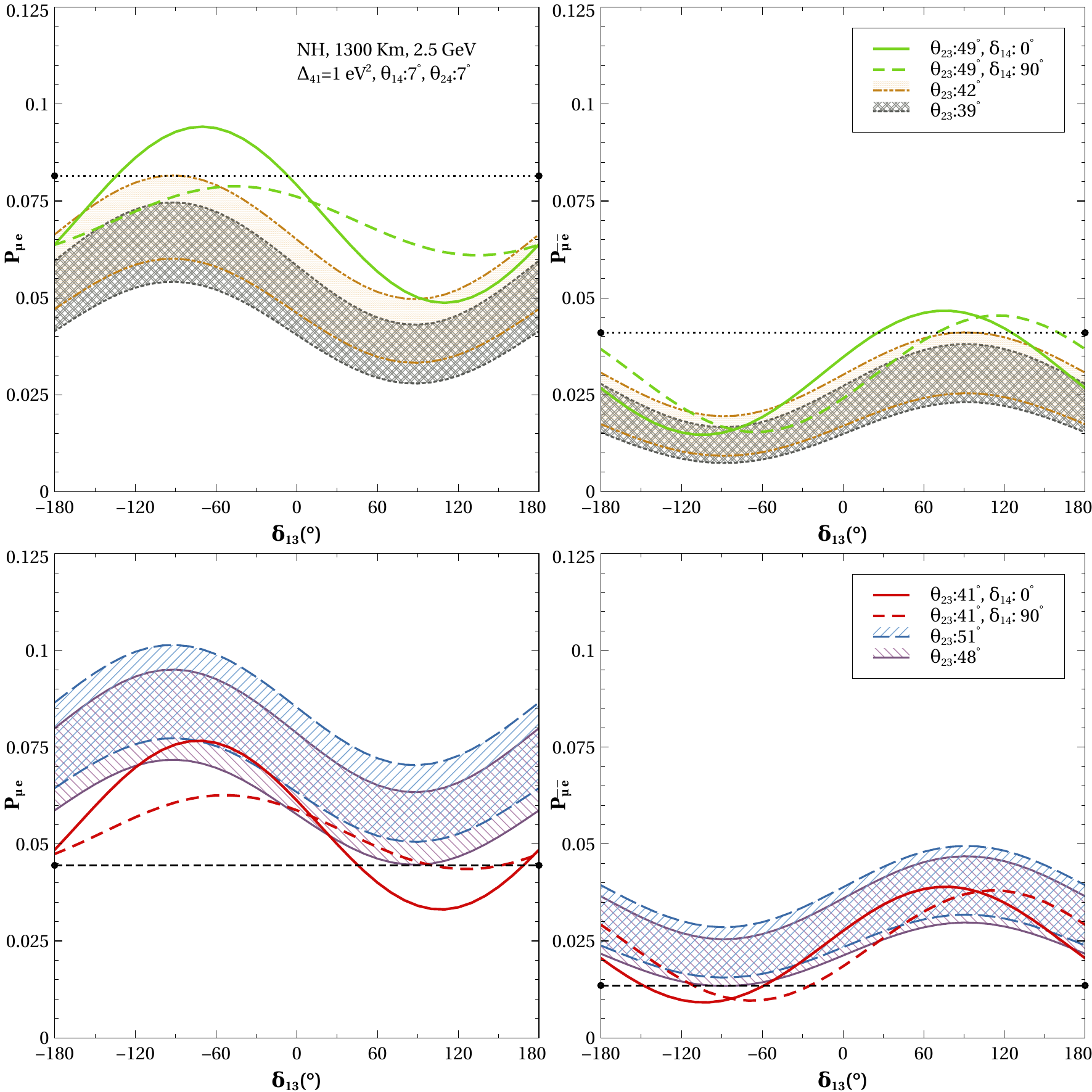}
		\caption{$ P_{\mu e}$ (left), and $P_{\bar{\mu}\bar{e}}$ (right) as a function of $\delta_{13}$ for variation of the phase $\delta_{14}$ at neutrino energy 2.5 GeV at 1300 km baseline for NH.}
		\label{fig:pme-pmeb-d13_2.5_oct_1300}
	\end{figure}

	In this section, we study the probabilities (GLoBES) as a function of the phases to understand the dependency of the degeneracy on these parameters. In \autoref{fig:pme_eng-oct}, we depict the appearance probability $P_{\mu e}$ for $\theta_{23} = 41^\circ$ (red), and $49^\circ$(blue) as a function of neutrino energy at 1300 km and 7000 km baselines. The bands correspond to the variation of $\delta_{13},\delta_{14}$. Two different sets of representative values of $\theta_{14},\theta_{24}$ are considered, e.g., $\theta_{14},\theta_{24} = 4^\circ$, which are allowed after MINOS+\cite{MINOS:2020iqj} bounds, and $\theta_{14},\theta_{24} = 7^\circ$, which are allowed by an earlier global fit\cite{Gariazzo:2017fdh} excluding the MINOS+ results. The significant observations are as follows,
    \begin{itemize}
        \item The probability bands of different octants overlap at 1300 km. While at 7000 km, a difference is observed between opposite octant bands. It shows that at a higher baseline, sensitivity for octant will be higher.

        \item The difference (overlap) between red and blue bands is more (lesser) for $4^\circ$ than $7^\circ$. It is obvious that with smaller sterile mixing angles, we will get better sensitivity.
    \end{itemize}
    From the above figures, we can observe that the variation in the phases can lead to overlap in the probabilities from opposite octants giving rise to degenerate solutions. Therefore, it is instructive to study the variation of the probabilities w.r.t. phases to understand for which values of these parameters degenerate solutions can occur. These plots are done at fixed energies. We choose this energy as 2.5 GeV for $P_{\mu e}$, at 1300 km, since first oscillation maxima occur at this energy as can be seen from \autoref{fig:pme_eng-oct}. The variation of the probabilities $P_{\mu e}$(left), $P_{\bar{\mu}\bar{e}}$(right) are shown as a function of phases $\delta_{13}$ (top), and $\delta_{14}$ (bottom) in \autoref{fig:pme-pmeb_2.5_oct_1300} for values of $\theta_{23}=39^\circ$(grey), $42^\circ$(orange), $48^\circ$(violet), $51^\circ$(blue) spanning over both octants. The curves for other values of $\theta_{23}$ will lie in between these ranges. The bands correspond to variation over the non-displayed phase $\delta_{14}(top)/\delta_{13}(bottom)$ over the range $-180^\circ:180^\circ$ respectively. Three horizontal iso-probability lines are drawn in \autoref{fig:pme-pmeb_2.5_oct_1300} to indicate the values of $\delta_{13}/\delta_{14}$ for which there are degeneracies (dot-dashed line) and there are no degeneracies (dotted, dashed lines) between the two octants. Note that in the probability vs $\delta_{13}$ plots for the three-generation case, there is a single curve for each $\theta_{23}$ whereas, in the presence of sterile neutrino, there are bands due to $\delta_{14}$ variation for a fixed $\theta_{23}$. We can infer the following points from \autoref{fig:pme-pmeb_2.5_oct_1300},
	\begin{itemize}       
	    \item The regions above the dotted line in the top panels indicate the values of $\delta_{13}$ for which there is no degeneracy in HO. This is around $\delta_{13}=-90^\circ(90^\circ)$ in $P_{\mu e}(P_{\bar{\mu}\bar{e}})$ channel. However, some portions of the blue and violet bands extend below the dotted lines in both figures and sometimes also overlap with the orange and violet bands, indicating that for these values of $\delta_{13}$, there are still degeneracies for certain values of $\delta_{14}$.
	    
	    \item Similarly, the regions below the dashed lines in the top panels signify the $\delta_{13}$ values devoid of degeneracy for $\theta_{23}$ in LO. This region for $P_{\mu e}(P_{\bar{\mu}\bar{e}})$ channel is around $\delta_{13}=90^\circ(-90^\circ)$. Here also, the portions of grey and orange bands above the dashed lines, as well as the portions coinciding with the blue and violet bands, indicate the existence of degeneracies at these values of $\delta_{13}$.
	    
	    \item From the top panels, we can clearly see a synergy between neutrino and anti-neutrino channels for octant degeneracy in both HO and LO. For instance, for HO (LO), the degeneracy is present around $\delta_{13}=90^\circ(-90^\circ)$ at $P_{\mu e}$ channel but absent for $P_{\bar{\mu}\bar{e}}$.
	    
	    \item In the bottom panels, the regions above the dotted line indicate that the no degeneracy region in HO lies around $\delta_{14}=-60^\circ (60^\circ)$ for $P_{\mu e}(P_{\bar{\mu}\bar{e}})$ channel. Note that the region has a larger spread in $\delta_{14}$ over $-180^\circ:95^\circ(-70^\circ:140^\circ)$ for $\theta_{23}=51^\circ$, and over $-180^\circ:65^\circ(-50^\circ:120^\circ)$ for $\theta_{23}=48^\circ$ in $P_{\mu e}(P_{\bar{\mu}\bar{e}})$ channel. Corresponding nondegenerate regions have a smaller spread in $\delta_{13}$ as seen from the top panel plots.
	    
	    \item There are regions below the dashed line, signifying no degeneracy in LO for the plots in the bottom panels. These regions occur around $\delta_{14}=-60^\circ (60^\circ)$ for $P_{\mu e}(P_{\bar{\mu}\bar{e}})$ channel. However, it is to be noted that unlike in the top panel, the non-degenerate region in LO is over the similar range of $\delta_{14}$ w.r.t HO as mentioned in the previous point. Therefore, we see in the neutrino (anti-neutrino) channel maximum sensitivity for both HO and LO is around $\delta_{14}=60^\circ (-60^\circ)$.

	    \item In the bottom panels, the probability bands are wider and the extent of overlap is higher around $-60^\circ(60^\circ)$ in $P_{\mu e}(P_{\bar{\mu}\bar{e}})$ channel. These give rise to WO-R$\delta_{14}$ degeneracies which are hard to resolve using neutrino plus anti-neutrino. The synergy between neutrino and anti-neutrino channels for octant degeneracy is less pronounced here.

        \item In the bottom panels for $P_{\mu e}$ channel, around $\delta_{14}=130^\circ$, there is a small region where there is no WO-R$\delta_{14}$ degeneracy between HO and LO for all values of $\delta_{13}$. For $P_{\bar\mu \bar{e}}$ channel there is a similar region with minimum degeneracy around $\delta_{14}=-130^\circ$.

        \item When the probability bands from HO (blue and violet) coincide with bands from LO (orange and grey) at the same $\delta_{13}/\delta_{14}$ values, those are examples of WO-R$\delta_{13}$/R$\delta_{14}$ degeneracies. While the regions of bands from opposite octants connected through iso-probability lines show WO-W$\delta_{13}$/W$\delta_{14}$ degeneracies.
	    
	\end{itemize}
	While performing $\chi^2$ analysis, we take fixed true values of parameters in one octant and marginalize $\chi^2$ over the relevant parameters in the opposite octant. Therefore, a better understanding of the octant degeneracy can be achieved if we keep $\theta_{23}$, and the phases constant in one octant and vary them in the opposite one. We replicate this in \autoref{fig:pme-pmeb-d13_2.5_oct_1300} where the probabilities in neutrino (left) and anti-neutrino (right) channels are drawn as a function of phase $\delta_{13}$. In the top [bottom] panel, the green [red] solid(dashed) line corresponds to $\theta_{23}=49^\circ[41^\circ]$ and $\delta_{14}=0^\circ(90^\circ)$. The grey and orange [violet and blue] bands correspond to $\theta_{23}=39^\circ,42^\circ[48^\circ,51^\circ]$ in LO[HO] for $\delta_{14}$ varying over $-180^\circ:180^\circ$. The horizontal iso-probability lines in the plots demarcate different degenerate and non-degenerate regions. The important points from \autoref{fig:pme-pmeb-d13_2.5_oct_1300} are as follows,
	\begin{itemize}
	    \item In the top panel, the region above the dotted line corresponds to no degeneracy. This region is around $\delta_{13}=-90^\circ(90^\circ)$ at $P_{\mu e}(P_{\bar{\mu}\bar{e}})$ channel for green solid ($\delta_{14}=0^\circ$) curve. However, the green dashed ($\delta_{14}=90^\circ$) curve has a non-degenerate region only in $P_{\bar{\mu}\bar{e}}$ channel around $\delta_{13}=90^\circ$. This suggests that for $\delta_{14}=0^\circ$, the octant sensitivity comes from both $P_{\mu e}$, and $P_{\bar{\mu}\bar{e}}$ channel around $\delta_{13}\sim0^\circ$ whereas for $\delta_{14}=90^\circ$ sensitivity comes only from $P_{\bar{\mu}\bar{e}}$ channel around $\delta_{13}\sim90^\circ$.
	    
	    \item For the bottom panel, the non-degenerate regions are below the dashed horizontal line. In $P_{\mu e}$ channel this region is around $\delta_{13}=120^\circ$ for $\delta_{14}=0^\circ$. A very small region for $\delta_{14}=90^\circ$ also extends below the dashed line. In $P_{\bar{\mu}\bar{e}}$ channel the region of no degeneracy lies around $\delta_{13}=-120^\circ(-60^\circ)$ for $\delta_{14}=0^\circ(90^\circ)$.
	\end{itemize}
	
	\begin{figure}[H]
		\centering
		\includegraphics[width=0.96\linewidth]{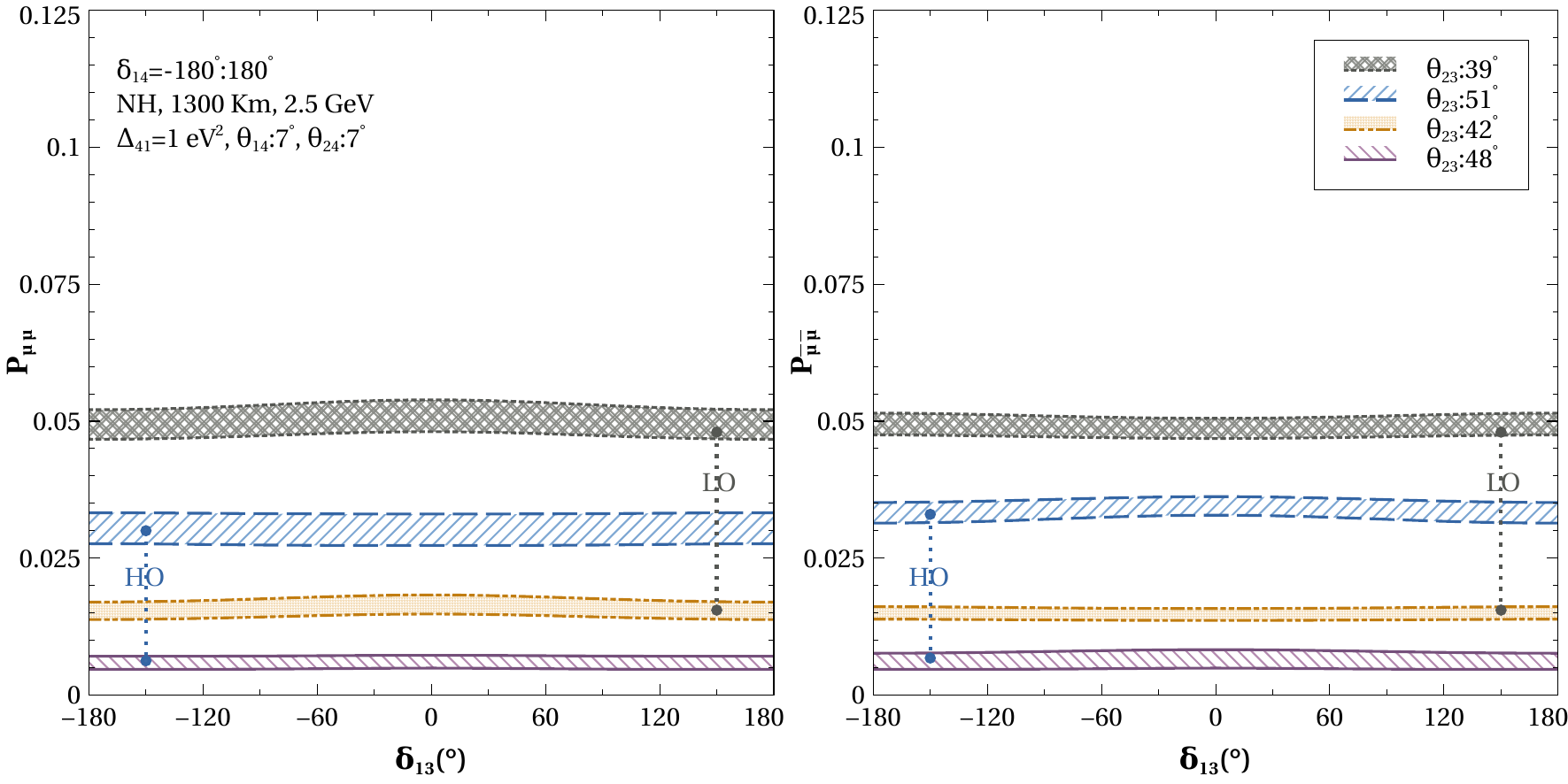}
		\includegraphics[width=0.96\linewidth]{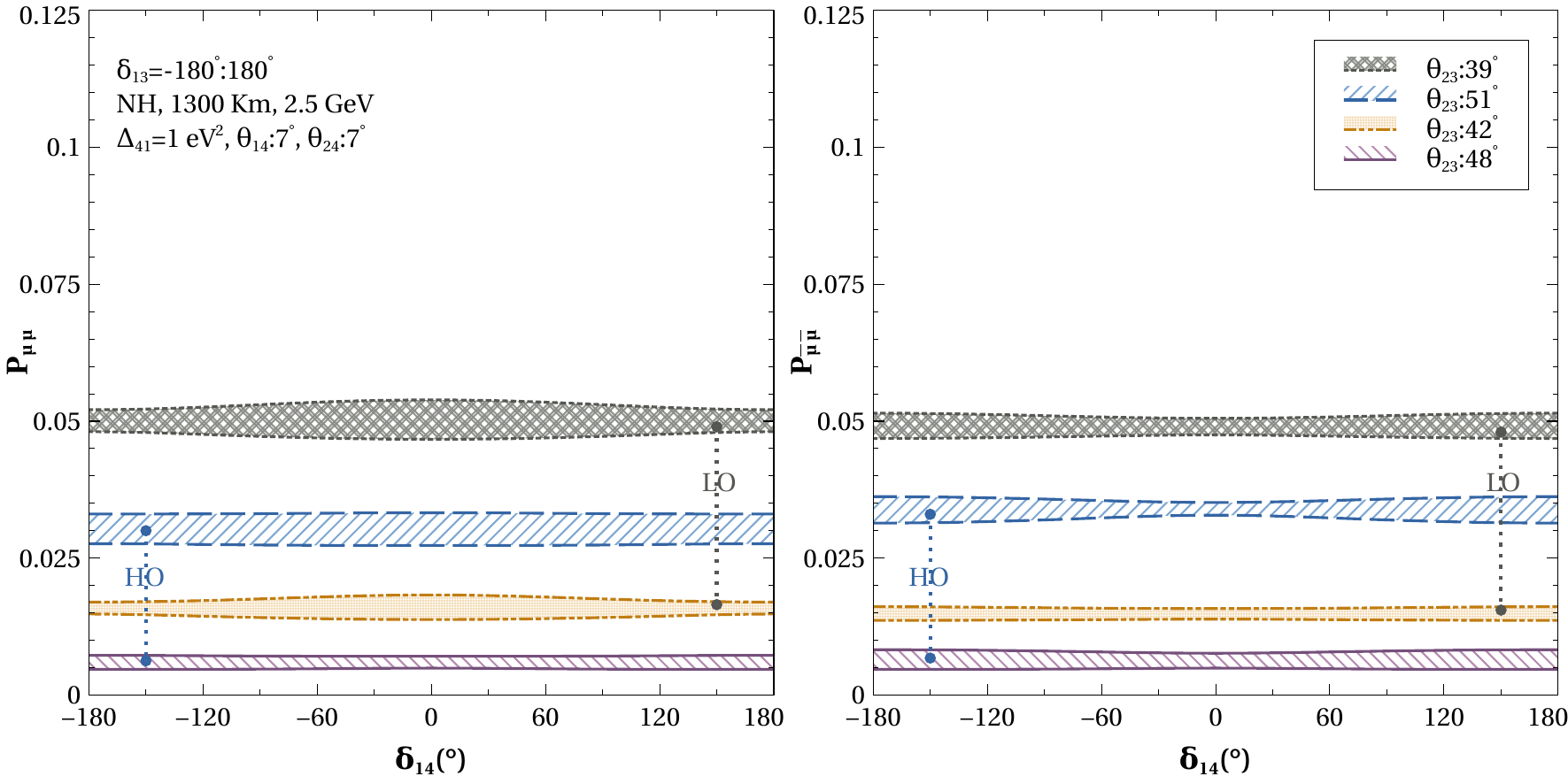}
		\caption{$ P_{\mu \mu}$ (left), and $P_{\bar{\mu}\bar{\mu}}$ (right) vs $\delta_{13}$ (top), $\delta_{14}$ (bottom) for variation of the respective another phase at neutrino energy 2.5 GeV at 1300 km baseline for NH.}
		\label{fig:pmm-pmmb_2.5_oct_1300}
	\end{figure}
	Now we focus on the disappearance channel probabilities $P_{\mu \mu}$ (left), and $P_{\bar\mu \bar\mu}$ (right) as a function of phases $\delta_{13}$(top panel), $\delta_{14}$(bottom panel) at 2.5 GeV in \autoref{fig:pmm-pmmb_2.5_oct_1300}. The following points may be noted,
	\begin{itemize}
	    \item The bands due to variation of $\delta_{13}/\delta_{14}$ are narrower than the ones for appearance channel. Hence, these bands are well separated from each other.
	    
	    \item The bands corresponding to $\theta_{23}=51^\circ$(blue) in HO comes in between the bands corresponding to $\theta_{23}=39^\circ$(grey) and $\theta_{23}=42^\circ$(yellow) in LO. On the other hand, the violet band corresponding to $\theta_{23}=48^\circ$ is outside the whole region of LO between the grey and yellow band.
	    This implies the presence (absence) of the octant degeneracy for $\theta_{23}=51^\circ(48^\circ)$ in $P_{\mu\mu}$ channel.
	    
	    \item Similarly, $\theta_{23}=39^\circ$ (grey) in LO demonstrates octant sensitivity since it lies outside the HO region between the blue and violet bands, but $\theta_{23}=42^\circ$ (yellow) lies within the HO region and therefore is not sensitive to the octant. A similar feature can also be seen from probability vs $\delta_{14}$ plots in the bottom panel. 
	  
    \end{itemize}
    We can conclude that for certain trues values of $\theta_{23}$, the $P_{\mu \mu}$ channel can contribute to the octant sensitivity at 1300 km.

\begin{figure}[H]
		\centering
		\includegraphics[width=0.96\linewidth]{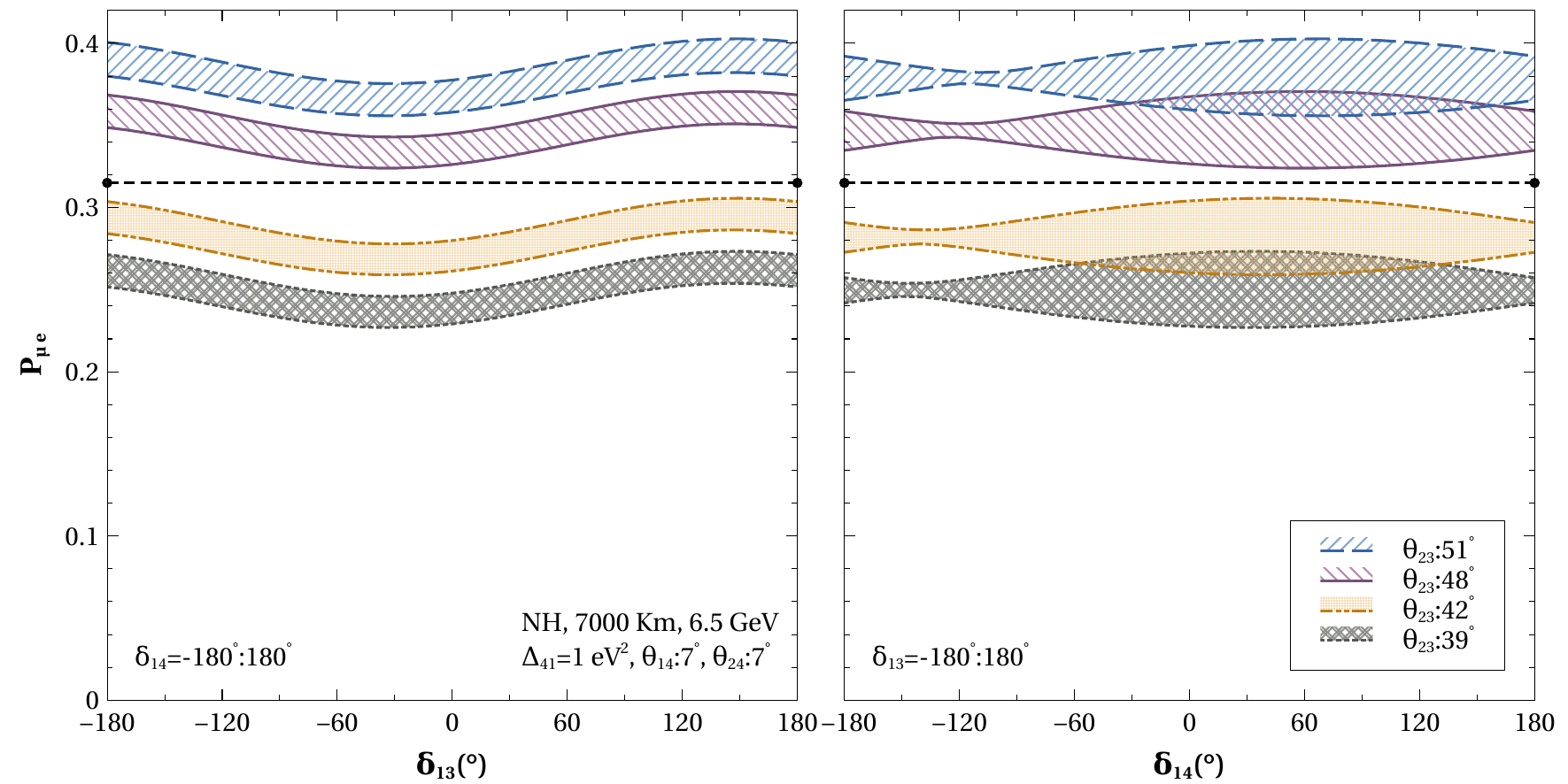}
		\caption{$ P_{\mu e}$ vs $\delta_{13}$(left), and $\delta_{14}$(right) for variation of the respective another phase at neutrino energy 6.5 GeV at 7000 km baseline for NH.}
		\label{fig:pme_6.5_oct_7k}
	\end{figure}

\begin{figure}[H]
		\centering
		\includegraphics[width=0.96\linewidth]{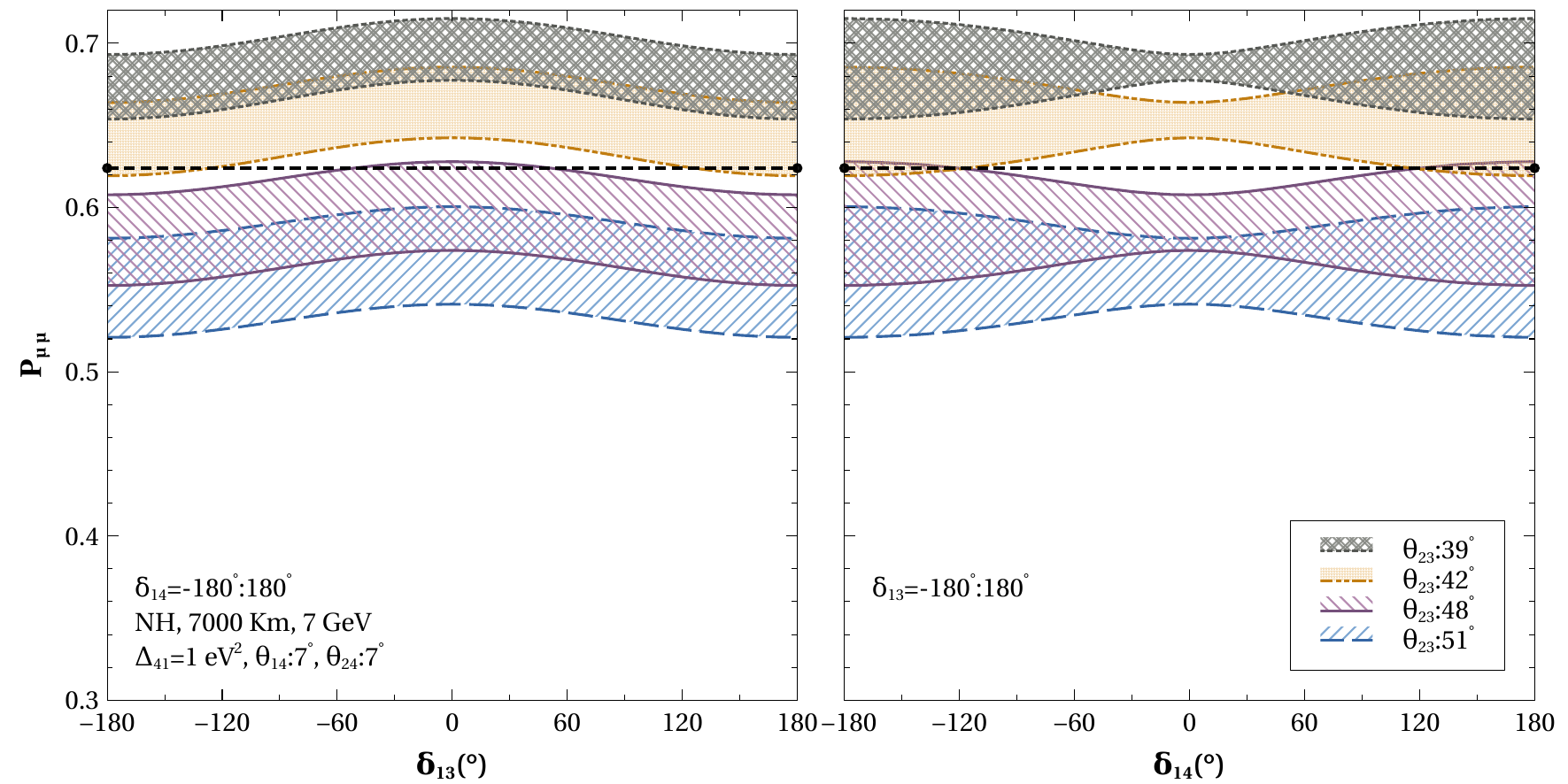}
		\caption{$ P_{\mu \mu}$ vs $\delta_{13}$(left), and $\delta_{14}$(right) for variation of the respective another phase at energy 7 GeV at 7000 km baseline for NH.}
		\label{fig:pmm_7_oct_7k}
	\end{figure}

	Next, we study the behaviour of the probabilities at a higher baseline of 7000 km, where the resonant matter effect comes into play. We observe the appearance probability $P_{\mu e}$ as a function of the phase $\delta_{13}$ (left), and $\delta_{14}$ while the respective other phase variation creates band at different values of $\theta_{23}=39^\circ,42^\circ,48^\circ,51^\circ$ spanning over both octants at energy maxima of 6.5 GeV in \autoref{fig:pme_6.5_oct_7k}. We see similar variations of the disappearance channel probability $P_{\mu \mu}$ at maxima energy of 7 GeV in \autoref{fig:pmm_7_oct_7k}. Energies of 6.5 GeV and 7 GeV are chosen as they correspond to the maxima in $P_{\mu e}, P_{\mu \mu}$ channels at this baseline, respectively. The effect of sterile mixing angles and phases on octant sensitivity in the $P_{\mu e}$ channel at other energies can be seen in  \autoref{fig:pme_eng-oct}. The following facts can be noted,
	\begin{itemize}
	    \item Unlike at 1300 km, the $P_{\mu e}$ probability bands of opposite octant at 7000 km are clearly separated. It suggests that even with the variation of phases and $\theta_{23}$ in both octants, the octant degeneracy can be clearly removed at higher baselines.
	    
	    \item In $P_{\mu \mu}$ channel, the LO and HO bands are mostly separated apart from the occurrence of WO-W$\delta_{13}$ (left panel), WO-R$\delta_{14}$/W$\delta_{14}$ (right panel) degeneracies respectively around $\delta_{13}$, $\delta_{14}$ values of $\pm150^\circ$ in a tiny region. This suggests contributions to the octant sensitivity also come from the $P_{\mu \mu}$ channel. The sensitivity of the octant in $P_{\mu\mu}$ comes from the first term in \eqref{Pmm24_1}, which has a more significant contribution at 7000 km than 1300 km as noted in \autoref{fig:pmm_terms} due to larger matter effect.
	\end{itemize}
	
\section{Experimental and Simulation Details of the LArTPC detector}
\label{sec:analysis}
    As a typical example for the long baseline analysis, we consider an experimental setup consisting of a near detector (ND) and far detector (FD) exposed to a megawatt-scale muon neutrino beam produced by Long Baseline Neutrino Facility (LBNF) at the Fermilab. The ND will be placed close to the source of the beam, while the FD, comprising a LArTPC detector of 40 kt will be installed 1300 km away. The large LArTPC detector at this depth will also collect atmospheric neutrinos. In this analysis, we have used beams coming from the accelerator as well as neutrinos generated in the atmosphere by cosmic ray interactions. The experimental setup considered in our work is similar to that proposed by the DUNE experiment\cite{DUNE:2020lwj}\cite{DUNE:2016hlj}.
\subsection{Events from beam neutrinos}
    We use a beam power of 1.2 MW, leading to a total exposure of $10\times 10^{21}$ POT. The neutrino beam simulation for the experiment has been carried out using the GLoBES\cite{Huber:2004ka} software with the most recent publicly available configuration file\cite{DUNE:2016ymp}. We assume experimental run time for 3.5 years each in the neutrino and the antineutrino mode with a total exposure of 280 kt-yr.

\begin{figure}[H]
		\centering
		\includegraphics[width=0.96\linewidth]{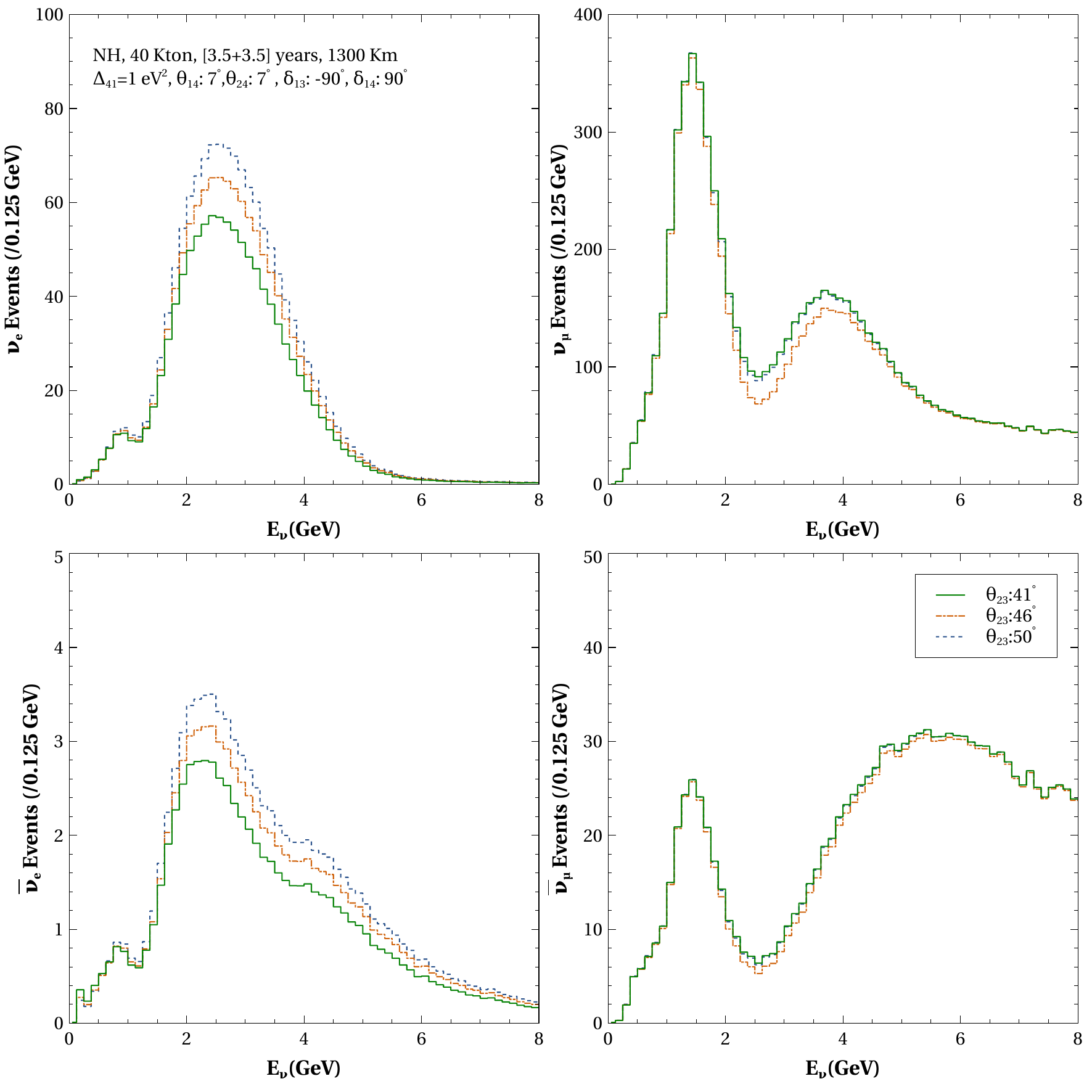}
		\caption{Electron (left) and muon neutrino (right) event spectrum for neutrinos (top) and anti-neutrinos (bottom) as a function of energy for true $\theta_{23}=41^\circ$(green) with true phases $\delta_{13}=-90^\circ,\delta_{14}=90^\circ$ at 1300 km for test values of $\theta_{23}=46^\circ$(blue) and $\theta_{23}=50^\circ$(orange) for NH.} 
		\label{fig:events_e_oct}
	\end{figure}

\begin{figure}[H]
		\centering
		\includegraphics[width=0.94\linewidth]{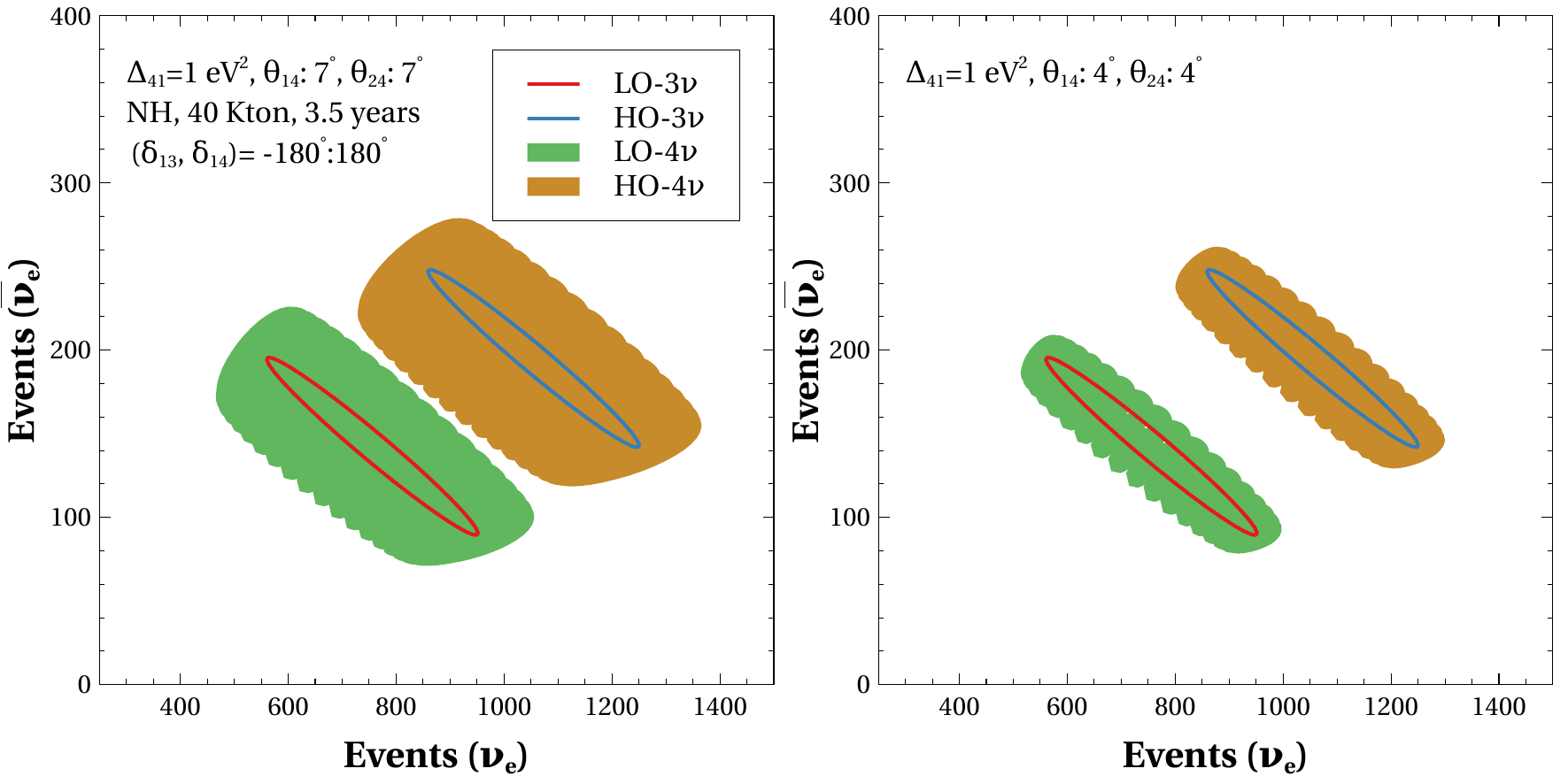}
		\caption{Bi-events plot in $\nu_e-\bar{\nu}_e$ plane for $\theta_{23}=41^\circ$(red, green), $49^\circ$(blue, yellow) at 1300 km with variation of phases $ \delta_{13},\delta_{14}$ corresponding to $\theta_{14},\theta_{24}=7^\circ$ (left), $4^\circ$ (right) for NH.}
		\label{fig:pme_bievents}
	\end{figure}

    We have plotted the electron and muon events spectrum for 1300 km baseline considering normal hierarchy with sterile mixing angle of $\theta_{14},\theta_{24}=7^\circ$ at fixed phases $\delta_{13}=-90^\circ,\delta_{14}=90^\circ$ in \autoref{fig:events_e_oct}. There are differences between the spectra of the events for the true value of $\theta_{23}=41^\circ$(green) in LO with the values of $\theta_{23}$ in HO for $46^\circ$(orange), $50^\circ$(blue). This is indicative of the octant sensitivity. It should be noted that although the green spectrum is closer to the orange one($46^\circ$) for electron events (left panels), for muon events (right panels) the green one is closer to the blue one($50^\circ$). This indicates that the maximum sensitivity occurs at different $\theta_{23}$ values in the opposite octant for electron and muon events. This will lead to the synergy between electron and muon events when we compute the combined octant sensitivity at $\chi^2$ level. The maximum difference in events is observed in the energy region of 2-4 GeV, where the spectra of the event have maxima in the case of both electrons and muons.
	We present bi-events plots in \autoref{fig:pme_bievents} considering the total no of electron neutrino and anti-neutrino events obtained by integrating over the full energy range. The elliptic regions are due to variations in the relevant phases over their full range. This figure shows that in the case of three flavour oscillation framework the ellipses for $\theta_{23}$ being in two different octants are well separated, showing no octant degeneracy with combined $\nu_e+\bar{\nu}_e$ events of 3.5+3.5 years with 40 kt LArTPC detector. Now, if in the presence of an additional sterile neutrino, these ellipses turn into blobs, a combination of many ellipses\cite{Agarwalla:2016xlg}. From this figure, we can see that the separation between the green(LO) and yellow (HO) regions increases with smaller values of sterile mixing angles $\theta_{14},\theta_{24}$, leading to an enhanced octant sensitivity.
	
\subsection{Events from atmospheric neutrinos}
    The atmospheric neutrino and anti-neutrino events are obtained by folding the relevant incident fluxes with the appropriate disappearance and appearance probabilities, charge current (CC) cross sections, detector efficiency, resolution, detector mass, and exposure time. 
     The $\mu^-$, and $e^{-}$ event rates in an energy bin of width $\mathrm{dE_\nu}$ and in a solid angle bin of width ${\mathrm{d \Omega_\nu}}$ are as follows,
    \begin{equation} \label{eq:muevent}
    \rm{ \frac{d^2 N_{\mu}}{d \Omega \;dE} = \frac{D_{eff}\Sigma}{2\pi} \left[\left(\frac{d^2 \Phi_\mu}{d \cos \theta \; dE}\right) P_{\mu\mu} +\left(\frac{d^2 \Phi_e}{d \cos \theta \; dE}\right)P_{e\mu}\right]}.
    \end{equation}
   \begin{equation} \label{eq:eevent}
    \mathrm{ \frac{d^2 N_e}{d \Omega \;dE} = \frac{D_{eff}\Sigma}{2\pi}\left[\left( \frac{d^2 \Phi_{\mu}}{d \cos \theta\; dE}\right)P_{\mu e} + \left(\frac{d^2 \Phi_e}{d \cos \theta \; dE}\right)P_{e e} \right]}
    \end{equation}

    Here ${\mathrm{\Phi_{\mu}}}$ and ${\mathrm{\Phi_{e}}}$ are the $\mathrm{\nu_\mu}$ and ${\mathrm{\nu_e}}$ atmospheric fluxes  respectively obtained from Honda et.al.\cite{PhysRevD.92.023004} at the Homestake site; $P_{\mu\mu}(P_{ee})$ and $P_{\mu e}$ are disappearance and appearance probabilities; $\rm{\Sigma}$ is the total charge current (CC) cross-section and $\rm{D_{eff}}$ is the detector efficiency. The $\mu^+$, and $e^{+}$ event rates are similar to the above expression with the fluxes, probabilities, and cross sections replaced by those for $\mathrm{\bar\nu_\mu}$ and ${\mathrm{\bar\nu_e}}$ respectively.
    For the LArTPC detector, the energy and angular resolution are implemented using the Gaussian resolution function as follows,
    \begin{equation} \label{eq:esmear}
    \mathrm{ R_{E_\nu}(E_t,E_m) = \frac{1}{\sqrt{2\pi}\sigma} \exp\left[-\frac{(E_m - E_t)^2}{2 \sigma^2}\right]}\,.
    \end{equation}
    \begin{equation} \label{eq:anglesmear}
    \rm{ R_{\theta_\nu}(\Omega_t, \Omega_m) = N \exp \left[ - \frac{(\theta_t -\theta_m)^2 + \sin^2 \theta_t ~(\phi_t - \phi_m)^2}{2 (\Delta\theta)^2} \right] } \,,
    \end{equation}
    where N is a normalization constant. Here, $\rm{E_m}$ (${\rm{\Omega_m}}$), and $\rm{E_t}$ (${\rm{\Omega_t}}$) denote the measured and true values of energy (zenith angle) respectively. The smearing width $\sigma$ is a function of the energy $\rm{E_t}$. The smearing function for the zenith angle is a bit more complicated because the direction of the incident neutrino is specified by two variables: the polar angle ${\rm{\theta_t}}$ and the azimuthal angle ${\rm{\phi_t}}$. We denote both these angles together by ${\rm{\Omega_t}}$.
    The measured direction of the neutrino, with polar angle ${\rm{\theta_m}}$ and azimuthal angle ${\rm{\phi_m}}$, which together we denote by ${\rm{\Omega_m}}$, is expected to be within a cone of half-angle $\Delta \theta$ of the true direction. Assumptions of the far detector (LArTPC) parameters are mentioned in \autoref{table:LAr-parameter}\cite{Barger:2014dfa}.
        \begin{table}[H]
		\centering
		\begin{tabular}{|c|c|}
			\hline
			Parameter uncertainty & Value \\\hline
			$\mu^{+/-}$ Angular  & $2.5^\circ$\\
			$e^{+/-}$ Angular  & $3.0^\circ$\\
			($\mu^{+/-}$, $e^{+/-}$) Energy & GLB files for each E bin \cite{DUNE:2016ymp} \\ Detection efficiency  &  GLB files for each E bin \cite{DUNE:2016ymp}\\
			Flux normalization & 20$\%$\\ Zenith angle dependence  & 5$\%$\\
			 Cross section  & 10$\%$\\
			Overall systematic  & 5$\%$\\
			Tilt  & 5$\%$\\
		    \hline
		\end{tabular}
		\caption{Assumptions of the LArTPC far detector parameters and uncertainties.}
		\label{table:LAr-parameter}
	\end{table}
    \subsubsection{Charge identification using muon capture in liquid argon}
\begin{figure}[H]
       \centering
       \includegraphics[width=0.96\linewidth]{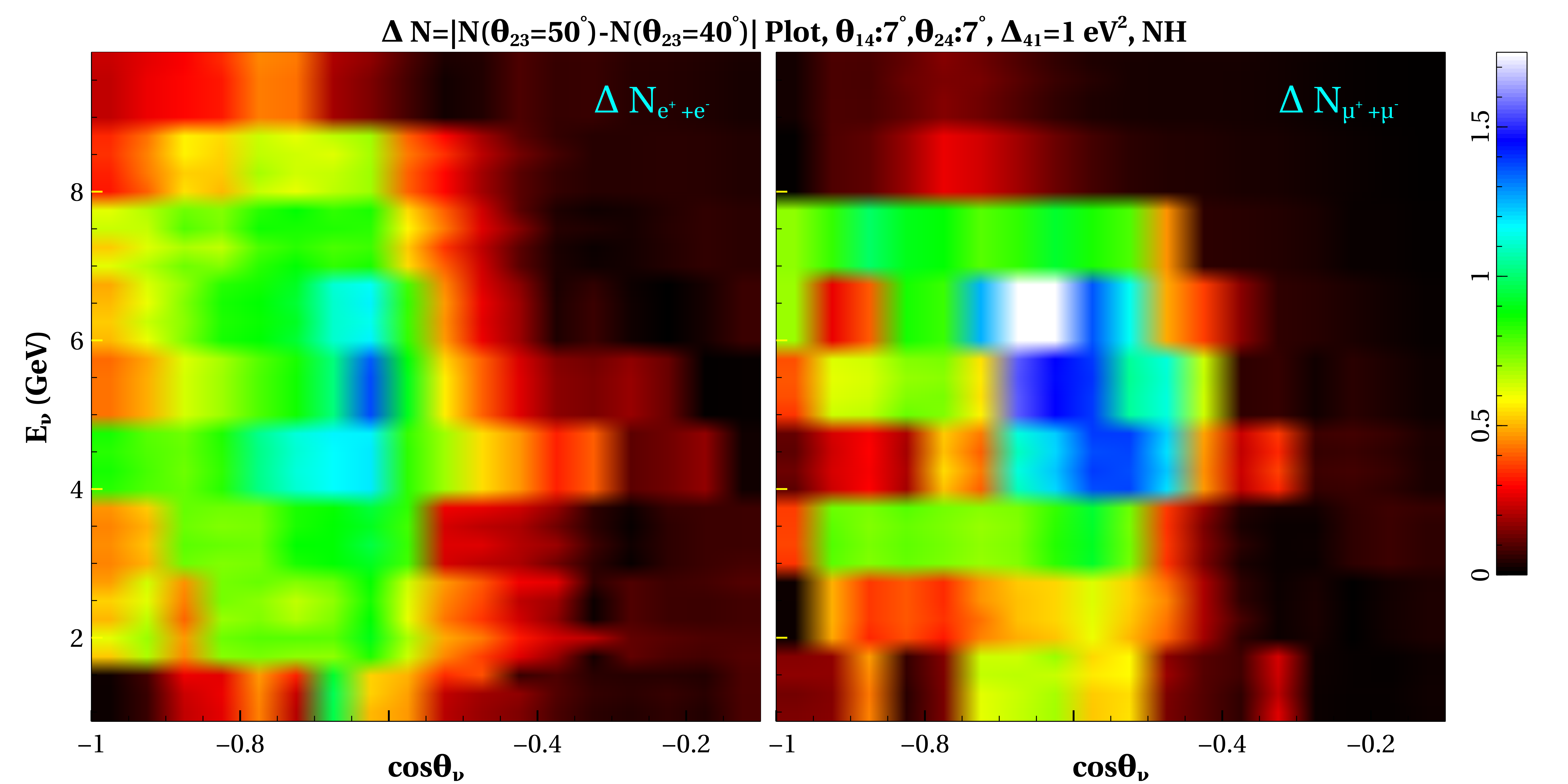}
       \caption{The difference of atmospheric events between HO and LO has been plotted in $E_\nu-\cos\theta_\nu$ plane for $e^+ + e^-$(left), and $\mu^+ +\mu^-$(right) events.}
       \label{fig:event_diff_nu-e-mu}
   \end{figure}

    Magnetizing the large 40 kt LArTPC detector is difficult and expensive, but the charge id of the muon can be identified using the capture vs decay process of the muon inside the argon as studied previously for the DUNE detector\cite{PhysRevD.100.093004}. We have implemented the charge id of the muon as follows: some fraction of the $\mu^{-}$ like events that undergo the capture process are identified using capture fraction efficiency, and the rest of the muons, as well as all the $\mu^{+}$ undergo muon decay.
    The lifetime of the muon resulting from the capture and decay processes can be written as,
   \begin{equation}
       \tau = \bigl(\frac{1}{\tau_{cap}} + \frac{Q}{\tau_{free}})^{-1}
   \end{equation}
   where $\tau_{cap}$ is the lifetime in the capture process, $\tau_{free}$ is the decay lifetime, and Q is the Huff correction factor\cite{PhysRevC.35.2212}. We can define $\mu^{-}$ capture fraction as,
   \begin{equation}
       \epsilon^{cap} = \frac{\tau}{\tau_{cap}} = 1- \frac{\tau}{\tau_{free}}
   \end{equation}
   We use the most precise value of $\mu^{-}$ lifetime in argon\cite{Themuonlifetime}, $\mu^{-}$ capture fraction becomes $\epsilon^{cap}$= 71.9$\%$. Electron charge identification is impossible at GeV energies and electron events are summed for each energy and angular bin.
   For the sensitivity calculation, the $\mu^{-}$ and $\mu^{+}$ are separated as follows: the $\mu^{-}$ events selected that undergo muon capture are given by,
   \begin{equation}
       {N_{i,j,\mu^{-}}}^{cap}= {\epsilon}^{cap}\times N_{\mu^{-}}
   \end{equation}
   and the remaining $\mu^{-}$ events are included within the $\mu^{+}$ event bin as follows,
   \begin{equation}
       N_{i,j,\mu^{+}}^{rest} = \bigl(1- {\epsilon}^{cap})N_{i,j,\mu^{-}} + N_{i,j,\mu^{+}} 
   \end{equation}
    In \autoref{fig:event_diff_nu-e-mu}, we show the absolute differences of atmospheric events between HO $\&$ LO in $E_{\nu}$-$\cos\theta_{\nu}$ plane for $\mu^{+}+\mu^{-}$ (left), and $e^{+} + e^{-}$ (right). This clearly shows that the difference is larger at the matter-resonance region, as observed from the probability oscillogram plot in \autoref{fig:pme_pmm_oscillogram}. The electron event spectrum shows a significant difference in the energy range of $2-8$ GeV for $\cos{\theta_\nu}$ range of $-0.5:-0.9$. The muon events also contribute, especially in a few parts of the energy range $3-8$ GeV for $\cos{\theta_\nu}$ range of $-0.5:-0.9$. This plot captures the octant sensitivity at different baselines and energies for fixed values of oscillation parameters.

    \subsection{$\chi^2$ analysis}
    The computation of $\chi^2$ is performed using the method of pulls. This method allows us to take into account the various statistical and systematic uncertainties in a straightforward way. The flux, cross sections and other systematic uncertainties are included by allowing these inputs to deviate from their standard values in the computation of the expected rate in the $\text{i-j}^{\text{th}}$ bin, ${\mathrm{ N^{th}_{ij} }}$. Let the ${\mathrm{k^{th} }}$ input deviate from its standard value by ${\mathrm{\sigma_k \;\xi_k }}$, where ${\mathrm{ \sigma_k }}$ is its uncertainty. Then the value of ${\mathrm{ N^{th}_{ij} }}$ with the modified inputs is given by,
    
    \begin{equation}\label{eqn:cij}
    \mathrm{ N^{th}_{ij} =  N^{th}_{ij}(std) + \sum^{npull}_{k=1}
    c_{ij}^k \xi_k }\,,
    \end{equation}
    
    where ${\mathrm{ N^{th}_{ij}(std) }}$ is the expected rate in the $\text{i-j}^{\text{th}}$ bin calculated with the standard values of the inputs and \textit{npull}=5 is the number of sources of uncertainty.
    The ${\mathrm{ \xi_k }}$'s are called the \emph{pull} variables and they determine the number of ${\mathrm{ \sigma}}$'s by which the ${\mathrm{ k^{th} }}$ input deviates from its standard value. In Eq.~(\ref{eqn:cij}), ${\mathrm{ c_{ij}^k }}$ is the change in  ${\mathrm{N^{th}_{ij} }}$ when the ${\mathrm{ k^{th} }}$ input is changed by ${\mathrm{ \sigma_k }}$ (i.e. by 1 standard deviation). Since the uncertainties in the inputs are not very large, we only consider changes in ${\mathrm{N^{th}_{ij} }}$ that is linear in ${\mathrm{ \xi_k }}$. Thus we have the modified $\chi^2$ as,
    \begin{equation}  \label{eqn:chisq}
    \mathrm{ {\chi^2(\xi_k)} = \sum_{i,j}\;
    \frac{\left[~N_{ij}^{th}(std) \;+\; \sum^{npull}_{k=1}\;
    c_{ij}^k\; \xi_k - N_{ij}^{ex}~\right]^2}{N_{ij}^{ex}} +
    \sum^{npull}_{k=1}\; \xi_k^2  }\,,
    \end{equation}
    where the additional ${\rm{\xi_k^2}}$-dependent term is the penalty imposed for moving the value of the ${\mathrm{k^{th}}}$ input away from its standard value by ${\rm{\sigma_k \;\xi_k}}$. The $\chi^2$ with pulls, which includes the effects of all theoretical and systematic uncertainties (as mentioned in \autoref{table:LAr-parameter}), is obtained by minimizing ${\rm{\chi^2(\xi_k)}}$ with respect to all the pulls ${\rm{\xi_k}}$ as follows,
    \begin{equation}  \label{eqn:chisqmin}
    {\mathrm{ \chi^2_{pull} = Min_{\xi_k}\left[
    \chi^2(\xi_k)\right] }}
    \end{equation}
    In the case of a LArTPC detector without charge-id and with change-id, $\chi^2$ is defined as,
    \begin{equation}
     \chi^{2}_{w/o\text{ }charge-id} = \chi^{2}_{\mu^{-} + \mu^{+}} + \chi^{2}_{e^{-}+ e^{+}}
    \end{equation}
    \begin{equation}
           \chi^{2}_{charge-id} = \chi^{2}_{\mu^{-}}  + \chi^{2}_{ \mu^{+}} + \chi^{2}_{e^{-}+ e^{+}}
    \end{equation}
    Finally, $\Delta{\chi^{2}}$ is marginalized over the oscillation parameters as mentioned in \autoref{table:chi-parameter}. 
    \begin{table}[H]
		\centering
		\begin{tabular}{|c|c|c|}
			\hline
			Parameter & True Value & Marginalization Range\\\hline
			$\theta_{12}$ & $33.47^\circ$ & N.A.\\
			$\theta_{13}$ & $8.54^\circ$ & N.A.\\
			$\theta_{23}$ & $49^\circ(41^\circ)$ & $39^\circ:44^\circ(46^\circ:51^\circ)$\\
                $\theta_{14},\theta_{24}$ (A)& $7^\circ$ & $3^\circ:9^\circ$\\
                $\theta_{14},\theta_{24}$ (B)& $4^\circ$ & $0^\circ:6^\circ$\\
			$\Delta_{21}$ & $7.42\times 10^{-5}$ eV$^2$ &N.A.\\
                $\Delta_{31}$ & $2.515\times 10^{-3}$ eV$^2$ & N.A.\\
			$\Delta_{41}$ & $1$ eV$^2$ & N.A.\\
			$\delta_{13},\delta_{14}$ & many & $-180^\circ:180^\circ$\\
		    \hline
		\end{tabular}
		\caption{True values of all the oscillation parameters and their range of marginalization. Two different sets of $\theta_{14},\theta_{24}$ are considered. Set A is according to Global fit. Set B is taken considering MINOS+ bounds.}
		\label{table:chi-parameter}
	\end{table}
\section{Results and Discussion}\label{sec:discussion}
\begin{figure}[H]
		\centering
		\includegraphics[width=0.93\linewidth]{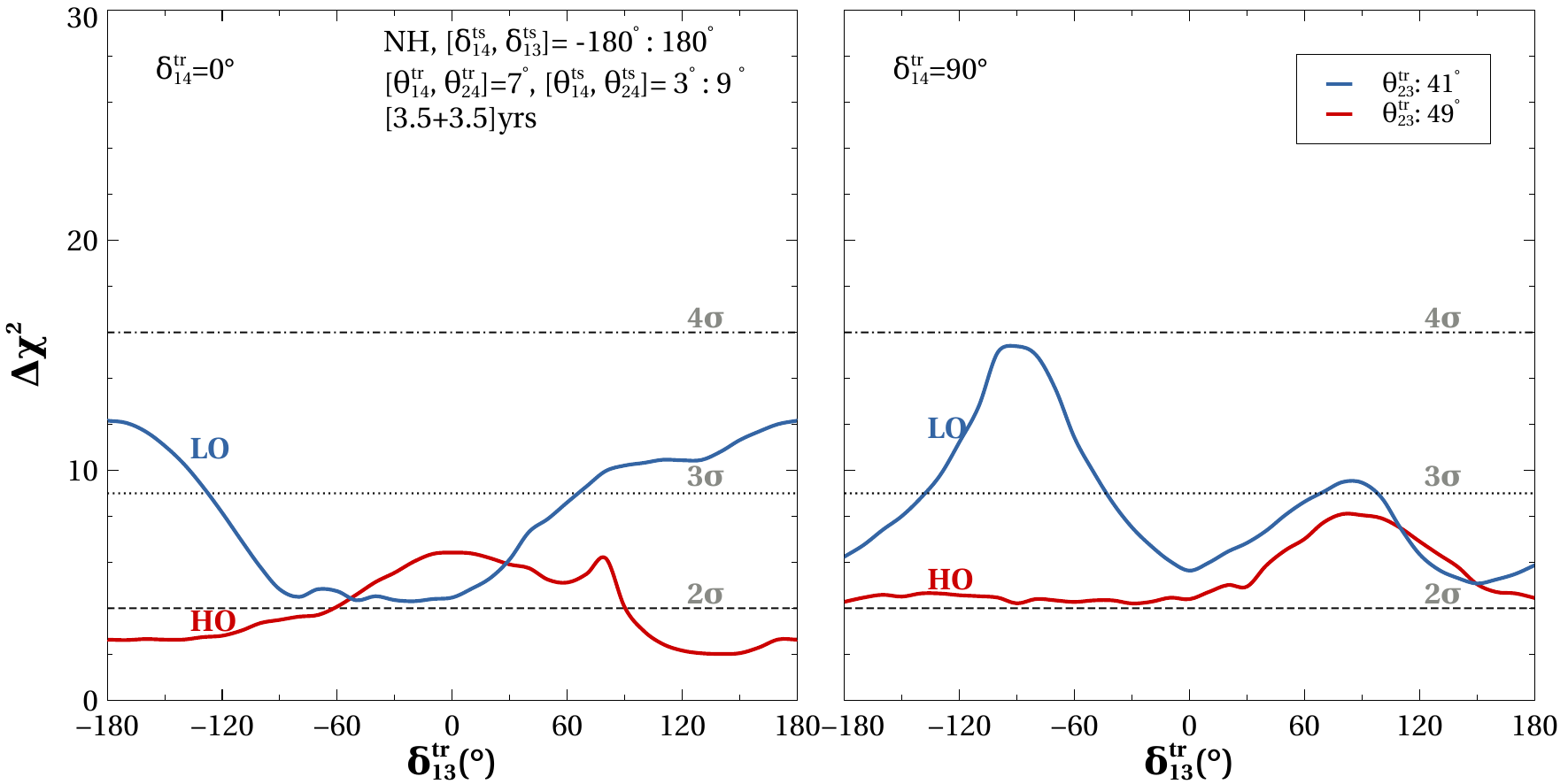}
		\caption{Sensitivity to the octant of $\theta_{23}$ with beam only analysis as a function of $\delta_{13}^{true}$ due to $\theta_{23}^{true}=41^\circ$ in LO(blue), and $49^\circ$in HO(red) for $\delta_{14}^{true}=0^\circ$ (left), $90^\circ$ (right).}
		\label{fig:chi_oct_sen_beam_dune}
    \end{figure}

    The results are demonstrated for beam only, atmospheric only, and a combination of both of these. We also explain the underlying degeneracies through the contour plots of octant sensitivity in $\delta_{13}-\delta_{14}$ test plane.
    In \autoref{fig:chi_oct_sen_beam_dune}, the sensitivity to the octant of $\theta_{23}$ degeneracy $(\Delta \chi^2)$ has been plotted as a function of true $\delta_{13}$ for NH. The marginalised $\Delta \chi^2$ values for true $\theta_{23}=41^\circ$(blue), $49^\circ$(red) have been shown for true $\delta_{14}=0^\circ$ (left panel), $90^\circ$ (right panel). The observable points are,
    \begin{itemize}
        \item The sensitivity of $\theta_{23}$ is prominently higher for LO as compared to HO for most of the $\delta_{13}^{\rm{true}}$ values. 

        \item The $\Delta \chi^2$ vs $\delta_{13}$ curve has strikingly different features for different $\delta_{14}^{true}$ values as can be seen from the two panels in \autoref{fig:chi_oct_sen_beam_dune}.

        \item For $\delta_{14}^{true}=0^\circ$ and LO the highest sensitivity comes around $\delta_{13}=\pm 120^\circ$. This feature can be understood from \autoref{fig:pme-pmeb-d13_2.5_oct_1300} which shows that there is no degeneracy in $P_{\mu e}(P_{\bar\mu \bar{e}})$ channel at $\delta_{13}=120^\circ(-120^\circ)$.

        \item On the other hand for $\delta_{14}^{true}=90^\circ$ the maximum sensitivity occurs for $\delta_{13}=-90^\circ$. From the red dashed curves depicted in the bottom panels of \autoref{fig:pme-pmeb-d13_2.5_oct_1300}, we can see that this sensitivity comes from $P_{\bar\mu \bar{e}}$ channel.

        \item For HO and $\delta_{14}^{true}=0^\circ$ the octant sensitivity is higher around the range $\delta_{13}=-60^\circ:60^\circ$. From the solid green curve drawn in the top panels of \autoref{fig:pme-pmeb-d13_2.5_oct_1300}, we can see that there is no degeneracy in the range $-120^\circ:0^\circ (0^\circ:120^\circ)$ comes from $P_{\mu e}(P_{\bar\mu \bar{e}})$ channel with a maximum difference between the HO curve and the LO band occurring at $\delta_{13}=-60^\circ(60^\circ)$.
        
        \item In case of $\delta_{14}=90^\circ$ in HO, the highest sensitivity is at $\delta_{13}=90^\circ$. From the top panel in \autoref{fig:pme-pmeb-d13_2.5_oct_1300}, it can be seen that is no degeneracy in $P_{\bar\mu \bar{e}}$ around $\delta_{13}=90^\circ$.
    \end{itemize}
    In the above discussion, we try to explain the salient features of \autoref{fig:chi_oct_sen_beam_dune} in terms of the probabilities plotted in \autoref{fig:pme-pmeb-d13_2.5_oct_1300} for an energy of 2.5 GeV. However, it should be borne in mind that the source has a broadband beam and contributions from other energy bins also influence the $\Delta\chi^2$.
    \begin{figure}[H]
		\centering
		\includegraphics[width=0.93\linewidth]{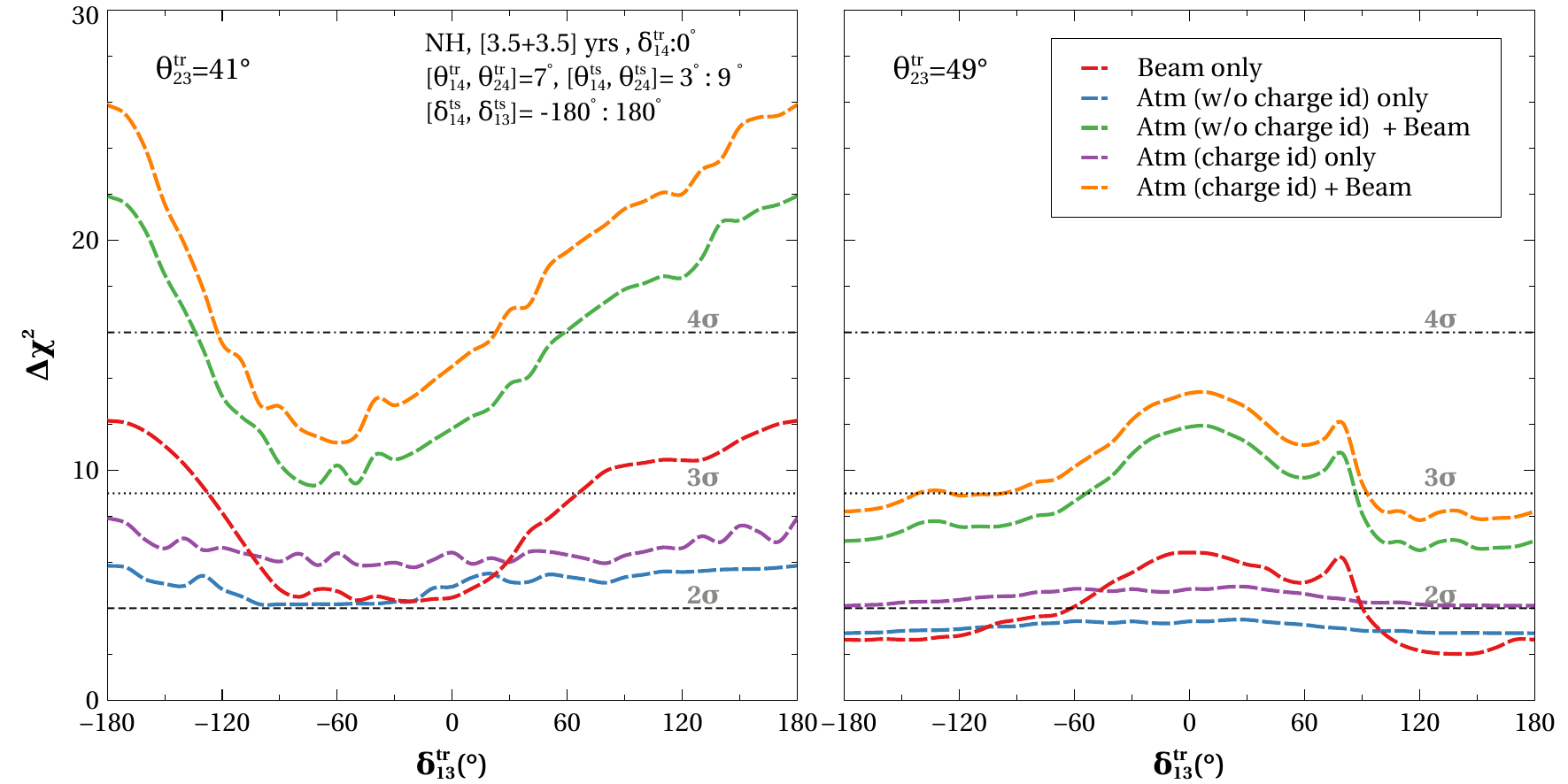}
		\caption{Sensitivity to the octant of $\theta_{23}$ as a function of $\delta_{13}^{\rm{true}}$ at $\delta_{14}^{\rm{true}}=0^\circ$ for $\theta_{23}^{\rm{true}}=41^\circ$ (left) and $49^\circ$ (right). The representative plots are shown for simulated data from beam only(red), atmospheric only w/o charge-id (blue), atmospheric only with charge-id (violet), beam+atmospheric w/o charge-id(green), and beam+atmospheric with charge-id(yellow) analysis with 280 kt-yr exposure.}
		\label{fig:chi_oct_sen_beam-atm_dune}
    \end{figure}
	In \autoref{fig:chi_oct_sen_beam-atm_dune}, we have shown the sensitivity to the octant of $\theta_{23}$ for atmospheric neutrinos without and with partial charge id of muon events(blue and violet curves respectively) as well as combining both beam and atmospheric data (green and orange curves) using the 40 kt far detector. In the figure, we also present the $\Delta\chi^2$ for beam only data (red curve). These plots are obtained for true values of $\theta_{23}=41^\circ$(left), $49^\circ$(right) respectively. Here are the observations from \autoref{fig:chi_oct_sen_beam-atm_dune},
	
    \begin{itemize}
        \item The sensitivity for atmospheric data is less than $2\sigma$ for HO and slightly higher than $2\sigma$ for LO for whole $\delta_{13}^{\rm{true}}$ parameter space.
        
        \item For the case including charge id, the sensitivity increases slightly. In matter $P_{\mu\mu}$, and $P_{\bar\mu \bar\mu}$ probabilities are very different due to the presence of resonant matter effect in $P_{\mu\mu}$ since we are considering normal hierarchy. This leads to a synergy when neutrino and anti-neutrino $\chi^2$ are added separately and enhances the sensitivity.
        
        \item Combining atmospheric and beam data, the sensitivity increases up to more than $4\sigma(3\sigma)$ for LO(HO) depending on the values of $\delta_{13}^{true}$.
        
        \item The $\Delta \chi^2$ for atmospheric data has very less dependence on $\delta_{13}^{true}$. Therefore in the combined case, the nature of $\Delta \chi^2$ is mostly dictated by the beam data.

    \end{itemize}
    
    \begin{table}[H]
		\centering
		\begin{tabular}{|c|c|c|c|c|c|}
			\hline
			$\theta_{23}$ & $\delta_{14}$ & Above $2\sigma$ & Above $3\sigma$ & Above $2\sigma$ & Above $3\sigma$\\\hline\multicolumn{4}{|c|}{Beam+Atmospheric w/o(with) charge-id}&\multicolumn{2}{c|}{Beam}\\\hline
			\multicolumn{2}{|c|}{True Value}&\multicolumn{4}{c|}{3.5+3.5 Years, $\theta_{14}=7^\circ,\theta_{24}=7^\circ$}\\\hline
			$41^\circ$ & $0^\circ$ & $100\%$(100\%) & $100\%$(100\%) & $100\%$ & $46\%$ \\
			$49^\circ$ & $0^\circ$ & $100\%$(100\%) & $38\%$(53\%) & $42\%$ & $0\%$ \\
			$41^\circ$ & $90^\circ$ & $100\%$(100\%) & $100\%$(100\%) & $100\%$ & $32\%$ \\
			$49^\circ$ & $90^\circ$ & $100\%$(100\%) & $30\%$(48\%) & $100\%$ & $0\%$ \\
		    \hline
		\end{tabular}
		\caption{The percentages of $\delta_{13}^{\rm{true}}$ parameter space that has $\chi^2$ value above $2\sigma,3\sigma$ for various combination of true values of $\theta_{23},\delta_{14}$ and $\theta_{14},\theta_{24}=7^\circ$ as seen in \autoref{fig:chi_oct_sen_beam_dune}, \autoref{fig:chi_oct_sen_beam-atm_dune}.}
		\label{table:chi_percentage}
	\end{table}
    
\begin{figure}[H]
		\centering
		\includegraphics[width=0.93\linewidth]{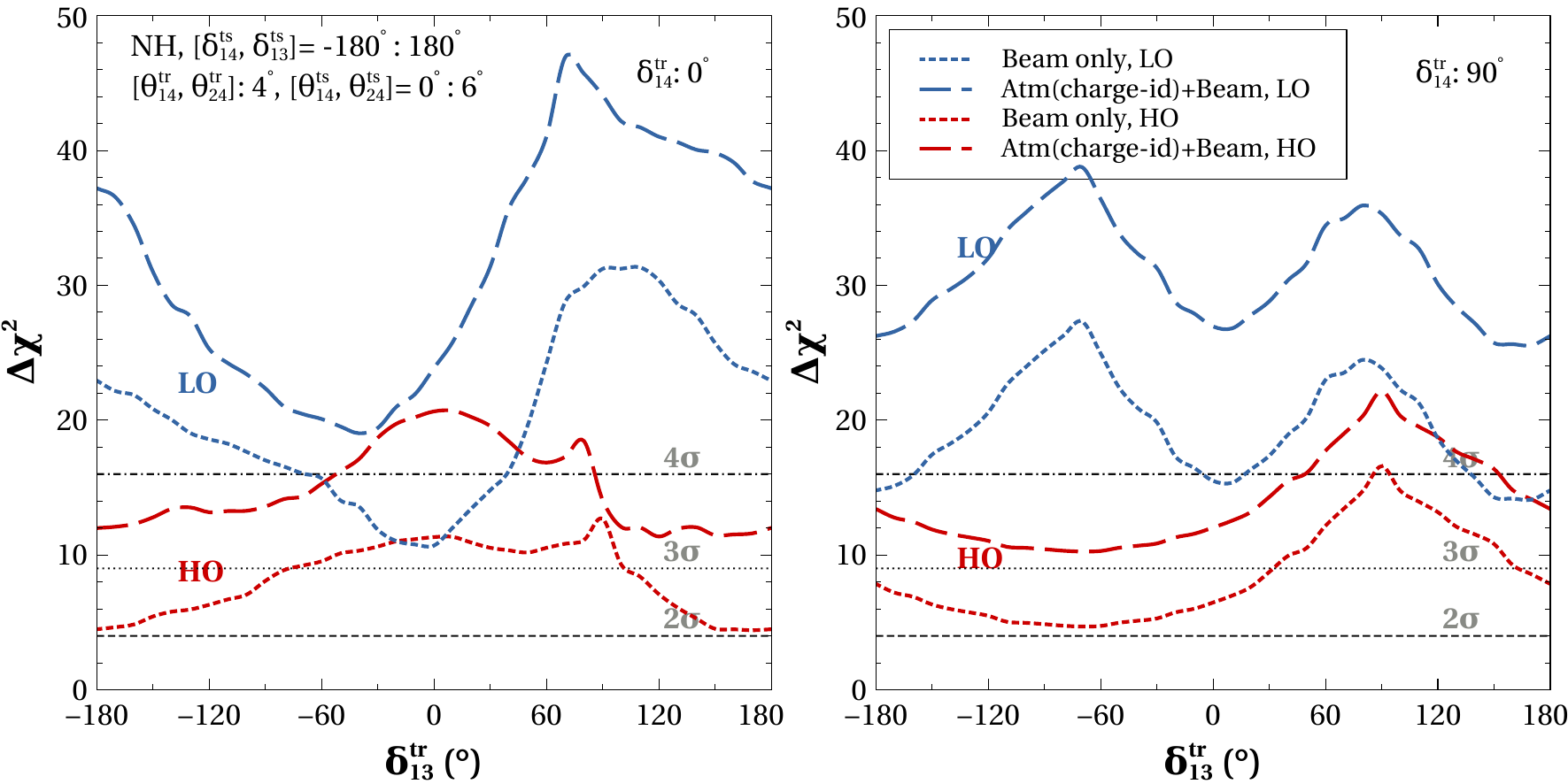}
		\caption{Sensitivity to the octant of $\theta_{23}$ with beam only (dotted) and beam+atmospheric with charge-id (dashed) analysis as a function of $\delta_{13}^{true}$ for true values of $\delta_{14}=0^\circ$ (left), $90^\circ$ (right). The representative plots are shown for true values of $\theta_{23}$ in HO (red), LO (blue), and $\theta_{14},\theta_{24}=4^\circ$.}
		\label{fig:chi_oct_beamatm_dune-446}
    \end{figure}

\begin{figure}[H]
		\centering
		\includegraphics[width=0.95\linewidth]{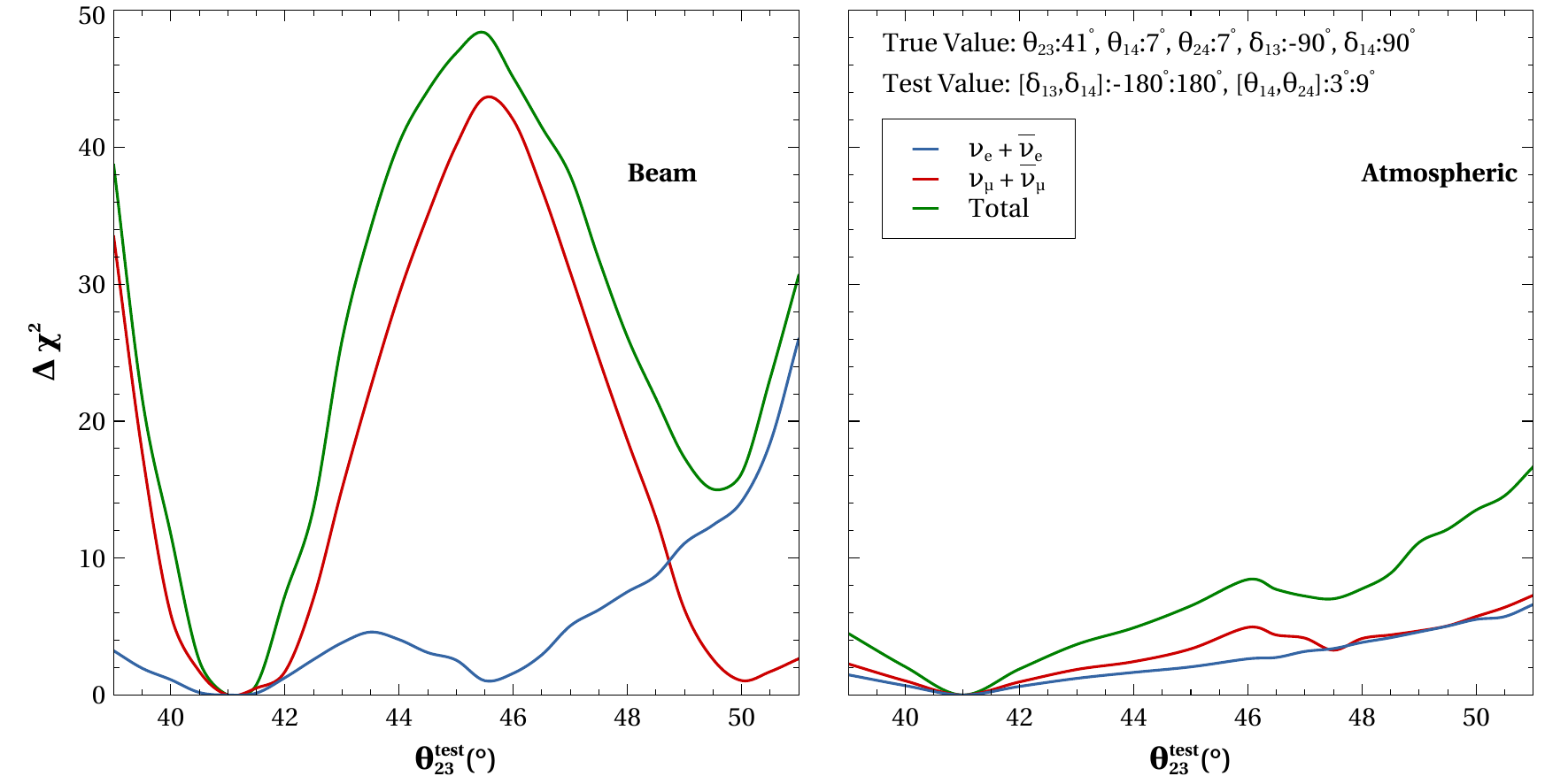}
		\caption{Octant sensitivity as a function of $\theta_{23}^{\rm{test}}$ from beam (left), and atmospheric (right) neutrinos using 280 kt-yr exposure of LArTPC detector with $\theta_{23}^{tr} = 41^{\circ},\delta_{13}^{tr}=-90^\circ, \delta_{14}^{tr}=90^\circ$.}
		\label{fig:chi_oct_sen_tr-41-90-90m}
    \end{figure}

\begin{figure}[H]
		\centering
		\includegraphics[width=0.9\linewidth]{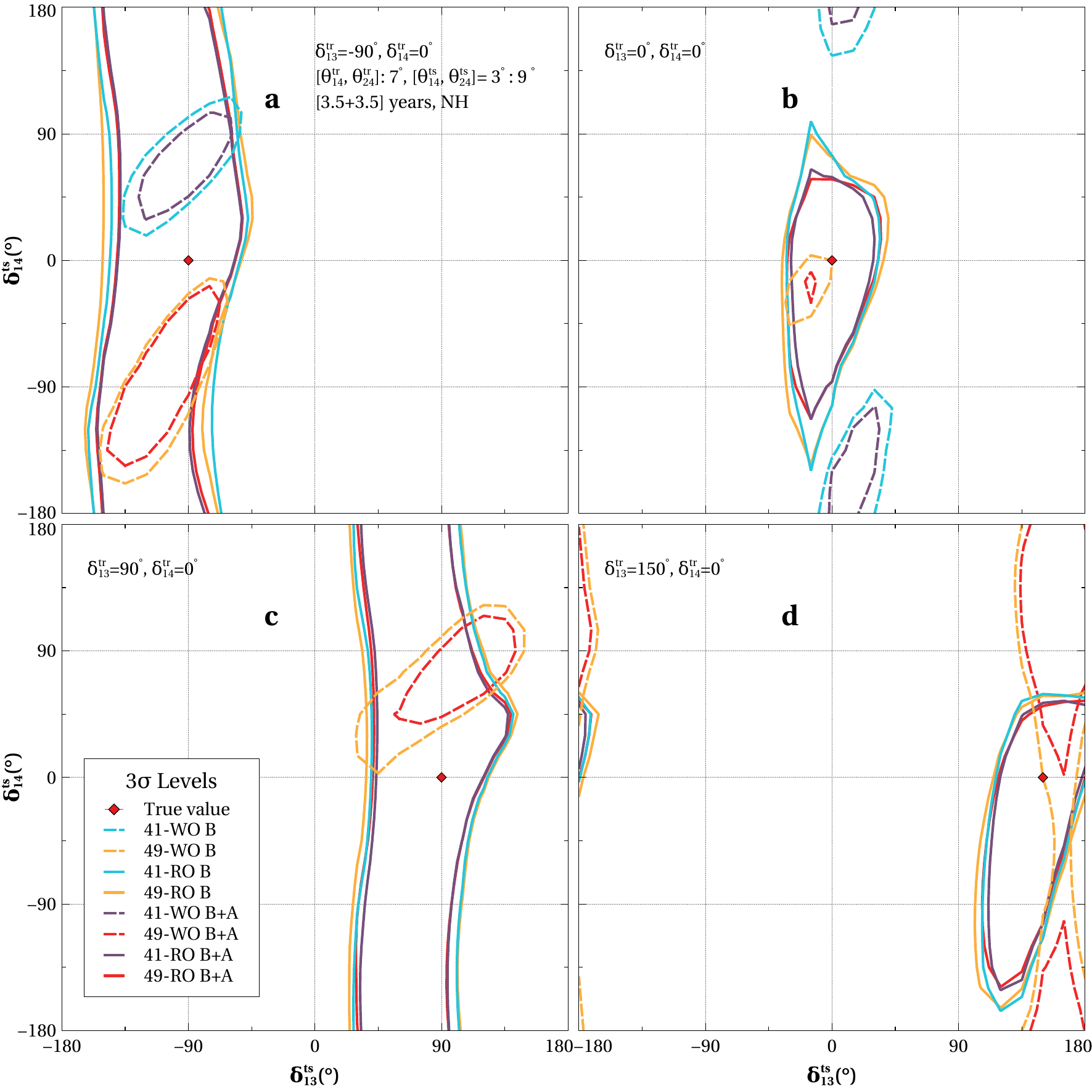}
		\caption{$3\sigma$ contour plot of sensitivity to the octant of $\theta_{23}$ in test $\delta_{13}-\delta_{14}$ plane with 7 years of data for $\delta_{14}^{true}=0^\circ$ and $\delta_{13}^{true}=-90^\circ,0^\circ,90^\circ,150^\circ$ in panels a,b,c,d respectively. The representative plots are shown for the true value of $\theta_{23}=41^\circ$ in LO (blue and violet) and $49^\circ$ in HO (yellow and red) for right octant solutions(solid) and wrong octant solutions(dashed) for simulated beam only (B) and beam+atmospheric (B+A) data.}
		\label{fig:chi_oct_sen_d13-d14-0}
    \end{figure}

    The percentage of values of $\delta_{13}^{\rm{true}}$ for which $\Delta\chi^2$ value of octant sensitivity for true value of $\theta_{14},\theta_{24}=7^\circ$ is above $2\sigma$, and $3\sigma$ are shown in the above \autoref{table:chi_percentage}.
    \begin{itemize}
        \item The percentage of values of the $\delta_{13}^{\rm{true}}$ for which $3 \sigma$ sensitivity is achieved, is higher for $\theta_{23}^{\rm{true}}$ in lower octant than in higher octant.
        
        \item The sensitivity for $\theta_{23}^{\rm{true}}=41^\circ$ (LO) is more than $3\sigma$ for $46\%(32\%)$ values of the $\delta_{13}^{\rm{true}}$ for $\delta_{14}^{\rm{true}}=0^\circ(90^\circ)$ with beam only data. However, in case of $\theta_{23}^{\rm{true}}=49^\circ$ (HO) $3\sigma$ sensitivity isn't observed for any values of $\delta_{13}^{\rm{true}}$ as $2\sigma$ sensitivity is achieved for $42\%(100\%)$ values of the $\delta_{13}^{true}$ for $\delta_{14}^{\rm{true}}=0^\circ(90^\circ)$.

        \item For the combination of both the beam and the atmospheric data (w/o charge-id), the sensitivity for $\theta_{23}=49^\circ$ increases to more than 3$\sigma$ for 38$\%(30\%)$ values of the $\delta_{13}^{\rm{true}}$ while for $41^\circ$ the whole $\delta_{13}^{\rm{true}}$ parameter space is allowed.
        
        \item When we use the combined data for beam, and atmospheric neutrinos with charge-id, the sensitivity improves further to provide more than 3$\sigma$ for all $\delta_{13}^{\rm{true}}$ values $\theta_{23}=41^\circ$ and for $53\%(48\%)$ of $\delta_{13}^{\rm{true}}$ values corresponding to $\delta_{14}^{\rm{true}}=0^\circ(90^\circ)$ for $\theta_{23}=49^\circ$.
    \end{itemize}
    In \autoref{fig:chi_oct_beamatm_dune-446}, the octant sensitivity is depicted as a function of $\delta_{13}^{true}$ corresponding to $\theta_{23}^{true}=41^\circ$ (blue) and $49^\circ$ (red) for true values of $\theta_{14},\theta_{24}=4^\circ$. In the left panel, $\delta_{14}^{true}$ is taken as $0^\circ$, and in the right panel, it is $90^\circ$. The dotted curves denote sensitivity for beam only cases, whereas, the dashed ones are for beam + atmospheric(with charge id) cases.
    We observe the following in \autoref{fig:chi_oct_beamatm_dune-446},
     \begin{itemize}
         \item An increase in the sensitivity in beam only and beam+atmospheric scenarios compared to the sensitivity obtained for the true value of $\theta_{14},\theta_{24}=7^\circ$ (\autoref{fig:chi_oct_sen_beam-atm_dune}).

         \item The sensitivity for $\theta_{23}=49^\circ$ is more than 3$\sigma$ irrespective of $\delta_{13}^{true}$ values when we consider the beam + atmospheric (with charge-id) analysis.

         \item For true value of $\theta_{23}=41^\circ$, the octant sensitivity is greater than 4$\sigma$ over the full range of $\delta_{13}^{true}$.
     \end{itemize} 
    The percentage of $\delta_{13}^{true}$ values for which more than 2$\sigma$, 3$\sigma$ octant sensitivity for true value of $\theta_{14},\theta_{24}=4^\circ$ is achieved have been enlisted in \autoref{table:chi_percentage2}.
    \begin{table}[H]
		\centering
		\begin{tabular}{|c|c|c|c|c|c|}
			\hline
			$\theta_{23}$ & $\delta_{14}$ & Above $2\sigma$ & Above $3\sigma$ & Above $2\sigma$ & Above $3\sigma$ \\\hline\multicolumn{4}{|c|}{Beam+Atmospheric with charge-id}&\multicolumn{2}{c|}{Beam}\\\hline
			\multicolumn{2}{|c|}{True Value}&\multicolumn{4}{c|}{3.5+3.5 Years, $\theta_{14}=4^\circ,\theta_{24}=4^\circ$}\\\hline
			$41^\circ$ & $0^\circ$ & $100\%$ & $100\%$ & $100\%$ & $100\%$\\
			$49^\circ$ & $0^\circ$ & $100\%$ & $100\%$ & $100\%$ & $50\%$\\
			$41^\circ$ & $90^\circ$ & $100\%$ & $100\%$ & $100\%$ & $75\%$\\
			$49^\circ$ & $90^\circ$ & $100\%$ & $100\%$ & $100\%$ & $36\%$\\
		    \hline
		\end{tabular}
		\caption{The percentages of $\delta_{13}^{\rm{true}}$ parameter space that has $\chi^2$ value above $2\sigma,3\sigma$ for various combination of true values of $\theta_{23},\delta_{14}$, and $\theta_{14},\theta_{24}=4^\circ$ as seen in \autoref{fig:chi_oct_beamatm_dune-446}}
		\label{table:chi_percentage2}
    \end{table}
    One of the noteworthy features of a liquid argon detector is its sensitivity to both electron and muon events. In order to explore if there is any synergy between these, we show in \autoref{fig:chi_oct_sen_tr-41-90-90m} how the value of $\chi^2$ for octant sensitivity from muon (red) and electron events (blue) varies with $\theta_{23}^{\rm{test}}$. These sensitivity curves are obtained using true values of $\theta_{23}=41^\circ, \delta_{13}=-90^\circ,\delta_{14}=90^\circ$ for beam (left) and atmospheric (right) neutrinos.
    The observations from \autoref{fig:chi_oct_sen_tr-41-90-90m} are as follows,
    \begin{itemize}
        \item In the case of beam neutrinos, the octant sensitivity for appearance channel increases with $\theta_{23}^{\rm{test}}$ whereas the sensitivity for disappearance channel mimics the nature of $\sin^2 2\theta_{23}$ with minima at $41^\circ$, and $50^\circ$. This different feature of octant sensitivity for $P_{\mu e}, P_{\mu\mu}$ channels can be seen in \autoref{fig:events_e_oct}. When we combine these two channels, the position of minimum sensitivity at $\theta_{23}^{\rm{test}}=50^\circ$ is still guided by muon events but due to the rising nature of electron $\chi^2$ a large octant sensitive contribution gets added and increases the overall value of the $\chi^2$.
        
        \item For atmospheric neutrinos, both muon and electron $\chi^2$ are similar. The muon $\chi^2$ is dictated by probabilities $P_{\mu\mu}, P_{e\mu}$, and the octant sensitivity coming from these channels is opposite, which dilutes the sensitivity for muons. On the other hand, for electron events, the octant sensitivity comes from only $P_{\mu e}$ since $P_{ee}$ doesn't depend on $\theta_{23}$. Therefore, even though atmospheric $\nu_\mu$ flux is almost twice as $\nu_e$ flux, both muon and electron events can give similar values of $\chi^2$. These features were also noted in three flavour case in \cite{Chatterjee:2013qus}.
    \end{itemize}
    In order to understand the $\theta_{23}$-$\delta_{13}$-$\delta_{14}$ degeneracies listed in \autoref{tab:8-deg}, we have provided the contour plots in $\delta_{13}$-$\delta_{14}$ plane showing the regions with octant sensitivity more than $3\sigma$. In \autoref{fig:chi_oct_sen_d13-d14-0}, the 3$\sigma$ contours are shown for the true value of sterile CP phase $\delta_{14}=0^\circ$ with four different true values of $\delta_{13}=-90^\circ,0^\circ,90^\circ,150^\circ$. In each panel, solid (dashed) lines represent the RO (WO) solutions. The blue, yellow (violet, red) correspond to contours from beam only (beam and atmospheric combined) analysis for $\theta_{23}^{true}=41^\circ,49^\circ$ respectively.
    
    \begin{table}[h!]
    	\centering
    	\begin{tabular}{|c|c|c|}
    		\hline
    		True $\delta_{13}$  &  True $\delta_{14}$ & Present Degeneracies\\\hline
    		$-90^\circ$ & $0^\circ$ & WO-R$\delta_{13}$-W$\delta_{14}$\\
    		$0^\circ$ & $0^\circ$ & WO-R$\delta_{13}$-R$\delta_{14}(49^\circ)$, WO-R$\delta_{13}$-W$\delta_{14}(41^\circ)$\\
    		$90^\circ$ & $0^\circ$ & WO-R$\delta_{13}$-W$\delta_{14}(49^\circ)$\\
    		$150^\circ$ & $0^\circ$ & WO-R$\delta_{13}$-R$\delta_{14}(49^\circ)$, WO-R$\delta_{13}$-W$\delta_{14}(49^\circ)$\\\hline
    	\end{tabular}
    	\caption{The degeneracies for different true value of $\delta_{13}$ with true $\delta_{14}=0^\circ$ as seen in \autoref{fig:chi_oct_sen_d13-d14-0}.}
    	\label{table:deg_countour-0}
    \end{table}
    \begin{figure}[H]
		\centering
		\includegraphics[width=0.9\linewidth]{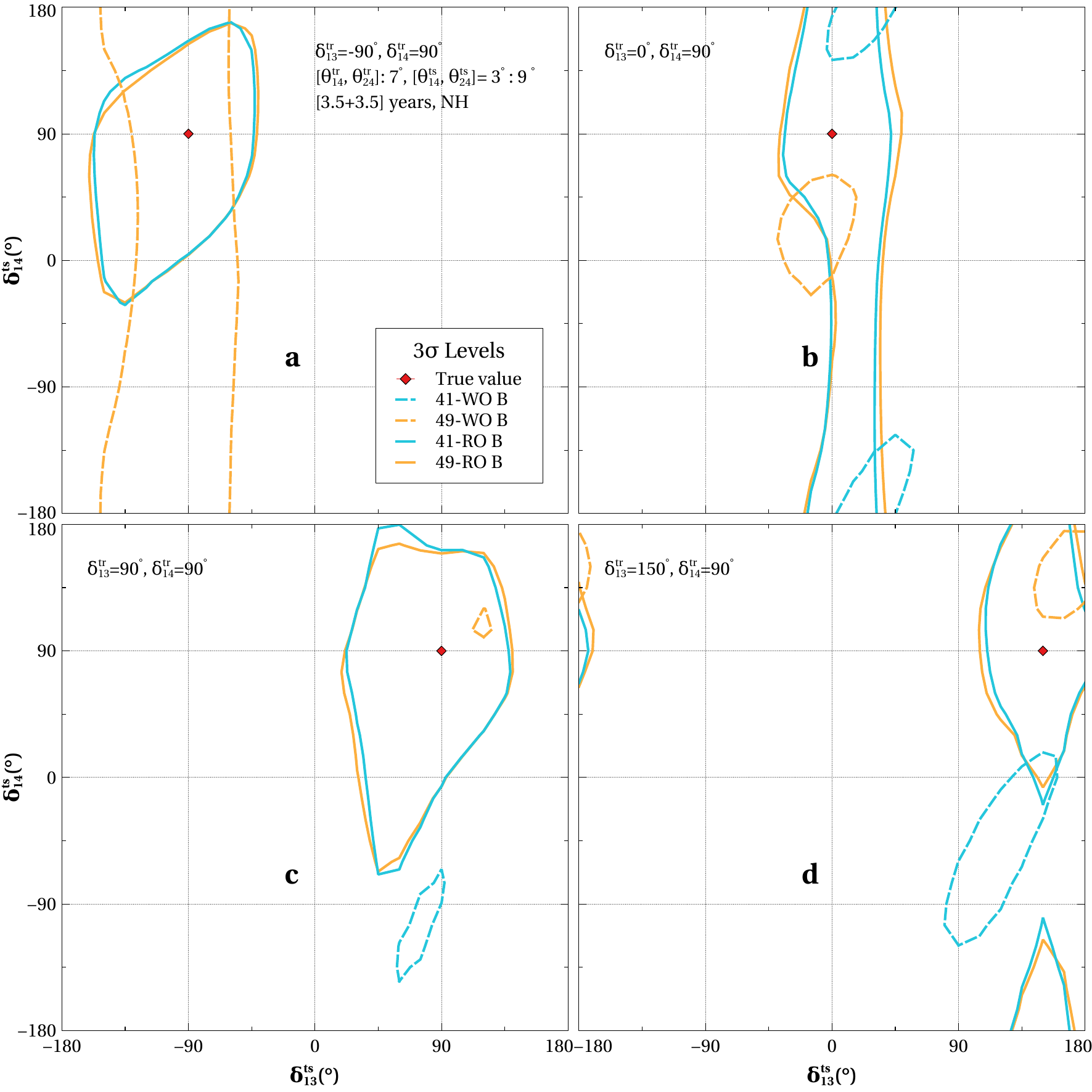}
		\caption{$3\sigma$ contour plot of sensitivity to the octant of $\theta_{23}$ in test $\delta_{13}-\delta_{14}$ plane with 7 years of beam only simulated data for $\delta_{14}^{true}=90^\circ$ and $\delta_{13}^{true}=-90^\circ,0^\circ,90^\circ,150^\circ$ in panels a,b,c,d respectively. The representative plots are shown for true value of $\theta_{23}=41^\circ$ in LO (blue) and $49^\circ$(yellow) in HO for right octant solutions(solid) and wrong octant solutions(dashed).}
		\label{fig:chi_oct_sen_d13-d14-90}
    \end{figure}
    \begin{table}[h!]
    	\centering
    	\begin{tabular}{|c|c|c|}
    		\hline
    		True $\delta_{13}$  &  True $\delta_{14}$ & Present Degeneracies\\\hline
    		$-90^\circ$ & $90^\circ$ & WO-R$\delta_{13}$-R$\delta_{14}(49^\circ)$, WO-R$\delta_{13}$-W$\delta_{14}(49^\circ)$ \\
    		$0^\circ$ & $90^\circ$ & WO-R$\delta_{13}$-W$\delta_{14}$\\
    		$90^\circ$ & $90^\circ$ & WO-W$\delta_{13}$-W$\delta_{14}(49^\circ)$, WO-R$\delta_{13}$-W$\delta_{14}(41^\circ)$\\
    		$150^\circ$ & $90^\circ$ &  WO-R$\delta_{13}$-W$\delta_{14}$\\\hline
    	\end{tabular}
    	\caption{The degeneracies for different true value of $\delta_{13}$ with true $\delta_{14}=90^\circ$ as seen in \autoref{fig:chi_oct_sen_d13-d14-90}.}
    	\label{table:deg_countour-90}
    \end{table}
    The noteworthy observations from \autoref{fig:chi_oct_sen_d13-d14-0} are as follows,
    
    \begin{itemize}
    	\item In panel "\textit{a}", the solid contours spanning the full range of $\delta_{14}$ indicate true solutions with poor precision in $\delta_{14}$ for both $\theta_{23}^{true}=41^\circ,49^\circ$. We also observe dashed contours indicating WO-R$\delta_{13}$-W$\delta_{14}$ solutions for both $\theta_{23}^{true}$.
    	
    	\item In panel "\textit{b}", the precision of the true solutions improves significantly. A small region of WO solutions for $\theta_{23}^{true}=49^\circ$ occurs adjacent to the true value. We also find WO-R$\delta_{13}$-W$\delta_{14}$ solutions for $\theta_{23}^{true}=41^\circ$.
    	
    	\item Comparing the true solutions in panels "\textit{c}", and "\textit{d}" but the precision of $\delta_{14}$ is notably better in "d". In these panels, WO solutions are present for only $\theta_{23}^{true}=49^\circ$. For $\theta_{23}^{true}=41^\circ$, the octant can be determined at more than $3\sigma$ sensitivity as seen from the solid blue curve in the left panel of \autoref{fig:chi_oct_sen_beam_dune} and hence WO solutions are not observed. In panel "c" we find WO-R$\delta_{13}$-W$\delta_{14}$ solution wheres the WO-R$\delta_{13}$ solutions are observed in panel "\textit{d}".
    	
    	\item Inclusion of atmospheric analysis shrinks all the contours improving octant sensitivity. The choice of $\delta_{13}^{true}$ affects the precision of RO solutions as well as the occurrence of degeneracies.
    \end{itemize}
    
    Similarly, we have plotted the $3\sigma$ contours in \autoref{fig:chi_oct_sen_d13-d14-90} showing WO (dashed), and RO (solid) solutions w.r.t. true values of $\theta_{23}=41^\circ$ (blue), and $49^\circ$ (yellow) for the true value of $\delta_{14}=90^\circ$ with  $\delta_{13}=-90^\circ,0^\circ,90^\circ,150^\circ$ using beam-only analysis. The observations from \autoref{fig:chi_oct_sen_d13-d14-90} are as follows,
    
    \begin{itemize}
    	\item In panel "\textit{a}", we see the WO-R$\delta_{13}$ solutions spanning the full range of $\delta_{14}$ for only $\theta_{23}^{true}=49^\circ$. We also find true solutions with notable precision in $\delta_{14}$ for both $\theta_{23}^{true}=41^\circ,49^\circ$ as compared to panel "\textit{a}" in \autoref{fig:chi_oct_sen_d13-d14-0}.
    	
    	\item In panel "\textit{b}", the precision of $\delta_{14}$ in true solutions deteriorates w.r.t panel "\textit{a}" covering the full $\delta_{14}$ range. We observe a small region of WO-R$\delta_{13}$-W$\delta_{14}$ solution for $\theta_{23}^{true}=49^\circ$, along with a bigger region of WO-R$\delta_{13}$-W$\delta_{14}$ solution for $\theta_{23}^{true}=41^\circ$.
    	
    	\item In panels "\textit{c}" and "\textit{d}", the true solutions show better precision in $\delta_{14}$ as compared to the same panels in \autoref{fig:chi_oct_sen_d13-d14-0}. We can also observe for $\theta_{23}^{true}=49^\circ$ a tine region of WO-W$\delta_{13}$-W$\delta_{14}$ in panel "\textit{c}" while in panel "\textit{d}" WO-R$\delta_{13}$-W$\delta_{14}$ solutions occur. There are WO-R$\delta_{13}$-W$\delta_{14}$ solutions for $\theta_{23}^{true}=41^\circ$ in both panel "\textit{c}", and "\textit{d}" but the region is smaller in "\textit{c}".
    	
    	\item Overall, we see the precision of the RO true solutions along with the size and type of WO contours depend on $\delta_{13}^{true}$ for fixed $\delta_{14}^{true}$.
    \end{itemize}
    The most common degeneracies seen in \autoref{fig:chi_oct_sen_d13-d14-0}, \autoref{fig:chi_oct_sen_d13-d14-90} are  WO-R$\delta_{13}$-R$\delta_{14}$, WO-R$\delta_{13}$-W$\delta_{14}$. It indicates that the presence of $\delta_{14}$ creates more problems in precise measurement of the octant of $\theta_{23}$. We also observe true solutions with poor precision in $\delta_{14}$. If we repeat the above analysis for true values of $\theta_{14},\theta_{24}=4^\circ$ along with marginalization in the range of $0-6^\circ$, the $3\sigma$ contours get smaller due to higher octant sensitivity.

	The regions under $3\sigma$ sensitivity in the contour plots of \autoref{fig:chi_oct_sen_d13-d14-0}, \autoref{fig:chi_oct_sen_d13-d14-90} can be understood using the difference in the probability plots in $\delta_{13}-\delta_{14}$ plane. We will mainly focus on the dominant $P_{\mu e}$ channel to understand the effect.
    \begin{figure}[H]
        \centering
        \includegraphics[width=0.74\linewidth]{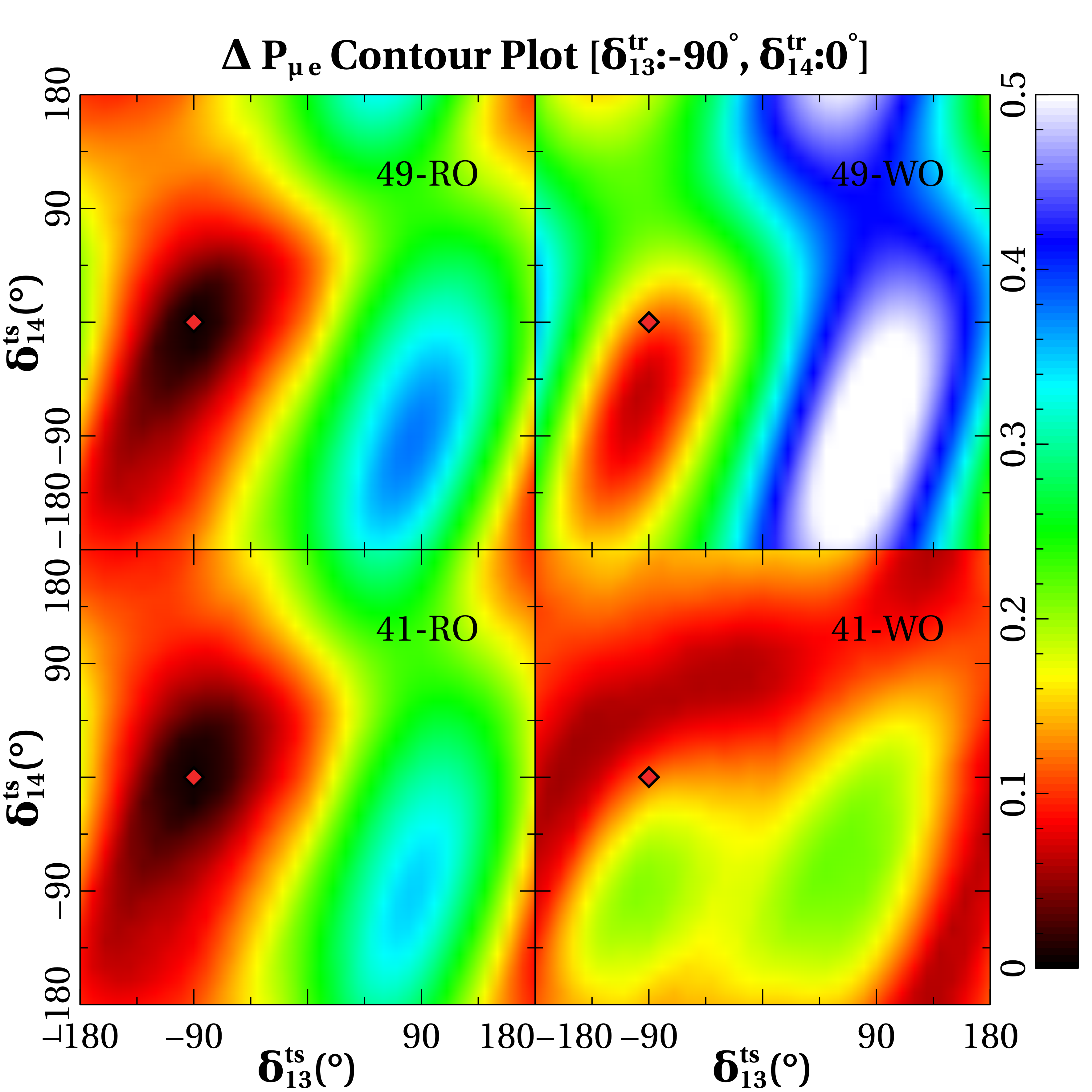}
        \caption{Contour plot in test $\delta_{13}-\delta_{14}$ plane showing the difference in probability $\Delta P_{\mu e}$ with $\theta_{23}$ being fixed at one octant while $\theta_{23}$ varies in the opposite octant for WO solutions (right) and in the same octant for RO solutions (left) at true values of $\delta_{13}=-90^\circ,\delta_{14}=0^\circ,\theta_{23}=49^\circ$(top), $41^\circ$(bottom). Black and dark red show the least differences, while blue and white show the highest. }
        \label{fig:delP_d23-14_90m-0}
    \end{figure}
     In \autoref{fig:delP_d23-14_90m-0}, the contour plot in test $\delta_{13}$-$\delta_{14}$ plane represents the difference between the probabilities $P_{\mu e}$ of opposite octants while varying the $\theta_{23}$ value only in same (left) /opposite (right) octant for the true $\theta_{23}=49^\circ$(top), $41^\circ$(bottom) with $\delta_{13}^{tr} = -90^\circ, \delta_{14}^{tr}= 0^\circ$ corresponding to panel "\textit{a}" of \autoref{fig:chi_oct_sen_d13-d14-0}. The understandings are as follows,
    \begin{itemize}
        \item First, we consider the right octant solutions in the panels at the left side column. It can be clearly seen that the black and darker red regions around the true value on the left side of $\delta_{13}-\delta_{14}$ plane where the difference in the probability is minimum in \autoref{fig:delP_d23-14_90m-0} is similar to the $3\sigma$ regions under the \textit{solid} curves in panel "\textit{a}" of \autoref{fig:chi_oct_sen_d13-d14-0}. These darker regions also indicate poor precision of $\delta_{14}$.

        \item For $49^\circ$-WO solution, minima arise in the darker red region including the true value in the top-right panel of \autoref{fig:delP_d23-14_90m-0} similar to the \textit{yellow-dashed} contour in panel "\textit{a}" of \autoref{fig:chi_oct_sen_d13-d14-0}. Similarly, for $41^{\circ}$-WO solutions in the bottom-right panel, the minimum difference is observed in the darker red region just above the true value similar to the \textit{blue-dashed} contour in panel "\textit{a}" of \autoref{fig:chi_oct_sen_d13-d14-0}. These darker red regions clearly show precise WO-R$\delta_{13}$-W$\delta_{14}$ degenerate solutions.
    \end{itemize}
    
\section{Conclusions}\label{sec:conclusion}
    In this work, we expound the possibility of determining the octant of $\theta_{23}$, in the 3+1 framework, assuming the presence of an extra sterile neutrino in addition to the three standard ones. We present our results for a beam based long baseline experiment as well as for atmospheric neutrinos considering a LArTPC detector. We also do a combined analysis of beam and atmospheric neutrinos and probe the synergies between these two options, which can result in an enhanced sensitivity. For the beam neutrinos, the typical baseline considered in our study is 1300 km, similar to that proposed by the DUNE collaboration.
    We provide the analytic expressions for oscillation probabilities in the presence of an extra sterile neutrino using the approximation that the mass squared difference $\Delta_{21}$ is zero. We show that these expressions match well with the numerical probabilities, especially in the resonance region. 

    We study in detail the different parameter degeneracies, emphasizing especially the influence of the phases $\delta_{13}, \delta_{14}$ in the determination of octant of $\theta_{23}$. This is done by plotting the probability curves for two different $\theta_{23}$ values belonging to the opposite octants-- (i) as a function of  $\delta_{13}/\delta_{14}$ for fixed energy and baseline, (ii) as a function of energy, for varying $\delta_{13},\delta_{14}$ at fixed baselines. We also illustrate (iii) the difference in the appearance and disappearance probabilities for two values $\theta_{23}$ belonging to opposite octants in the $\cos \theta_\nu -E$ plane.

    We perform a $\chi^2$ analysis and show that for a set of true values of sterile parameters, one can achieve more than $3\sigma$ octant sensitivity depending upon the true value of $\delta_{13}$ using beam neutrinos. The representative true values of the sterile neutrino parameters considered by us correspond to $\Delta_{41}=1$ eV$^2$, $[\theta_{14},\theta_{24}]=7^\circ$ and $4^\circ$, $\delta_{14}=0^\circ$, and $90^\circ$, $\theta_{34}=0^\circ$. For true values of $\theta_{14},\theta_{24}=7^\circ$, $\theta_{23}=41^\circ(49^\circ)$, and $\delta_{14}=90^\circ$ one gets more than $3\sigma$ sensitivity for $51\%(18\%)$ of the $\delta_{13}^{true}$ space. On the other hand for true values of $\theta_{14},\theta_{24}=4^\circ$, the sensitivity for $\theta_{23}=41^\circ(49^\circ)$, and $\delta_{14}=90^\circ$ reaches more than $3\sigma$ sensitivity for $75\%(36\%)$ of the $\delta_{13}^{true}$ space. It can be noted that greater sensitivity is obtained when true values of $\theta_{14},\theta_{24}$ are smaller.
    
    
    
    In case of $\theta_{14},\theta_{24}=7^\circ$, combining the beam and the atmospheric neutrinos (with charge-id), we can obtain $3\sigma$ sensitivity in the $100\%(48\%)$ of the $\delta_{13}^{true}$ space for $\theta_{23}=41^\circ(49^\circ),\delta_{14}=90^\circ$. However, the sensitivity for $\theta_{23}=41^\circ(49^\circ),\delta_{14}=90^\circ$ is over $3\sigma$ for entire range of $\delta_{13}^{true}$ when $\theta_{14},\theta_{24}=4^\circ$.
    
    At fixed hierarchy, there can be a total of 8-fold degeneracies (\autoref{tab:8-deg}) with at least one of the parameters - octant of $\theta_{23}$, $\delta_{13}$, $\delta_{14}$ assuming a wrong value. We also identify the extra degeneracies due to the presence of $\delta_{14}$ assuming the normal hierarchy and summarise these in \autoref{table:deg_countour-0}, and \autoref{table:deg_countour-90}. We can conclude that the presence of the phase $\delta_{14}$ leads to the occurrence of new degeneracies that hinder the discovery of the octant of $\theta_{23}$ precisely.
    
    In summary, the combination of the beam and the atmospheric neutrinos provides promising results using a LArTPC detector in the presence of an eV scale sterile neutrino.
    
\acknowledgments
   We want to express special acknowledgement to Jaydeep Dutta, Sushant Raut, and Samiron Roy for their involvement in the earlier stages. We acknowledge discussions with Monojit Ghosh. SP acknowledges Kaustav Chakraborty for help in learning numerical analysis using GLoBES. AC acknowledges the University of Pittsburgh, where initial work has been done. AC also acknowledges the Ramanujan Fellowship (RJF/2021/000157), of the Science and Engineering Research Board of the Department of Science and Technology, Government of India. SG acknowledges the J.C. Bose Fellowship (JCB/2020/000011) of the Science and Engineering Research Board of the Department of Science and Technology, Government of India. It is to be noted that this work has been done solely by the authors and is not representative of the DUNE collaboration.

\appendix

\section{Probability calculation using Cayley Hamilton formalism}
We will now find out the analytic probability using the Cayley-Hamilton formalism\cite{Ohlsson:1999xb, Kamo:2002sj, Akhmedov:2004ny}. We calculate the time evolution operator and do not introduce auxiliary matter mixing angles.

The flavour eigenstates $ \psi_\alpha $ and mass eigenstates $ \psi_i $ are related as
\begin{equation}
	\psi_i=\sum_{j=e,\mu,\tau,s} U^\star_{\alpha j}\psi_j
\end{equation}
where $ U_{\alpha j} $ is component of unitary mixing matrix corresponding to mixing between $\psi_\alpha,\psi_j$,
\begin{equation}
	U=\tilde{R}_{34}(\theta_{34},\delta_{34})R_{24}(\theta_{24})\tilde{R}_{14}(\theta_{14},\delta_{14})R_{23}(\theta_{23})\tilde{R}_{13}(\theta_{13},\delta_{13})R_{12}(\theta_{12})
\end{equation}
The Schrodinger equation in mass basis is given as,
\begin{equation}\label{Scr_mass}
	\iota \dfrac{d}{dt}\psi_m (t)= \mathcal{H}_m \psi_m(t)
\end{equation} 
where total Hamiltonian $\mathcal{H}_m$ in mass basis, and interaction Hamiltonian $V_f$ in flavour basis are given as follows, 
\begin{eqnarray}
	\mathcal{H}_m=H_m + U^{-1}V_fU\\
	V_f=H_{int}=diag(2A',0,0,A') 
\end{eqnarray}

Equation \eqref{Scr_mass} gives the solution with time evolution operator $ e^{-\iota \mathcal{H}_m t} $ as,
\begin{equation}\label{sol_mass}
	\psi_m(t)=e^{-\iota \mathcal{H}_m t} \psi_m(0)
\end{equation}
We get the solution in terms of distance $ L $ travelled by neutrinos in time $ t $ as,
\begin{equation}\label{sol_mass2}
	\psi_m(L)=\psi_m(t=L)=e^{-\iota \mathcal{H}_m t} \psi_m(0)\equiv U_m(L)\psi_m(0)
\end{equation}
Solution in flavour state $ \psi_f $ is expressed at $ t=L $ as,
\begin{equation}
  \begin{split}
	\psi_f(L)&=U \phi_m(L)=U e^{-\iota \mathcal{H}_m t} U^{-1}U\psi_m(0)\\
 &=U e^{-\iota \mathcal{H}_m t} U^{-1} \psi_f(0)\equiv U_f(L) \psi_f(0)
\end{split}
\end{equation}

We will calculate the time evolution operator, i.e., the exponential of the matrix $ \mathcal{H}_m $ using the Cayley-Hamilton theorem. We construct a traceless matrix out of $ \mathcal{H}_m $ as ,
\begin{equation}
	\mathcal{H}_m=T+\dfrac{1}{4}(tr \mathcal{H}_m)I
\end{equation}
The time evolution operator is then redefined as,
\begin{equation}
	U_m(L)=e^{-\iota \mathcal{H}_m  L}=\phi e^{-\iota T L}
\end{equation}
The elements of the traceless matrix $ T $ in mass basis are as follows,
\begin{widetext}    
\begin{eqnarray}
	T_{11}&=&A\left[-\cos ^2\theta_{12} \left(2 \sin \theta_{13} \cos \theta_{13} \sin \theta_{14} \sin \theta_{23} \sin \theta_{24} \cos \theta_{24} \cos (\delta_{13}-\delta_{14})+\cos 2\theta_{23} \sin ^2\theta_{24}\right)\right.\nonumber\\&&\left.+2 \sin \theta_{12} \cos \theta_{12} \cos \theta_{23} \sin \theta_{24} (\cos \delta_{13} \sin \theta_{13} \sin \theta_{23} \sin \theta_{24}-\cos \delta_{14} \cos \theta_{13} \sin \theta_{14} \cos \theta_{24})\right.\nonumber\\&&\left.+\cos ^2\theta_{12} \cos ^2\theta_{13} \left(2-\sin ^2\theta_{24} \left(\sin ^2\theta_{14}+\sin ^2\theta_{23}\right)-\sin ^2\theta_{14}\right)\right.\nonumber\\&&\left.+\cos ^2\theta_{23} \sin ^2\theta_{24}\right]-\frac{3 A}{4}+\frac{1}{4} (-\Delta_{21}-\Delta_{31}-\Delta_{41})\\
	T_{12}&=&A\left[-\sin \theta_{12} \cos \theta_{12} \left(2 \sin \theta_{13} \cos \theta_{13} \sin \theta_{14} \sin \theta_{23} \sin \theta_{24} \cos \theta_{24} \cos (\delta_{13}-\delta_{14})+\cos 2\theta_{23} \sin ^2\theta_{24}\right)\right.\nonumber\\&&\left.-\sin \theta_{13} \sin \theta_{23} \cos \theta_{23} \sin ^2\theta_{24} \left(e^{-i \delta_{13}} \cos ^2\theta_{12}-e^{i \delta_{13}} \sin ^2\theta_{12}\right)\right.\nonumber\\&&\left.+\cos \theta_{13} \sin \theta_{14} \cos \theta_{23} \sin \theta_{24} \cos \theta_{24} \left(e^{-i \delta_{14}} \cos ^2\theta_{12}-e^{i \delta_{14}} \sin ^2\theta_{12}\right)\right.\nonumber\\&&\left.+\sin \theta_{12} \cos \theta_{12} \cos ^2\theta_{13} \left(2-\sin ^2\theta_{24} \left(\sin ^2\theta_{14}+\sin ^2\theta_{23}\right)-\sin ^2\theta_{14}\right)\right]\\
	T_{13}&=&A\left[-e^{i \delta_{14}-2 i \delta_{13}} \cos \theta_{12} \sin ^2\theta_{13} \sin \theta_{14} \sin \theta_{23} \sin \theta_{24} \cos \theta_{24}\right.\nonumber\\&&\left.- e^{i \delta_{14}-i \delta_{13}} \sin \theta_{12} \sin \theta_{13} \sin \theta_{14} \cos \theta_{23} \sin \theta_{24} \cos \theta_{24}\right.\nonumber\\&&\left.+ e^{-i \delta_{14}} \cos \theta_{12} \cos ^2\theta_{13} \sin \theta_{14} \sin \theta_{23} \sin \theta_{24} \cos \theta_{24}\right.\nonumber\\&&\left.
	+e^{-i \delta_{13}} \cos \theta_{12} \sin \theta_{13} \cos \theta_{13} \left(2-\sin ^2\theta_{24} \left(\sin ^2\theta_{14}+\sin ^2\theta_{23}\right)-\sin ^2\theta_{14}\right)\right]\nonumber\\&&-A \sin \theta_{12} \cos \theta_{13} \sin \theta_{23} \cos \theta_{23} \sin ^2\theta_{24}\\
	T_{14}&=&A\left[ e^{-i \delta_{13}} \cos \theta_{12} \sin \theta_{13} \cos \theta_{14} \sin \theta_{23} \sin \theta_{24} \cos \theta_{24}\right.\nonumber\\&&\left.+ e^{-i \delta_{14}} \cos \theta_{12} \cos \theta_{13} \sin \theta_{14} \cos \theta_{14} \left(2-\cos ^2\theta_{24}\right)+ \sin \theta_{12} \cos \theta_{14} \cos \theta_{23} \sin \theta_{24} \cos \theta_{24}\right]
\end{eqnarray}
\begin{eqnarray}
	T_{22}&=&A\left[-\sin ^2\theta_{12} \left(2 \sin \theta_{13} \cos \theta_{13} \sin \theta_{14} \sin \theta_{23} \sin \theta_{24} \cos \theta_{24} \cos (\delta_{13}-\delta_{14})+\cos 2\theta_{23} \sin ^2\theta_{24}\right)\right.\nonumber\\&&\left.+2 \sin \theta_{12} \cos \theta_{12} \cos \theta_{23} \sin \theta_{24} (\cos \delta_{14} \cos \theta_{13} \sin \theta_{14} \cos \theta_{24}-\cos \delta_{13} \sin \theta_{13} \sin \theta_{23} \sin \theta_{24})\right.\nonumber\\&&\left.+ \sin ^2\theta_{12} \cos ^2\theta_{13} \left(2-\sin ^2\theta_{24} \left(\sin ^2\theta_{14}+\sin ^2\theta_{23}\right)-\sin ^2\theta_{14}\right)\right.\nonumber\\&&\left.+ \cos ^2\theta_{23} \sin ^2\theta_{24}\right]-\frac{3 A}{4}+\frac{1}{4} (3 \Delta_{21}-\Delta_{31}-\Delta_{41})\\	
	T_{23}&=&A \left[-e^{i \delta_{14}-2 i \delta_{13}} \sin \theta_{12} \sin ^2\theta_{13} \sin \theta_{14} \sin \theta_{23} \sin \theta_{24} \cos \theta_{24}\right.\nonumber\\&&\left.+e^{i \delta_{14}-i \delta_{13}} \cos \theta_{12} \sin \theta_{13} \sin \theta_{14} \cos \theta_{23} \sin \theta_{24} \cos \theta_{24}\right.\nonumber\\&&\left.+e^{-i \delta_{14}} \sin \theta_{12} \cos ^2\theta_{13} \sin \theta_{14} \sin \theta_{23} \sin \theta_{24} \cos \theta_{24}\right.\nonumber\\&&\left.+ e^{-i \delta_{13}} \sin \theta_{12} \sin \theta_{13} \cos \theta_{13} \left(2-\sin ^2\theta_{24} \left(\sin ^2\theta_{14}+\sin ^2\theta_{23}\right)-\sin ^2\theta_{14}\right)\right]\nonumber\\&&+A \cos \theta_{12} \cos \theta_{13} \sin \theta_{23} \cos \theta_{23} \sin ^2\theta_{24}\\
	T_{24}&=&A\left[e^{-i \delta_{13}} \sin \theta_{12} \sin \theta_{13} \cos \theta_{14} \sin \theta_{23} \sin \theta_{24} \cos \theta_{24}\right.\nonumber\\&&\left.+e^{-i \delta_{14}} \sin \theta_{12} \cos \theta_{13} \sin \theta_{14} \cos \theta_{14} \left(2-\cos ^2\theta_{24}\right)\right.\nonumber\\&&\left.- \cos \theta_{12} \cos \theta_{14} \cos \theta_{23} \sin \theta_{24} \cos \theta_{24}\right]
\end{eqnarray}
\begin{eqnarray}
	T_{33}&=&\,A \left[2  \sin \theta_{13} \cos \theta_{13} \sin \theta_{14} \sin \theta_{23} \sin \theta_{24} \cos \theta_{24} \cos (\delta_{13}-\delta_{14})\right.\nonumber\\&& \left.+  \sin ^2\theta_{13} \left(2-\sin ^2\theta_{24} \left(\sin ^2\theta_{14}+\sin ^2\theta_{23}\right)-\sin ^2\theta_{14}\right)+ \sin ^2\theta_{23} \sin ^2\theta_{24}\right]\nonumber\\&&-\frac{3A}{4}+\frac{1}{4} (-\Delta_{21}+3 \Delta_{31}-\Delta_{41})\\
	T_{34}&=&A\left[e^{i \delta_{13}-i \delta_{14}} \sin \theta_{13} \sin \theta_{14} \cos \theta_{14} \left(2-\cos ^2\theta_{24}\right)\right.\nonumber\\&& \left.-\cos \theta_{13} \cos \theta_{14} \sin \theta_{23} \sin \theta_{24} \cos \theta_{24}\right]\\
	T_{44}&=&A\left[\cos ^2\theta_{14} \cos ^2\theta_{24}+2 A \sin ^2\theta_{14}\right]-\frac{3 A}{4}+\frac{1}{4} (-\Delta_{21}-\Delta_{31}+3 \Delta_{41})
\end{eqnarray}
\end{widetext}
Cayley-Hamilton theorem is used to get the form of the time evolution operator $ e^{-\iota T L} $. We need to solve the characteristic equation of matrix  $ T $ given by,
\begin{equation}
	\lambda^4 + c_3\lambda^3 +c_2\lambda^2 +c_1\lambda + c_0=0
\end{equation}
to obtain the energy eigenvalues $ \lambda $ where the constants are defined as follows,
\begin{widetext}
\begin{eqnarray}
	c_0&=&\frac{A^2}{128} \Delta_{41}^2 \left(8 \sin ^2\theta_{14}+29\right)+\nonumber\\&&\frac{A}{64}\left[\left(-\Delta_{31}^3+2 \Delta_{31}^2 \Delta_{41}+3 \Delta_{31} \Delta_{41}^2\right) \sin 2\theta_{13} \sin \theta_{14} \sin \theta_{23} \sin 2\theta_{24} \cos (\delta_{13}-\delta_{14})\right.\nonumber\\&&\left.+\Delta_{31}^3 \left(3-4 \cos ^2\theta_{13} \sin ^2\theta_{23} \sin ^2\theta_{24}-4 Q \sin ^2\theta_{13}\right)\right.\nonumber\\&&\left.+\Delta_{31}^2 \Delta_{41} \left(8 \cos ^2\theta_{13} \sin ^2\theta_{23} \sin ^2\theta_{24}+12 \cos ^2\theta_{24}-4 Q \left(2 \cos ^2\theta_{13}+1\right)+9\right)\right.\nonumber\\&&\left.+\Delta_{31} \Delta_{41}^2 \left(12 \cos ^2\theta_{13} \sin ^2\theta_{23} \sin ^2\theta_{24}+8 \cos ^2\theta_{24}+4 Q \left(1-3 \cos ^2\theta_{13}\right)+1\right)\right.\nonumber\\&&\left.-\Delta_{41}^3 \left(4 \cos ^2\theta_{24}-4 Q+5\right)\right]+\frac{\Delta_{21}}{64} \left(\Delta_{31}^3-5 \Delta_{31}^2 \Delta_{41}-5 \Delta_{31} \Delta_{41}^2+\Delta_{41}^3\right)\nonumber\\&&+\left(-\frac{3 \Delta_{31}^4}{256}+\frac{\Delta_{31}^3 \Delta_{41}}{64}+\frac{7 \Delta_{31}^2 \Delta_{41}^2}{128}+\frac{\Delta_{31} \Delta_{41}^3}{64}-\frac{3 \Delta_{41}^4}{256}\right)
\end{eqnarray}
\begin{eqnarray}
	c_1&=&\frac{1}{8}A^2 \Delta_{41}\left(5-7\sin^2\theta_{14}\right)+\frac{A}{8}\Delta_{31}^2\left(3-4\sin^2\theta_{23}\sin^2\theta_{24}\cos^2\theta_{13}-4 Q\sin^2\theta_{13}\right)\nonumber\\&&-\frac{A}{8}\Delta_{41}^2\left(5+4\cos^2\theta_{24}-4 Q\right)+\frac{A}{16}\Delta_{31}\Delta_{41}\left(4+8\cos^2\theta_{24}-5 P\cos^2\theta_{13}\right)\nonumber\\&&+\frac{A}{4}\left(-\text{$\Delta_{31}$}^2+ \text{$\Delta_{31}$} \text{$\Delta_{41}$}\right) \sin 2\theta_{13} \sin\theta_{14} \sin\theta_{23} \sin 2\theta_{24} \cos(\delta_{13}-\delta_{14})\nonumber\\&&+\frac{\Delta_{21}}{8}\left(\Delta_{41}-\Delta_{31}\right)^2 +\frac{1}{8}\left(-\Delta_{31}^3+\Delta_{31}^2\Delta_{41}+\Delta_{31}\Delta_{41}^2-\Delta_{41}^3\right)
\end{eqnarray}
\begin{eqnarray}
	c_2&=& \frac{A}{4}\Delta_{31}\left(3-4\sin^2\theta_{23}\sin^2\theta_{24}\cos^2\theta_{13}-4 Q\sin^2\theta_{13}\right)-\frac{A}{4}\Delta_{41}\left(5+4\cos^2\theta_{24}-4 Q\right)\nonumber\\&&-\frac{11}{8}A^2-\frac{A}{2}\Delta_{31} \sin 2\theta_{13} \sin\theta_{14} \sin\theta_{23} \sin 2\theta_{24} \cos(\delta_{13}-\delta_{14})\nonumber\\&&+\frac{\Delta_{21}}{4}\left(\Delta_{41}+\Delta_{31}\right)+\frac{1}{8}\left(-3\Delta_{31}^2+2\Delta_{31}\Delta_{41}-3\Delta_{41}^2\right)
\end{eqnarray}
\begin{eqnarray}
	c_3=\text{Trace(T)}=0
\end{eqnarray}
\end{widetext}
\begin{eqnarray}
	P=2-\sin^2\theta_{14}-\sin^2\theta_{24}\left(\sin^2\theta_{14}+\sin^2\theta_{23}\right)\\
	Q=2-\sin^2\theta_{14}-\sin^2\theta_{24}\sin^2\theta_{14}
\end{eqnarray}
The energy eigenvalues are as follows,
\begin{align}
	\lambda_{1,2}=-\frac{1}{2}\left[\sqrt{-c_2+t_0}\pm\sqrt{-c_2-t_0-2\sqrt{-4c_0+t_0^2}}\right]\\
	\lambda_{3,4}=-\frac{1}{2}\left[-\sqrt{-c_2+t_0}\pm\sqrt{-c_2-t_0+2\sqrt{-4c_0+t_0^2}}\right]
\end{align}
where $t_0  $ is a real root of the following equation,
\begin{equation}
	t^3-c_2t^2-4c_0t+4c_0c_2-c_1^2=0
\end{equation}
The general form of probability is given by 
\begin{eqnarray}
	P_{\alpha\beta}=\sum_{a=1}^{4}\sum_{b=1}^{4}(\tilde{B}_a)_{\alpha\beta}(\tilde{B}_b)_{\alpha\beta}^\star e^{-\iota L(\lambda_a-\lambda_b)}
\end{eqnarray}
\begin{figure}[H]
	\centering
	\includegraphics[width=0.48\linewidth]{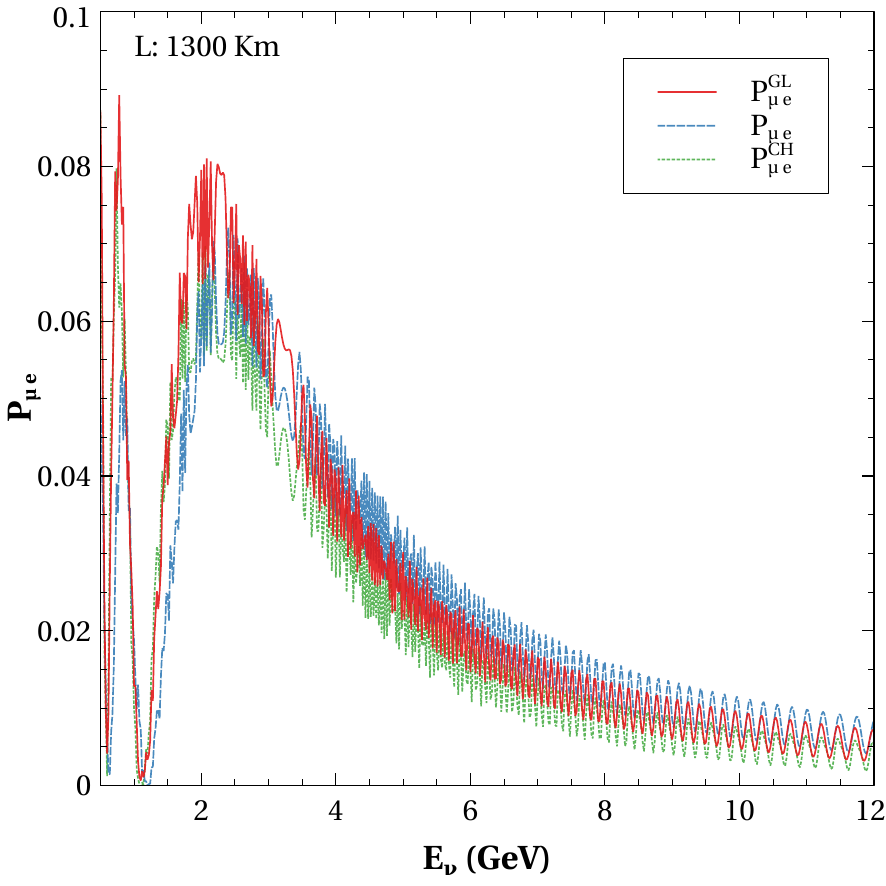}
	\includegraphics[width=0.48\linewidth]{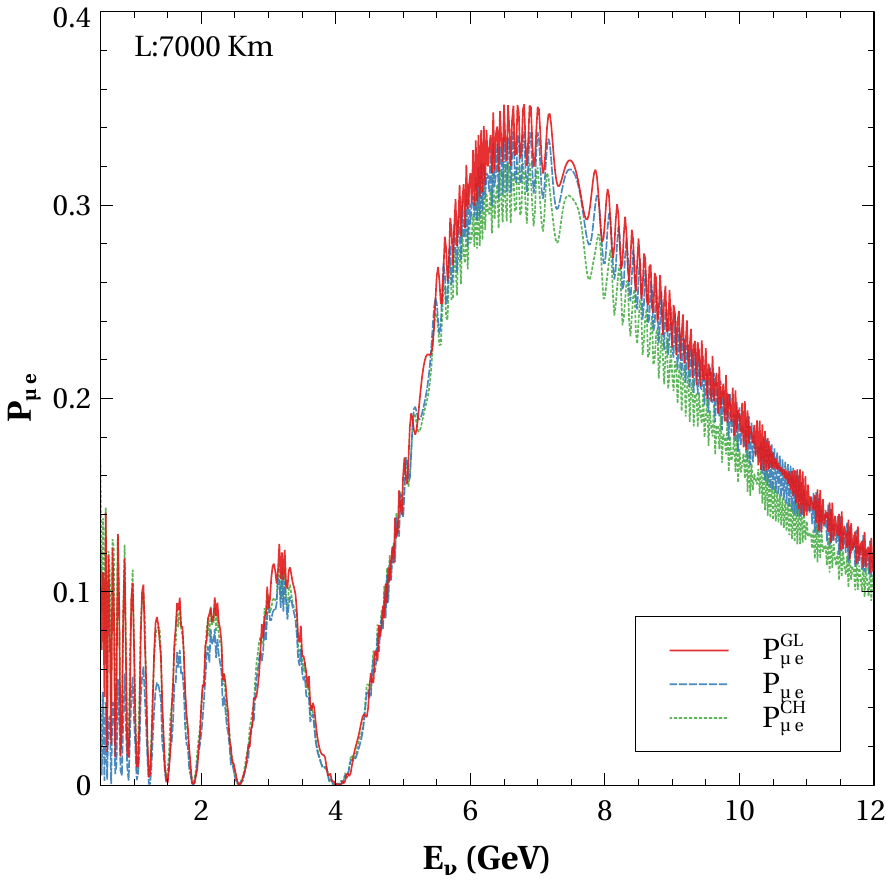}
	\caption{Comparison of the probabilities using GLoBES $P_{\mu e}^{GL}$ (red), the Cayley Hamilton method $ P_{\mu e}^{CH}$ (green), and TMSD approx. $P_{\mu e}$ (blue) at 1300 km(left), 7000 km(right) baseline.}
	\label{fig:pme_ch_comp}
\end{figure}
where,
\begin{equation}
(\tilde{B}_a)_{\alpha\beta}=\frac{(c_1+c_2\lambda_a+\lambda_a^3)\delta_{\alpha \beta}+(c_2+\lambda_a^2)\tilde{T}_{\alpha\beta}+\lambda_a\tilde{T}^2_{\alpha\beta}+\tilde{T}_{\alpha\beta}^3}{4\lambda_a^3+c_1+2c_2\lambda_a}
\end{equation}
where components of $T,T^2,T^3$ in flavour basis are defined as following,
\begin{eqnarray}
\tilde{T}_{\alpha\beta}=<\alpha|UTU^{-1}|\beta>, \tilde{T}_{\alpha\beta}^{2,3}=<\alpha|UT^{2,3}U^{-1}|\beta> 
\end{eqnarray}

In \autoref{fig:pme_ch_comp}, we see that the Cayley Hamilton probabilities at lower energies show a better match with numerical probabilities evaluated using GLoBES, whereas at higher energies, especially at resonance region, the TMSD probabilities match better, as was also seen in \autoref{fig:pme_terms}, and \autoref{fig:pmm_terms}.

\bibliographystyle{apsrev4-1}
\bibliography{references}

\begin{thebibliography}{97}%
\makeatletter
\providecommand \@ifxundefined [1]{%
 \@ifx{#1\undefined}
}%
\providecommand \@ifnum [1]{%
 \ifnum #1\expandafter \@firstoftwo
 \else \expandafter \@secondoftwo
 \fi
}%
\providecommand \@ifx [1]{%
 \ifx #1\expandafter \@firstoftwo
 \else \expandafter \@secondoftwo
 \fi
}%
\providecommand \natexlab [1]{#1}%
\providecommand \enquote  [1]{``#1''}%
\providecommand \bibnamefont  [1]{#1}%
\providecommand \bibfnamefont [1]{#1}%
\providecommand \citenamefont [1]{#1}%
\providecommand \href@noop [0]{\@secondoftwo}%
\providecommand \href [0]{\begingroup \@sanitize@url \@href}%
\providecommand \@href[1]{\@@startlink{#1}\@@href}%
\providecommand \@@href[1]{\endgroup#1\@@endlink}%
\providecommand \@sanitize@url [0]{\catcode `\\12\catcode `\$12\catcode
  `\&12\catcode `\#12\catcode `\^12\catcode `\_12\catcode `\%12\relax}%
\providecommand \@@startlink[1]{}%
\providecommand \@@endlink[0]{}%
\providecommand \url  [0]{\begingroup\@sanitize@url \@url }%
\providecommand \@url [1]{\endgroup\@href {#1}{\urlprefix }}%
\providecommand \urlprefix  [0]{URL }%
\providecommand \Eprint [0]{\href }%
\providecommand \doibase [0]{http://dx.doi.org/}%
\providecommand \selectlanguage [0]{\@gobble}%
\providecommand \bibinfo  [0]{\@secondoftwo}%
\providecommand \bibfield  [0]{\@secondoftwo}%
\providecommand \translation [1]{[#1]}%
\providecommand \BibitemOpen [0]{}%
\providecommand \bibitemStop [0]{}%
\providecommand \bibitemNoStop [0]{.\EOS\space}%
\providecommand \EOS [0]{\spacefactor3000\relax}%
\providecommand \BibitemShut  [1]{\csname bibitem#1\endcsname}%
\let\auto@bib@innerbib\@empty
\bibitem [{\citenamefont {Esteban}\ \emph {et~al.}(2017)\citenamefont
  {Esteban}, \citenamefont {Gonzalez-Garcia}, \citenamefont {Maltoni},
  \citenamefont {Martinez-Soler},\ and\ \citenamefont
  {Schwetz}}]{Esteban:2016qun}%
  \BibitemOpen
  \bibfield  {author} {\bibinfo {author} {\bibfnamefont {I.}~\bibnamefont
  {Esteban}}, \bibinfo {author} {\bibfnamefont {M.~C.}\ \bibnamefont
  {Gonzalez-Garcia}}, \bibinfo {author} {\bibfnamefont {M.}~\bibnamefont
  {Maltoni}}, \bibinfo {author} {\bibfnamefont {I.}~\bibnamefont
  {Martinez-Soler}}, \ and\ \bibinfo {author} {\bibfnamefont {T.}~\bibnamefont
  {Schwetz}},\ }\href {\doibase 10.1007/JHEP01(2017)087} {\bibfield  {journal}
  {\bibinfo  {journal} {JHEP}\ }\textbf {\bibinfo {volume} {01}},\ \bibinfo
  {pages} {087} (\bibinfo {year} {2017})},\ \Eprint
  {http://arxiv.org/abs/1611.01514} {arXiv:1611.01514 [hep-ph]} \BibitemShut
  {NoStop}%
\bibitem [{\citenamefont {de~Salas}\ \emph {et~al.}(2021)\citenamefont
  {de~Salas}, \citenamefont {Forero}, \citenamefont {Gariazzo}, \citenamefont
  {Mart\'\i{}nez-Mirav\'e}, \citenamefont {Mena}, \citenamefont {Ternes},
  \citenamefont {T\'ortola},\ and\ \citenamefont {Valle}}]{deSalas:2020pgw}%
  \BibitemOpen
  \bibfield  {author} {\bibinfo {author} {\bibfnamefont {P.~F.}\ \bibnamefont
  {de~Salas}}, \bibinfo {author} {\bibfnamefont {D.~V.}\ \bibnamefont
  {Forero}}, \bibinfo {author} {\bibfnamefont {S.}~\bibnamefont {Gariazzo}},
  \bibinfo {author} {\bibfnamefont {P.}~\bibnamefont {Mart\'\i{}nez-Mirav\'e}},
  \bibinfo {author} {\bibfnamefont {O.}~\bibnamefont {Mena}}, \bibinfo {author}
  {\bibfnamefont {C.~A.}\ \bibnamefont {Ternes}}, \bibinfo {author}
  {\bibfnamefont {M.}~\bibnamefont {T\'ortola}}, \ and\ \bibinfo {author}
  {\bibfnamefont {J.~W.~F.}\ \bibnamefont {Valle}},\ }\href {\doibase
  10.1007/JHEP02(2021)071} {\bibfield  {journal} {\bibinfo  {journal} {JHEP}\
  }\textbf {\bibinfo {volume} {02}},\ \bibinfo {pages} {071} (\bibinfo {year}
  {2021})},\ \Eprint {http://arxiv.org/abs/2006.11237} {arXiv:2006.11237
  [hep-ph]} \BibitemShut {NoStop}%
\bibitem [{\citenamefont {Capozzi}\ \emph {et~al.}(2020)\citenamefont
  {Capozzi}, \citenamefont {Valentino}, \citenamefont {Lisi}, \citenamefont
  {Marrone}, \citenamefont {Melchiorri},\ and\ \citenamefont
  {Palazzo}}]{Capozzi_2020}%
  \BibitemOpen
  \bibfield  {author} {\bibinfo {author} {\bibfnamefont {F.}~\bibnamefont
  {Capozzi}}, \bibinfo {author} {\bibfnamefont {E.~D.}\ \bibnamefont
  {Valentino}}, \bibinfo {author} {\bibfnamefont {E.}~\bibnamefont {Lisi}},
  \bibinfo {author} {\bibfnamefont {A.}~\bibnamefont {Marrone}}, \bibinfo
  {author} {\bibfnamefont {A.}~\bibnamefont {Melchiorri}}, \ and\ \bibinfo
  {author} {\bibfnamefont {A.}~\bibnamefont {Palazzo}},\ }\href {\doibase
  10.1103/physrevd.101.116013} {\bibfield  {journal} {\bibinfo  {journal}
  {Physical Review D}\ }\textbf {\bibinfo {volume} {101}} (\bibinfo {year}
  {2020}),\ 10.1103/physrevd.101.116013}\BibitemShut {NoStop}%
\bibitem [{\citenamefont {Ghosh}\ \emph
  {et~al.}(2016{\natexlab{a}})\citenamefont {Ghosh}, \citenamefont {Ghoshal},
  \citenamefont {Goswami}, \citenamefont {Nath},\ and\ \citenamefont
  {Raut}}]{Ghosh:2015ena}%
  \BibitemOpen
  \bibfield  {author} {\bibinfo {author} {\bibfnamefont {M.}~\bibnamefont
  {Ghosh}}, \bibinfo {author} {\bibfnamefont {P.}~\bibnamefont {Ghoshal}},
  \bibinfo {author} {\bibfnamefont {S.}~\bibnamefont {Goswami}}, \bibinfo
  {author} {\bibfnamefont {N.}~\bibnamefont {Nath}}, \ and\ \bibinfo {author}
  {\bibfnamefont {S.~K.}\ \bibnamefont {Raut}},\ }\href {\doibase
  10.1103/PhysRevD.93.013013} {\bibfield  {journal} {\bibinfo  {journal} {Phys.
  Rev. D}\ }\textbf {\bibinfo {volume} {93}},\ \bibinfo {pages} {013013}
  (\bibinfo {year} {2016}{\natexlab{a}})},\ \Eprint
  {http://arxiv.org/abs/1504.06283} {arXiv:1504.06283 [hep-ph]} \BibitemShut
  {NoStop}%
\bibitem [{\citenamefont {Abe}\ \emph {et~al.}(2020)\citenamefont {Abe} \emph
  {et~al.}}]{T2K:2019bcf}%
  \BibitemOpen
  \bibfield  {author} {\bibinfo {author} {\bibfnamefont {K.}~\bibnamefont
  {Abe}} \emph {et~al.} (\bibinfo {collaboration} {T2K}),\ }\href {\doibase
  10.1038/s41586-020-2177-0} {\bibfield  {journal} {\bibinfo  {journal}
  {Nature}\ }\textbf {\bibinfo {volume} {580}},\ \bibinfo {pages} {339}
  (\bibinfo {year} {2020})},\ \bibinfo {note} {[Erratum: Nature 583, E16
  (2020)]},\ \Eprint {http://arxiv.org/abs/1910.03887} {arXiv:1910.03887
  [hep-ex]} \BibitemShut {NoStop}%
\bibitem [{\citenamefont {Ayres}\ \emph {et~al.}(2007)\citenamefont {Ayres}
  \emph {et~al.}}]{NOvA:2007rmc}%
  \BibitemOpen
  \bibfield  {author} {\bibinfo {author} {\bibfnamefont {D.~S.}\ \bibnamefont
  {Ayres}} \emph {et~al.} (\bibinfo {collaboration} {NOvA}),\ }\href {\doibase
  10.2172/935497} {\  (\bibinfo {year} {2007}),\ 10.2172/935497}\BibitemShut
  {NoStop}%
\bibitem [{\citenamefont {Abi}\ \emph {et~al.}(2020{\natexlab{a}})\citenamefont
  {Abi} \emph {et~al.}}]{DUNE:2020ypp}%
  \BibitemOpen
  \bibfield  {author} {\bibinfo {author} {\bibfnamefont {B.}~\bibnamefont
  {Abi}} \emph {et~al.} (\bibinfo {collaboration} {DUNE}),\ }\href@noop {} {\
  (\bibinfo {year} {2020}{\natexlab{a}})},\ \Eprint
  {http://arxiv.org/abs/2002.03005} {arXiv:2002.03005 [hep-ex]} \BibitemShut
  {NoStop}%
\bibitem [{\citenamefont {Alekou}\ \emph {et~al.}(2022)\citenamefont {Alekou}
  \emph {et~al.}}]{Alekou:2022emd}%
  \BibitemOpen
  \bibfield  {author} {\bibinfo {author} {\bibfnamefont {A.}~\bibnamefont
  {Alekou}} \emph {et~al.},\ }\href {\doibase 10.1140/epjs/s11734-022-00664-w}
  {\  (\bibinfo {year} {2022}),\ 10.1140/epjs/s11734-022-00664-w},\ \Eprint
  {http://arxiv.org/abs/2206.01208} {arXiv:2206.01208 [hep-ex]} \BibitemShut
  {NoStop}%
\bibitem [{\citenamefont {Abe}\ \emph {et~al.}(2015)\citenamefont {Abe} \emph
  {et~al.}}]{Hyper-KamiokandeProto-:2015xww}%
  \BibitemOpen
  \bibfield  {author} {\bibinfo {author} {\bibfnamefont {K.}~\bibnamefont
  {Abe}} \emph {et~al.} (\bibinfo {collaboration} {Hyper-Kamiokande Proto-}),\
  }\href {\doibase 10.1093/ptep/ptv061} {\bibfield  {journal} {\bibinfo
  {journal} {PTEP}\ }\textbf {\bibinfo {volume} {2015}},\ \bibinfo {pages}
  {053C02} (\bibinfo {year} {2015})},\ \Eprint
  {http://arxiv.org/abs/1502.05199} {arXiv:1502.05199 [hep-ex]} \BibitemShut
  {NoStop}%
\bibitem [{\citenamefont {Brunner}(2018)}]{Brunner:2017zcc}%
  \BibitemOpen
  \bibfield  {author} {\bibinfo {author} {\bibfnamefont {J.}~\bibnamefont
  {Brunner}} (\bibinfo {collaboration} {KM3Net}),\ }\href {\doibase
  10.22323/1.307.0057} {\bibfield  {journal} {\bibinfo  {journal} {PoS}\
  }\textbf {\bibinfo {volume} {NEUTEL2017}},\ \bibinfo {pages} {057} (\bibinfo
  {year} {2018})}\BibitemShut {NoStop}%
\bibitem [{\citenamefont {Aartsen}\ \emph {et~al.}(2017)\citenamefont {Aartsen}
  \emph {et~al.}}]{IceCube:2016xxt}%
  \BibitemOpen
  \bibfield  {author} {\bibinfo {author} {\bibfnamefont {M.~G.}\ \bibnamefont
  {Aartsen}} \emph {et~al.} (\bibinfo {collaboration} {IceCube}),\ }\href
  {\doibase 10.1088/1361-6471/44/5/054006} {\bibfield  {journal} {\bibinfo
  {journal} {J. Phys. G}\ }\textbf {\bibinfo {volume} {44}},\ \bibinfo {pages}
  {054006} (\bibinfo {year} {2017})},\ \Eprint
  {http://arxiv.org/abs/1607.02671} {arXiv:1607.02671 [hep-ex]} \BibitemShut
  {NoStop}%
\bibitem [{\citenamefont {Ahmed}\ \emph {et~al.}(2017)\citenamefont {Ahmed}
  \emph {et~al.}}]{ICAL:2015stm}%
  \BibitemOpen
  \bibfield  {author} {\bibinfo {author} {\bibfnamefont {S.}~\bibnamefont
  {Ahmed}} \emph {et~al.} (\bibinfo {collaboration} {ICAL}),\ }\href {\doibase
  10.1007/s12043-017-1373-4} {\bibfield  {journal} {\bibinfo  {journal}
  {Pramana}\ }\textbf {\bibinfo {volume} {88}},\ \bibinfo {pages} {79}
  (\bibinfo {year} {2017})},\ \Eprint {http://arxiv.org/abs/1505.07380}
  {arXiv:1505.07380 [physics.ins-det]} \BibitemShut {NoStop}%
\bibitem [{\citenamefont {Ghosh}\ \emph
  {et~al.}(2016{\natexlab{b}})\citenamefont {Ghosh}, \citenamefont {Goswami},\
  and\ \citenamefont {Raut}}]{Ghosh:2014rna}%
  \BibitemOpen
  \bibfield  {author} {\bibinfo {author} {\bibfnamefont {M.}~\bibnamefont
  {Ghosh}}, \bibinfo {author} {\bibfnamefont {S.}~\bibnamefont {Goswami}}, \
  and\ \bibinfo {author} {\bibfnamefont {S.~K.}\ \bibnamefont {Raut}},\ }\href
  {\doibase 10.1140/epjc/s10052-016-3962-7} {\bibfield  {journal} {\bibinfo
  {journal} {Eur. Phys. J. C}\ }\textbf {\bibinfo {volume} {76}},\ \bibinfo
  {pages} {114} (\bibinfo {year} {2016}{\natexlab{b}})},\ \Eprint
  {http://arxiv.org/abs/1412.1744} {arXiv:1412.1744 [hep-ph]} \BibitemShut
  {NoStop}%
\bibitem [{\citenamefont {Barger}\ \emph {et~al.}(2014)\citenamefont {Barger},
  \citenamefont {Bhattacharya}, \citenamefont {Chatterjee}, \citenamefont
  {Gandhi}, \citenamefont {Marfatia},\ and\ \citenamefont
  {Masud}}]{Barger:2013rha}%
  \BibitemOpen
  \bibfield  {author} {\bibinfo {author} {\bibfnamefont {V.}~\bibnamefont
  {Barger}}, \bibinfo {author} {\bibfnamefont {A.}~\bibnamefont
  {Bhattacharya}}, \bibinfo {author} {\bibfnamefont {A.}~\bibnamefont
  {Chatterjee}}, \bibinfo {author} {\bibfnamefont {R.}~\bibnamefont {Gandhi}},
  \bibinfo {author} {\bibfnamefont {D.}~\bibnamefont {Marfatia}}, \ and\
  \bibinfo {author} {\bibfnamefont {M.}~\bibnamefont {Masud}},\ }\href
  {\doibase 10.1103/PhysRevD.89.011302} {\bibfield  {journal} {\bibinfo
  {journal} {Phys. Rev. D}\ }\textbf {\bibinfo {volume} {89}},\ \bibinfo
  {pages} {011302} (\bibinfo {year} {2014})},\ \Eprint
  {http://arxiv.org/abs/1307.2519} {arXiv:1307.2519 [hep-ph]} \BibitemShut
  {NoStop}%
\bibitem [{\citenamefont {Ghosh}\ \emph
  {et~al.}(2014{\natexlab{a}})\citenamefont {Ghosh}, \citenamefont {Ghoshal},
  \citenamefont {Goswami},\ and\ \citenamefont {Raut}}]{Ghosh:2013yon}%
  \BibitemOpen
  \bibfield  {author} {\bibinfo {author} {\bibfnamefont {M.}~\bibnamefont
  {Ghosh}}, \bibinfo {author} {\bibfnamefont {P.}~\bibnamefont {Ghoshal}},
  \bibinfo {author} {\bibfnamefont {S.}~\bibnamefont {Goswami}}, \ and\
  \bibinfo {author} {\bibfnamefont {S.~K.}\ \bibnamefont {Raut}},\ }\href
  {\doibase 10.1103/PhysRevD.89.011301} {\bibfield  {journal} {\bibinfo
  {journal} {Phys. Rev. D}\ }\textbf {\bibinfo {volume} {89}},\ \bibinfo
  {pages} {011301} (\bibinfo {year} {2014}{\natexlab{a}})},\ \Eprint
  {http://arxiv.org/abs/1306.2500} {arXiv:1306.2500 [hep-ph]} \BibitemShut
  {NoStop}%
\bibitem [{\citenamefont {Chatterjee}\ \emph {et~al.}(2013)\citenamefont
  {Chatterjee}, \citenamefont {Ghoshal}, \citenamefont {Goswami},\ and\
  \citenamefont {Raut}}]{Chatterjee:2013qus}%
  \BibitemOpen
  \bibfield  {author} {\bibinfo {author} {\bibfnamefont {A.}~\bibnamefont
  {Chatterjee}}, \bibinfo {author} {\bibfnamefont {P.}~\bibnamefont {Ghoshal}},
  \bibinfo {author} {\bibfnamefont {S.}~\bibnamefont {Goswami}}, \ and\
  \bibinfo {author} {\bibfnamefont {S.~K.}\ \bibnamefont {Raut}},\ }\href
  {\doibase 10.1007/JHEP06(2013)010} {\bibfield  {journal} {\bibinfo  {journal}
  {JHEP}\ }\textbf {\bibinfo {volume} {06}},\ \bibinfo {pages} {010} (\bibinfo
  {year} {2013})},\ \Eprint {http://arxiv.org/abs/1302.1370} {arXiv:1302.1370
  [hep-ph]} \BibitemShut {NoStop}%
\bibitem [{\citenamefont {Ghosh}\ \emph
  {et~al.}(2014{\natexlab{b}})\citenamefont {Ghosh}, \citenamefont {Ghoshal},
  \citenamefont {Goswami},\ and\ \citenamefont {Raut}}]{Ghosh:2014dba}%
  \BibitemOpen
  \bibfield  {author} {\bibinfo {author} {\bibfnamefont {M.}~\bibnamefont
  {Ghosh}}, \bibinfo {author} {\bibfnamefont {P.}~\bibnamefont {Ghoshal}},
  \bibinfo {author} {\bibfnamefont {S.}~\bibnamefont {Goswami}}, \ and\
  \bibinfo {author} {\bibfnamefont {S.~K.}\ \bibnamefont {Raut}},\ }\href
  {\doibase 10.1016/j.nuclphysb.2014.04.013} {\bibfield  {journal} {\bibinfo
  {journal} {Nucl. Phys. B}\ }\textbf {\bibinfo {volume} {884}},\ \bibinfo
  {pages} {274} (\bibinfo {year} {2014}{\natexlab{b}})},\ \Eprint
  {http://arxiv.org/abs/1401.7243} {arXiv:1401.7243 [hep-ph]} \BibitemShut
  {NoStop}%
\bibitem [{\citenamefont {Ghosh}\ \emph {et~al.}(2013)\citenamefont {Ghosh},
  \citenamefont {Thakore},\ and\ \citenamefont {Choubey}}]{Ghosh:2012px}%
  \BibitemOpen
  \bibfield  {author} {\bibinfo {author} {\bibfnamefont {A.}~\bibnamefont
  {Ghosh}}, \bibinfo {author} {\bibfnamefont {T.}~\bibnamefont {Thakore}}, \
  and\ \bibinfo {author} {\bibfnamefont {S.}~\bibnamefont {Choubey}},\ }\href
  {\doibase 10.1007/JHEP04(2013)009} {\bibfield  {journal} {\bibinfo  {journal}
  {JHEP}\ }\textbf {\bibinfo {volume} {04}},\ \bibinfo {pages} {009} (\bibinfo
  {year} {2013})},\ \Eprint {http://arxiv.org/abs/1212.1305} {arXiv:1212.1305
  [hep-ph]} \BibitemShut {NoStop}%
\bibitem [{\citenamefont {Chakraborty}\ \emph {et~al.}(2019)\citenamefont
  {Chakraborty}, \citenamefont {Goswami}, \citenamefont {Gupta},\ and\
  \citenamefont {Thakore}}]{Chakraborty:2019jlv}%
  \BibitemOpen
  \bibfield  {author} {\bibinfo {author} {\bibfnamefont {K.}~\bibnamefont
  {Chakraborty}}, \bibinfo {author} {\bibfnamefont {S.}~\bibnamefont
  {Goswami}}, \bibinfo {author} {\bibfnamefont {C.}~\bibnamefont {Gupta}}, \
  and\ \bibinfo {author} {\bibfnamefont {T.}~\bibnamefont {Thakore}},\ }\href
  {\doibase 10.1007/JHEP05(2019)137} {\bibfield  {journal} {\bibinfo  {journal}
  {JHEP}\ }\textbf {\bibinfo {volume} {05}},\ \bibinfo {pages} {137} (\bibinfo
  {year} {2019})},\ \Eprint {http://arxiv.org/abs/1902.02963} {arXiv:1902.02963
  [hep-ph]} \BibitemShut {NoStop}%
\bibitem [{\citenamefont {Fukasawa}\ \emph {et~al.}(2017)\citenamefont
  {Fukasawa}, \citenamefont {Ghosh},\ and\ \citenamefont
  {Yasuda}}]{Fukasawa:2016yue}%
  \BibitemOpen
  \bibfield  {author} {\bibinfo {author} {\bibfnamefont {S.}~\bibnamefont
  {Fukasawa}}, \bibinfo {author} {\bibfnamefont {M.}~\bibnamefont {Ghosh}}, \
  and\ \bibinfo {author} {\bibfnamefont {O.}~\bibnamefont {Yasuda}},\ }\href
  {\doibase 10.1016/j.nuclphysb.2017.02.008} {\bibfield  {journal} {\bibinfo
  {journal} {Nucl. Phys. B}\ }\textbf {\bibinfo {volume} {918}},\ \bibinfo
  {pages} {337} (\bibinfo {year} {2017})},\ \Eprint
  {http://arxiv.org/abs/1607.03758} {arXiv:1607.03758 [hep-ph]} \BibitemShut
  {NoStop}%
\bibitem [{\citenamefont {Ghosh}(2016)}]{Ghosh:2015tan}%
  \BibitemOpen
  \bibfield  {author} {\bibinfo {author} {\bibfnamefont {M.}~\bibnamefont
  {Ghosh}},\ }\href {\doibase 10.1103/PhysRevD.93.073003} {\bibfield  {journal}
  {\bibinfo  {journal} {Phys. Rev. D}\ }\textbf {\bibinfo {volume} {93}},\
  \bibinfo {pages} {073003} (\bibinfo {year} {2016})},\ \Eprint
  {http://arxiv.org/abs/1512.02226} {arXiv:1512.02226 [hep-ph]} \BibitemShut
  {NoStop}%
\bibitem [{\citenamefont {Hill}(1995)}]{Hill:1995gf}%
  \BibitemOpen
  \bibfield  {author} {\bibinfo {author} {\bibfnamefont {J.~E.}\ \bibnamefont
  {Hill}},\ }\href {\doibase 10.1103/PhysRevLett.75.2654} {\bibfield  {journal}
  {\bibinfo  {journal} {Phys. Rev. Lett.}\ }\textbf {\bibinfo {volume} {75}},\
  \bibinfo {pages} {2654} (\bibinfo {year} {1995})},\ \Eprint
  {http://arxiv.org/abs/hep-ex/9504009} {arXiv:hep-ex/9504009} \BibitemShut
  {NoStop}%
\bibitem [{\citenamefont {Aguilar-Arevalo}\ \emph {et~al.}(2020)\citenamefont
  {Aguilar-Arevalo} \emph {et~al.}}]{Aguilar-Arevalo:2020nvw}%
  \BibitemOpen
  \bibfield  {author} {\bibinfo {author} {\bibfnamefont {A.}~\bibnamefont
  {Aguilar-Arevalo}} \emph {et~al.} (\bibinfo {collaboration} {MiniBooNE}),\
  }\href@noop {} {\  (\bibinfo {year} {2020})},\ \Eprint
  {http://arxiv.org/abs/2006.16883} {arXiv:2006.16883 [hep-ex]} \BibitemShut
  {NoStop}%
\bibitem [{\citenamefont {Abrams}\ \emph {et~al.}(1989)\citenamefont {Abrams}
  \emph {et~al.}}]{Abrams:1989yk}%
  \BibitemOpen
  \bibfield  {author} {\bibinfo {author} {\bibfnamefont {G.~S.}\ \bibnamefont
  {Abrams}} \emph {et~al.},\ }\href {\doibase 10.1103/PhysRevLett.63.2173}
  {\bibfield  {journal} {\bibinfo  {journal} {Phys. Rev. Lett.}\ }\textbf
  {\bibinfo {volume} {63}},\ \bibinfo {pages} {2173} (\bibinfo {year}
  {1989})}\BibitemShut {NoStop}%
\bibitem [{\citenamefont {Giunti}\ and\ \citenamefont
  {Laveder}(2011)}]{Giunti:2010zu}%
  \BibitemOpen
  \bibfield  {author} {\bibinfo {author} {\bibfnamefont {C.}~\bibnamefont
  {Giunti}}\ and\ \bibinfo {author} {\bibfnamefont {M.}~\bibnamefont
  {Laveder}},\ }\href {\doibase 10.1103/PhysRevC.83.065504} {\bibfield
  {journal} {\bibinfo  {journal} {Phys. Rev. C}\ }\textbf {\bibinfo {volume}
  {83}},\ \bibinfo {pages} {065504} (\bibinfo {year} {2011})},\ \Eprint
  {http://arxiv.org/abs/1006.3244} {arXiv:1006.3244 [hep-ph]} \BibitemShut
  {NoStop}%
\bibitem [{\citenamefont {Acero}\ \emph {et~al.}(2009)\citenamefont {Acero},
  \citenamefont {Giunti},\ and\ \citenamefont {Laveder}}]{Acero:2008zz}%
  \BibitemOpen
  \bibfield  {author} {\bibinfo {author} {\bibfnamefont {M.~A.}\ \bibnamefont
  {Acero}}, \bibinfo {author} {\bibfnamefont {C.}~\bibnamefont {Giunti}}, \
  and\ \bibinfo {author} {\bibfnamefont {M.}~\bibnamefont {Laveder}},\ }\href
  {\doibase 10.1016/j.nuclphysbps.2009.02.050} {\bibfield  {journal} {\bibinfo
  {journal} {Nucl. Phys. B Proc. Suppl.}\ }\textbf {\bibinfo {volume} {188}},\
  \bibinfo {pages} {211} (\bibinfo {year} {2009})}\BibitemShut {NoStop}%
\bibitem [{\citenamefont {Barinov}\ \emph {et~al.}(2021)\citenamefont {Barinov}
  \emph {et~al.}}]{https://doi.org/10.48550/arxiv.2109.11482}%
  \BibitemOpen
  \bibfield  {author} {\bibinfo {author} {\bibfnamefont {V.~V.}\ \bibnamefont
  {Barinov}} \emph {et~al.},\ }\href {\doibase 10.48550/ARXIV.2109.11482}
  {\enquote {\bibinfo {title} {Results from the baksan experiment on sterile
  transitions (best)},}\ } (\bibinfo {year} {2021})\BibitemShut {NoStop}%
\bibitem [{\citenamefont {Huber}(2011)}]{Huber_2011}%
  \BibitemOpen
  \bibfield  {author} {\bibinfo {author} {\bibfnamefont {P.}~\bibnamefont
  {Huber}},\ }\href {\doibase 10.1103/physrevc.84.024617} {\bibfield  {journal}
  {\bibinfo  {journal} {Physical Review C}\ }\textbf {\bibinfo {volume} {84}}
  (\bibinfo {year} {2011}),\ 10.1103/physrevc.84.024617}\BibitemShut {NoStop}%
\bibitem [{\citenamefont {Mueller}\ \emph {et~al.}(2011)\citenamefont
  {Mueller}, \citenamefont {Lhuillier}, \citenamefont {Fallot}, \citenamefont
  {Letourneau}, \citenamefont {Cormon}, \citenamefont {Fechner}, \citenamefont
  {Giot}, \citenamefont {Lasserre}, \citenamefont {Martino}, \citenamefont
  {Mention}, \citenamefont {Porta},\ and\ \citenamefont
  {Yermia}}]{Mueller_2011}%
  \BibitemOpen
  \bibfield  {author} {\bibinfo {author} {\bibfnamefont {T.~A.}\ \bibnamefont
  {Mueller}}, \bibinfo {author} {\bibfnamefont {D.}~\bibnamefont {Lhuillier}},
  \bibinfo {author} {\bibfnamefont {M.}~\bibnamefont {Fallot}}, \bibinfo
  {author} {\bibfnamefont {A.}~\bibnamefont {Letourneau}}, \bibinfo {author}
  {\bibfnamefont {S.}~\bibnamefont {Cormon}}, \bibinfo {author} {\bibfnamefont
  {M.}~\bibnamefont {Fechner}}, \bibinfo {author} {\bibfnamefont
  {L.}~\bibnamefont {Giot}}, \bibinfo {author} {\bibfnamefont {T.}~\bibnamefont
  {Lasserre}}, \bibinfo {author} {\bibfnamefont {J.}~\bibnamefont {Martino}},
  \bibinfo {author} {\bibfnamefont {G.}~\bibnamefont {Mention}}, \bibinfo
  {author} {\bibfnamefont {A.}~\bibnamefont {Porta}}, \ and\ \bibinfo {author}
  {\bibfnamefont {F.}~\bibnamefont {Yermia}},\ }\href {\doibase
  10.1103/physrevc.83.054615} {\bibfield  {journal} {\bibinfo  {journal}
  {Physical Review C}\ }\textbf {\bibinfo {volume} {83}} (\bibinfo {year}
  {2011}),\ 10.1103/physrevc.83.054615}\BibitemShut {NoStop}%
\bibitem [{\citenamefont {Svirida}\ \emph {et~al.}(2019)\citenamefont {Svirida}
  \emph {et~al.}}]{Svirida:2019kbq}%
  \BibitemOpen
  \bibfield  {author} {\bibinfo {author} {\bibfnamefont {D.}~\bibnamefont
  {Svirida}} \emph {et~al.},\ }\href {\doibase 10.22323/1.337.0066} {\bibfield
  {journal} {\bibinfo  {journal} {PoS}\ }\textbf {\bibinfo {volume}
  {NOW2018}},\ \bibinfo {pages} {066} (\bibinfo {year} {2019})}\BibitemShut
  {NoStop}%
\bibitem [{\citenamefont {Danilov}\ and\ \citenamefont
  {Skrobova}(2021)}]{https://doi.org/10.48550/arxiv.2112.13413}%
  \BibitemOpen
  \bibfield  {author} {\bibinfo {author} {\bibfnamefont {M.}~\bibnamefont
  {Danilov}}\ and\ \bibinfo {author} {\bibfnamefont {N.}~\bibnamefont
  {Skrobova}},\ }\href {\doibase 10.48550/ARXIV.2112.13413} {\enquote {\bibinfo
  {title} {New results from the danss experiment},}\ } (\bibinfo {year}
  {2021})\BibitemShut {NoStop}%
\bibitem [{\citenamefont {Ko}\ \emph {et~al.}(2017)\citenamefont {Ko} \emph
  {et~al.}}]{PhysRevLett.118.121802}%
  \BibitemOpen
  \bibfield  {author} {\bibinfo {author} {\bibfnamefont {Y.~J.}\ \bibnamefont
  {Ko}} \emph {et~al.} (\bibinfo {collaboration} {NEOS Collaboration}),\ }\href
  {\doibase 10.1103/PhysRevLett.118.121802} {\bibfield  {journal} {\bibinfo
  {journal} {Phys. Rev. Lett.}\ }\textbf {\bibinfo {volume} {118}},\ \bibinfo
  {pages} {121802} (\bibinfo {year} {2017})}\BibitemShut {NoStop}%
\bibitem [{\citenamefont {Almaz\'an}\ \emph {et~al.}(2020)\citenamefont
  {Almaz\'an} \emph {et~al.}}]{PhysRevD.102.052002}%
  \BibitemOpen
  \bibfield  {author} {\bibinfo {author} {\bibfnamefont {H.}~\bibnamefont
  {Almaz\'an}} \emph {et~al.} (\bibinfo {collaboration} {STEREO
  Collaboration}),\ }\href {\doibase 10.1103/PhysRevD.102.052002} {\bibfield
  {journal} {\bibinfo  {journal} {Phys. Rev. D}\ }\textbf {\bibinfo {volume}
  {102}},\ \bibinfo {pages} {052002} (\bibinfo {year} {2020})}\BibitemShut
  {NoStop}%
\bibitem [{\citenamefont {Andriamirado}\ \emph {et~al.}(2021)\citenamefont
  {Andriamirado} \emph {et~al.}}]{PhysRevD.103.032001}%
  \BibitemOpen
  \bibfield  {author} {\bibinfo {author} {\bibfnamefont {M.}~\bibnamefont
  {Andriamirado}} \emph {et~al.} (\bibinfo {collaboration} {PROSPECT
  Collaboration}),\ }\href {\doibase 10.1103/PhysRevD.103.032001} {\bibfield
  {journal} {\bibinfo  {journal} {Phys. Rev. D}\ }\textbf {\bibinfo {volume}
  {103}},\ \bibinfo {pages} {032001} (\bibinfo {year} {2021})}\BibitemShut
  {NoStop}%
\bibitem [{\citenamefont {Minotti}(2022)}]{Minotti:2022yae}%
  \BibitemOpen
  \bibfield  {author} {\bibinfo {author} {\bibfnamefont {A.}~\bibnamefont
  {Minotti}},\ }\href {\doibase 10.22323/1.402.0246} {\bibfield  {journal}
  {\bibinfo  {journal} {PoS}\ }\textbf {\bibinfo {volume} {NuFact2021}},\
  \bibinfo {pages} {246} (\bibinfo {year} {2022})}\BibitemShut {NoStop}%
\bibitem [{\citenamefont {Goswami}(1997)}]{Goswami:1995yq}%
  \BibitemOpen
  \bibfield  {author} {\bibinfo {author} {\bibfnamefont {S.}~\bibnamefont
  {Goswami}},\ }\href {\doibase 10.1103/PhysRevD.55.2931} {\bibfield  {journal}
  {\bibinfo  {journal} {Phys. Rev. D}\ }\textbf {\bibinfo {volume} {55}},\
  \bibinfo {pages} {2931} (\bibinfo {year} {1997})},\ \Eprint
  {http://arxiv.org/abs/hep-ph/9507212} {arXiv:hep-ph/9507212} \BibitemShut
  {NoStop}%
\bibitem [{\citenamefont {Dentler}\ \emph {et~al.}(2018)\citenamefont
  {Dentler}, \citenamefont {Hern\'andez-Cabezudo}, \citenamefont {Kopp},
  \citenamefont {Machado}, \citenamefont {Maltoni}, \citenamefont
  {Martinez-Soler},\ and\ \citenamefont {Schwetz}}]{Dentler:2018sju}%
  \BibitemOpen
  \bibfield  {author} {\bibinfo {author} {\bibfnamefont {M.}~\bibnamefont
  {Dentler}}, \bibinfo {author} {\bibfnamefont {A.}~\bibnamefont
  {Hern\'andez-Cabezudo}}, \bibinfo {author} {\bibfnamefont {J.}~\bibnamefont
  {Kopp}}, \bibinfo {author} {\bibfnamefont {P.~A.~N.}\ \bibnamefont
  {Machado}}, \bibinfo {author} {\bibfnamefont {M.}~\bibnamefont {Maltoni}},
  \bibinfo {author} {\bibfnamefont {I.}~\bibnamefont {Martinez-Soler}}, \ and\
  \bibinfo {author} {\bibfnamefont {T.}~\bibnamefont {Schwetz}},\ }\href
  {\doibase 10.1007/JHEP08(2018)010} {\bibfield  {journal} {\bibinfo  {journal}
  {JHEP}\ }\textbf {\bibinfo {volume} {08}},\ \bibinfo {pages} {010} (\bibinfo
  {year} {2018})},\ \Eprint {http://arxiv.org/abs/1803.10661} {arXiv:1803.10661
  [hep-ph]} \BibitemShut {NoStop}%
\bibitem [{\citenamefont {Abazajian}\ \emph {et~al.}(2012)\citenamefont
  {Abazajian} \emph {et~al.}}]{Abazajian:2012ys}%
  \BibitemOpen
  \bibfield  {author} {\bibinfo {author} {\bibfnamefont {K.~N.}\ \bibnamefont
  {Abazajian}} \emph {et~al.},\ }\href@noop {} {\  (\bibinfo {year} {2012})},\
  \Eprint {http://arxiv.org/abs/1204.5379} {arXiv:1204.5379 [hep-ph]}
  \BibitemShut {NoStop}%
\bibitem [{\citenamefont {Adamson}\ \emph {et~al.}(2019)\citenamefont {Adamson}
  \emph {et~al.}}]{MINOS:2017cae}%
  \BibitemOpen
  \bibfield  {author} {\bibinfo {author} {\bibfnamefont {P.}~\bibnamefont
  {Adamson}} \emph {et~al.} (\bibinfo {collaboration} {MINOS+}),\ }\href
  {\doibase 10.1103/PhysRevLett.122.091803} {\bibfield  {journal} {\bibinfo
  {journal} {Phys. Rev. Lett.}\ }\textbf {\bibinfo {volume} {122}},\ \bibinfo
  {pages} {091803} (\bibinfo {year} {2019})},\ \Eprint
  {http://arxiv.org/abs/1710.06488} {arXiv:1710.06488 [hep-ex]} \BibitemShut
  {NoStop}%
\bibitem [{\citenamefont {Abe}\ \emph {et~al.}(2019)\citenamefont {Abe} \emph
  {et~al.}}]{T2K:2019efw}%
  \BibitemOpen
  \bibfield  {author} {\bibinfo {author} {\bibfnamefont {K.}~\bibnamefont
  {Abe}} \emph {et~al.} (\bibinfo {collaboration} {T2K}),\ }\href {\doibase
  10.1103/PhysRevD.99.071103} {\bibfield  {journal} {\bibinfo  {journal} {Phys.
  Rev. D}\ }\textbf {\bibinfo {volume} {99}},\ \bibinfo {pages} {071103}
  (\bibinfo {year} {2019})},\ \Eprint {http://arxiv.org/abs/1902.06529}
  {arXiv:1902.06529 [hep-ex]} \BibitemShut {NoStop}%
\bibitem [{\citenamefont {Vannerom}\ \emph {et~al.}(2022)\citenamefont
  {Vannerom}, \citenamefont {Fischer}, \citenamefont {Conrad}, \citenamefont
  {Blot},\ and\ \citenamefont {Arguelles}}]{Vannerom:2022cpf}%
  \BibitemOpen
  \bibfield  {author} {\bibinfo {author} {\bibfnamefont {D.}~\bibnamefont
  {Vannerom}}, \bibinfo {author} {\bibfnamefont {L.}~\bibnamefont {Fischer}},
  \bibinfo {author} {\bibfnamefont {J.}~\bibnamefont {Conrad}}, \bibinfo
  {author} {\bibfnamefont {S.}~\bibnamefont {Blot}}, \ and\ \bibinfo {author}
  {\bibfnamefont {C.}~\bibnamefont {Arguelles}},\ }\href {\doibase
  10.22323/1.380.0299} {\bibfield  {journal} {\bibinfo  {journal} {PoS}\
  }\textbf {\bibinfo {volume} {PANIC2021}},\ \bibinfo {pages} {299} (\bibinfo
  {year} {2022})}\BibitemShut {NoStop}%
\bibitem [{\citenamefont {Arg\"uelles}\ \emph {et~al.}(2022)\citenamefont
  {Arg\"uelles}, \citenamefont {Esteban}, \citenamefont {Hostert},
  \citenamefont {Kelly}, \citenamefont {Kopp}, \citenamefont {Machado},
  \citenamefont {Martinez-Soler},\ and\ \citenamefont
  {Perez-Gonzalez}}]{Arguelles:2021meu}%
  \BibitemOpen
  \bibfield  {author} {\bibinfo {author} {\bibfnamefont {C.~A.}\ \bibnamefont
  {Arg\"uelles}}, \bibinfo {author} {\bibfnamefont {I.}~\bibnamefont
  {Esteban}}, \bibinfo {author} {\bibfnamefont {M.}~\bibnamefont {Hostert}},
  \bibinfo {author} {\bibfnamefont {K.~J.}\ \bibnamefont {Kelly}}, \bibinfo
  {author} {\bibfnamefont {J.}~\bibnamefont {Kopp}}, \bibinfo {author}
  {\bibfnamefont {P.~A.~N.}\ \bibnamefont {Machado}}, \bibinfo {author}
  {\bibfnamefont {I.}~\bibnamefont {Martinez-Soler}}, \ and\ \bibinfo {author}
  {\bibfnamefont {Y.~F.}\ \bibnamefont {Perez-Gonzalez}},\ }\href {\doibase
  10.1103/PhysRevLett.128.241802} {\bibfield  {journal} {\bibinfo  {journal}
  {Phys. Rev. Lett.}\ }\textbf {\bibinfo {volume} {128}},\ \bibinfo {pages}
  {241802} (\bibinfo {year} {2022})},\ \Eprint
  {http://arxiv.org/abs/2111.10359} {arXiv:2111.10359 [hep-ph]} \BibitemShut
  {NoStop}%
\bibitem [{\citenamefont {Hausner}(2022)}]{Hausner:2022fli}%
  \BibitemOpen
  \bibfield  {author} {\bibinfo {author} {\bibfnamefont {H.}~\bibnamefont
  {Hausner}},\ }\emph {\bibinfo {title} {{Sterile Neutrino Search with the NOvA
  Detectors}}},\ \href@noop {} {Ph.D. thesis},\ \bibinfo  {school} {Wisconsin
  U., Madison, SAL} (\bibinfo {year} {2022})\BibitemShut {NoStop}%
\bibitem [{\citenamefont {Hu}(2021)}]{Hu:2020uvx}%
  \BibitemOpen
  \bibfield  {author} {\bibinfo {author} {\bibfnamefont {Z.}~\bibnamefont {Hu}}
  (\bibinfo {collaboration} {MINOS, MINOS+, Daya Bay, Bugey-3}),\ }\href
  {\doibase 10.22323/1.390.0201} {\bibfield  {journal} {\bibinfo  {journal}
  {PoS}\ }\textbf {\bibinfo {volume} {ICHEP2020}},\ \bibinfo {pages} {201}
  (\bibinfo {year} {2021})}\BibitemShut {NoStop}%
\bibitem [{\citenamefont {Adamson}\ \emph {et~al.}(2020)\citenamefont {Adamson}
  \emph {et~al.}}]{MINOS:2020iqj}%
  \BibitemOpen
  \bibfield  {author} {\bibinfo {author} {\bibfnamefont {P.}~\bibnamefont
  {Adamson}} \emph {et~al.} (\bibinfo {collaboration} {MINOS+, Daya Bay}),\
  }\href {\doibase 10.1103/PhysRevLett.125.071801} {\bibfield  {journal}
  {\bibinfo  {journal} {Phys. Rev. Lett.}\ }\textbf {\bibinfo {volume} {125}},\
  \bibinfo {pages} {071801} (\bibinfo {year} {2020})},\ \Eprint
  {http://arxiv.org/abs/2002.00301} {arXiv:2002.00301 [hep-ex]} \BibitemShut
  {NoStop}%
\bibitem [{\citenamefont {Gariazzo}\ \emph {et~al.}(2017)\citenamefont
  {Gariazzo}, \citenamefont {Giunti}, \citenamefont {Laveder},\ and\
  \citenamefont {Li}}]{Gariazzo:2017fdh}%
  \BibitemOpen
  \bibfield  {author} {\bibinfo {author} {\bibfnamefont {S.}~\bibnamefont
  {Gariazzo}}, \bibinfo {author} {\bibfnamefont {C.}~\bibnamefont {Giunti}},
  \bibinfo {author} {\bibfnamefont {M.}~\bibnamefont {Laveder}}, \ and\
  \bibinfo {author} {\bibfnamefont {Y.~F.}\ \bibnamefont {Li}},\ }\href
  {\doibase 10.1007/JHEP06(2017)135} {\bibfield  {journal} {\bibinfo  {journal}
  {JHEP}\ }\textbf {\bibinfo {volume} {06}},\ \bibinfo {pages} {135} (\bibinfo
  {year} {2017})},\ \Eprint {http://arxiv.org/abs/1703.00860} {arXiv:1703.00860
  [hep-ph]} \BibitemShut {NoStop}%
\bibitem [{\citenamefont {{MicroBooNE Collaboration}}\ \emph
  {et~al.}(2021)\citenamefont {{MicroBooNE Collaboration}}, \citenamefont
  {Abratenko} \emph {et~al.}}]{https://doi.org/10.48550/arxiv.2110.14054}%
  \BibitemOpen
  \bibfield  {author} {\bibinfo {author} {\bibnamefont {{MicroBooNE
  Collaboration}}}, \bibinfo {author} {\bibfnamefont {P.}~\bibnamefont
  {Abratenko}},  \emph {et~al.},\ }\href {\doibase 10.48550/ARXIV.2110.14054}
  {\enquote {\bibinfo {title} {Search for an excess of electron neutrino
  interactions in microboone using multiple final state topologies},}\ }
  (\bibinfo {year} {2021})\BibitemShut {NoStop}%
\bibitem [{\citenamefont {Abratenko}\ \emph {et~al.}(2022)\citenamefont
  {Abratenko} \emph {et~al.}}]{MicroBooNE:2022wdf}%
  \BibitemOpen
  \bibfield  {author} {\bibinfo {author} {\bibfnamefont {P.}~\bibnamefont
  {Abratenko}} \emph {et~al.} (\bibinfo {collaboration} {MicroBooNE}),\
  }\href@noop {} {\  (\bibinfo {year} {2022})},\ \Eprint
  {http://arxiv.org/abs/2210.10216} {arXiv:2210.10216 [hep-ex]} \BibitemShut
  {NoStop}%
\bibitem [{\citenamefont {Argüelles}\ \emph {et~al.}(2021)\citenamefont
  {Argüelles}, \citenamefont {Esteban}, \citenamefont {Hostert}, \citenamefont
  {Kelly}, \citenamefont {Kopp}, \citenamefont {Machado}, \citenamefont
  {Martinez-Soler},\ and\ \citenamefont
  {Perez-Gonzalez}}]{https://doi.org/10.48550/arxiv.2111.10359}%
  \BibitemOpen
  \bibfield  {author} {\bibinfo {author} {\bibfnamefont {C.~A.}\ \bibnamefont
  {Argüelles}}, \bibinfo {author} {\bibfnamefont {I.}~\bibnamefont {Esteban}},
  \bibinfo {author} {\bibfnamefont {M.}~\bibnamefont {Hostert}}, \bibinfo
  {author} {\bibfnamefont {K.~J.}\ \bibnamefont {Kelly}}, \bibinfo {author}
  {\bibfnamefont {J.}~\bibnamefont {Kopp}}, \bibinfo {author} {\bibfnamefont
  {P.~A.~N.}\ \bibnamefont {Machado}}, \bibinfo {author} {\bibfnamefont
  {I.}~\bibnamefont {Martinez-Soler}}, \ and\ \bibinfo {author} {\bibfnamefont
  {Y.~F.}\ \bibnamefont {Perez-Gonzalez}},\ }\href {\doibase
  10.48550/ARXIV.2111.10359} {\enquote {\bibinfo {title} {Microboone and the
  $\nu_e$ interpretation of the miniboone low-energy excess},}\ } (\bibinfo
  {year} {2021})\BibitemShut {NoStop}%
\bibitem [{\citenamefont {Aguilar}\ \emph {et~al.}(2022)\citenamefont {Aguilar}
  \emph {et~al.}}]{https://doi.org/10.48550/arxiv.2201.01724}%
  \BibitemOpen
  \bibfield  {author} {\bibinfo {author} {\bibfnamefont {A.}~\bibnamefont
  {Aguilar}} \emph {et~al.},\ }\href {\doibase 10.48550/ARXIV.2201.01724}
  {\enquote {\bibinfo {title} {Miniboone and microboone joint fit to a 3+1
  sterile neutrino scenario},}\ } (\bibinfo {year} {2022})\BibitemShut
  {NoStop}%
\bibitem [{\citenamefont {Böser}\ \emph {et~al.}(2020)\citenamefont {Böser},
  \citenamefont {Buck}, \citenamefont {Giunti}, \citenamefont {Lesgourgues},
  \citenamefont {Ludhova}, \citenamefont {Mertens}, \citenamefont {Schukraft},\
  and\ \citenamefont {Wurm}}]{B_ser_2020}%
  \BibitemOpen
  \bibfield  {author} {\bibinfo {author} {\bibfnamefont {S.}~\bibnamefont
  {Böser}}, \bibinfo {author} {\bibfnamefont {C.}~\bibnamefont {Buck}},
  \bibinfo {author} {\bibfnamefont {C.}~\bibnamefont {Giunti}}, \bibinfo
  {author} {\bibfnamefont {J.}~\bibnamefont {Lesgourgues}}, \bibinfo {author}
  {\bibfnamefont {L.}~\bibnamefont {Ludhova}}, \bibinfo {author} {\bibfnamefont
  {S.}~\bibnamefont {Mertens}}, \bibinfo {author} {\bibfnamefont
  {A.}~\bibnamefont {Schukraft}}, \ and\ \bibinfo {author} {\bibfnamefont
  {M.}~\bibnamefont {Wurm}},\ }\href {\doibase 10.1016/j.ppnp.2019.103736}
  {\bibfield  {journal} {\bibinfo  {journal} {Progress in Particle and Nuclear
  Physics}\ }\textbf {\bibinfo {volume} {111}},\ \bibinfo {pages} {103736}
  (\bibinfo {year} {2020})}\BibitemShut {NoStop}%
\bibitem [{\citenamefont {Chu}\ \emph {et~al.}(2018)\citenamefont {Chu},
  \citenamefont {Dasgupta}, \citenamefont {Dentler}, \citenamefont {Kopp},\
  and\ \citenamefont {Saviano}}]{Chu:2018gxk}%
  \BibitemOpen
  \bibfield  {author} {\bibinfo {author} {\bibfnamefont {X.}~\bibnamefont
  {Chu}}, \bibinfo {author} {\bibfnamefont {B.}~\bibnamefont {Dasgupta}},
  \bibinfo {author} {\bibfnamefont {M.}~\bibnamefont {Dentler}}, \bibinfo
  {author} {\bibfnamefont {J.}~\bibnamefont {Kopp}}, \ and\ \bibinfo {author}
  {\bibfnamefont {N.}~\bibnamefont {Saviano}},\ }\href {\doibase
  10.1088/1475-7516/2018/11/049} {\bibfield  {journal} {\bibinfo  {journal}
  {JCAP}\ }\textbf {\bibinfo {volume} {11}},\ \bibinfo {pages} {049} (\bibinfo
  {year} {2018})},\ \Eprint {http://arxiv.org/abs/1806.10629} {arXiv:1806.10629
  [hep-ph]} \BibitemShut {NoStop}%
\bibitem [{\citenamefont {Goswami}\ \emph {et~al.}(2022)\citenamefont
  {Goswami}, \citenamefont {K.~N.}, \citenamefont {Mukherjee},\ and\
  \citenamefont {Narendra}}]{Goswami:2021eqy}%
  \BibitemOpen
  \bibfield  {author} {\bibinfo {author} {\bibfnamefont {S.}~\bibnamefont
  {Goswami}}, \bibinfo {author} {\bibfnamefont {V.}~\bibnamefont {K.~N.}},
  \bibinfo {author} {\bibfnamefont {A.}~\bibnamefont {Mukherjee}}, \ and\
  \bibinfo {author} {\bibfnamefont {N.}~\bibnamefont {Narendra}},\ }\href
  {\doibase 10.1103/PhysRevD.105.095040} {\bibfield  {journal} {\bibinfo
  {journal} {Phys. Rev. D}\ }\textbf {\bibinfo {volume} {105}},\ \bibinfo
  {pages} {095040} (\bibinfo {year} {2022})},\ \Eprint
  {http://arxiv.org/abs/2111.14719} {arXiv:2111.14719 [hep-ph]} \BibitemShut
  {NoStop}%
\bibitem [{\citenamefont {Aker}\ \emph {et~al.}(2022)\citenamefont {Aker} \emph
  {et~al.}}]{KATRIN:2022spi}%
  \BibitemOpen
  \bibfield  {author} {\bibinfo {author} {\bibfnamefont {M.}~\bibnamefont
  {Aker}} \emph {et~al.} (\bibinfo {collaboration} {KATRIN}),\ }\href@noop {}
  {\  (\bibinfo {year} {2022})},\ \Eprint {http://arxiv.org/abs/2207.06337}
  {arXiv:2207.06337 [nucl-ex]} \BibitemShut {NoStop}%
\bibitem [{\citenamefont {Antonello}\ \emph {et~al.}(2015)\citenamefont
  {Antonello} \emph {et~al.}}]{MicroBooNE:2015bmn}%
  \BibitemOpen
  \bibfield  {author} {\bibinfo {author} {\bibfnamefont {M.}~\bibnamefont
  {Antonello}} \emph {et~al.} (\bibinfo {collaboration} {MicroBooNE, LAr1-ND,
  ICARUS-WA104}),\ }\href@noop {} {\  (\bibinfo {year} {2015})},\ \Eprint
  {http://arxiv.org/abs/1503.01520} {arXiv:1503.01520 [physics.ins-det]}
  \BibitemShut {NoStop}%
\bibitem [{\citenamefont {Ajimura}\ \emph {et~al.}(2021)\citenamefont {Ajimura}
  \emph {et~al.}}]{JSNS2:2021hyk}%
  \BibitemOpen
  \bibfield  {author} {\bibinfo {author} {\bibfnamefont {S.}~\bibnamefont
  {Ajimura}} \emph {et~al.} (\bibinfo {collaboration} {JSNS2}),\ }\href
  {\doibase 10.1016/j.nima.2021.165742} {\bibfield  {journal} {\bibinfo
  {journal} {Nucl. Instrum. Meth. A}\ }\textbf {\bibinfo {volume} {1014}},\
  \bibinfo {pages} {165742} (\bibinfo {year} {2021})},\ \Eprint
  {http://arxiv.org/abs/2104.13169} {arXiv:2104.13169 [physics.ins-det]}
  \BibitemShut {NoStop}%
\bibitem [{\citenamefont {Babu}\ \emph {et~al.}(2023)\citenamefont {Babu},
  \citenamefont {Brdar}, \citenamefont {de~Gouv\^ea},\ and\ \citenamefont
  {Machado}}]{Babu:2022non}%
  \BibitemOpen
  \bibfield  {author} {\bibinfo {author} {\bibfnamefont {K.~S.}\ \bibnamefont
  {Babu}}, \bibinfo {author} {\bibfnamefont {V.}~\bibnamefont {Brdar}},
  \bibinfo {author} {\bibfnamefont {A.}~\bibnamefont {de~Gouv\^ea}}, \ and\
  \bibinfo {author} {\bibfnamefont {P.~A.~N.}\ \bibnamefont {Machado}},\ }\href
  {\doibase 10.1103/PhysRevD.107.015017} {\bibfield  {journal} {\bibinfo
  {journal} {Phys. Rev. D}\ }\textbf {\bibinfo {volume} {107}},\ \bibinfo
  {pages} {015017} (\bibinfo {year} {2023})},\ \Eprint
  {http://arxiv.org/abs/2209.00031} {arXiv:2209.00031 [hep-ph]} \BibitemShut
  {NoStop}%
\bibitem [{\citenamefont {Hardin}\ \emph {et~al.}(2022)\citenamefont {Hardin},
  \citenamefont {Martinez-Soler}, \citenamefont {Diaz}, \citenamefont {Jin},
  \citenamefont {Kamp}, \citenamefont {Arg\"uelles}, \citenamefont {Conrad},\
  and\ \citenamefont {Shaevitz}}]{Hardin:2022muu}%
  \BibitemOpen
  \bibfield  {author} {\bibinfo {author} {\bibfnamefont {J.~M.}\ \bibnamefont
  {Hardin}}, \bibinfo {author} {\bibfnamefont {I.}~\bibnamefont
  {Martinez-Soler}}, \bibinfo {author} {\bibfnamefont {A.}~\bibnamefont
  {Diaz}}, \bibinfo {author} {\bibfnamefont {M.}~\bibnamefont {Jin}}, \bibinfo
  {author} {\bibfnamefont {N.~W.}\ \bibnamefont {Kamp}}, \bibinfo {author}
  {\bibfnamefont {C.~A.}\ \bibnamefont {Arg\"uelles}}, \bibinfo {author}
  {\bibfnamefont {J.~M.}\ \bibnamefont {Conrad}}, \ and\ \bibinfo {author}
  {\bibfnamefont {M.~H.}\ \bibnamefont {Shaevitz}},\ }\href@noop {} {\
  (\bibinfo {year} {2022})},\ \Eprint {http://arxiv.org/abs/2211.02610}
  {arXiv:2211.02610 [hep-ph]} \BibitemShut {NoStop}%
\bibitem [{\citenamefont {Agarwalla}\ \emph {et~al.}(2017)\citenamefont
  {Agarwalla}, \citenamefont {Chatterjee},\ and\ \citenamefont
  {Palazzo}}]{Agarwalla:2016xlg}%
  \BibitemOpen
  \bibfield  {author} {\bibinfo {author} {\bibfnamefont {S.~K.}\ \bibnamefont
  {Agarwalla}}, \bibinfo {author} {\bibfnamefont {S.~S.}\ \bibnamefont
  {Chatterjee}}, \ and\ \bibinfo {author} {\bibfnamefont {A.}~\bibnamefont
  {Palazzo}},\ }\href {\doibase 10.1103/PhysRevLett.118.031804} {\bibfield
  {journal} {\bibinfo  {journal} {Phys. Rev. Lett.}\ }\textbf {\bibinfo
  {volume} {118}},\ \bibinfo {pages} {031804} (\bibinfo {year} {2017})},\
  \Eprint {http://arxiv.org/abs/1605.04299} {arXiv:1605.04299 [hep-ph]}
  \BibitemShut {NoStop}%
\bibitem [{\citenamefont {Kumar~Agarwalla}\ \emph {et~al.}(2013)\citenamefont
  {Kumar~Agarwalla}, \citenamefont {Prakash},\ and\ \citenamefont
  {Uma~Sankar}}]{KumarAgarwalla:2013fko}%
  \BibitemOpen
  \bibfield  {author} {\bibinfo {author} {\bibfnamefont {S.}~\bibnamefont
  {Kumar~Agarwalla}}, \bibinfo {author} {\bibfnamefont {S.}~\bibnamefont
  {Prakash}}, \ and\ \bibinfo {author} {\bibfnamefont {S.}~\bibnamefont
  {Uma~Sankar}},\ }\href {\doibase 10.22323/1.180.0534} {\bibfield  {journal}
  {\bibinfo  {journal} {PoS}\ }\textbf {\bibinfo {volume} {EPS-HEP2013}},\
  \bibinfo {pages} {534} (\bibinfo {year} {2013})}\BibitemShut {NoStop}%
\bibitem [{\citenamefont {Ghosh}\ \emph {et~al.}(2017)\citenamefont {Ghosh},
  \citenamefont {Gupta}, \citenamefont {Matthews}, \citenamefont {Sharma},\
  and\ \citenamefont {Williams}}]{Ghosh:2017atj}%
  \BibitemOpen
  \bibfield  {author} {\bibinfo {author} {\bibfnamefont {M.}~\bibnamefont
  {Ghosh}}, \bibinfo {author} {\bibfnamefont {S.}~\bibnamefont {Gupta}},
  \bibinfo {author} {\bibfnamefont {Z.~M.}\ \bibnamefont {Matthews}}, \bibinfo
  {author} {\bibfnamefont {P.}~\bibnamefont {Sharma}}, \ and\ \bibinfo {author}
  {\bibfnamefont {A.~G.}\ \bibnamefont {Williams}},\ }\href {\doibase
  10.1103/PhysRevD.96.075018} {\bibfield  {journal} {\bibinfo  {journal} {Phys.
  Rev. D}\ }\textbf {\bibinfo {volume} {96}},\ \bibinfo {pages} {075018}
  (\bibinfo {year} {2017})},\ \Eprint {http://arxiv.org/abs/1704.04771}
  {arXiv:1704.04771 [hep-ph]} \BibitemShut {NoStop}%
\bibitem [{\citenamefont {Choubey}\ \emph {et~al.}(2019)\citenamefont
  {Choubey}, \citenamefont {Dutta},\ and\ \citenamefont
  {Pramanik}}]{Choubey_2019}%
  \BibitemOpen
  \bibfield  {author} {\bibinfo {author} {\bibfnamefont {S.}~\bibnamefont
  {Choubey}}, \bibinfo {author} {\bibfnamefont {D.}~\bibnamefont {Dutta}}, \
  and\ \bibinfo {author} {\bibfnamefont {D.}~\bibnamefont {Pramanik}},\ }\href
  {\doibase 10.1140/epjc/s10052-019-7479-8} {\bibfield  {journal} {\bibinfo
  {journal} {The European Physical Journal C}\ }\textbf {\bibinfo {volume}
  {79}} (\bibinfo {year} {2019}),\ 10.1140/epjc/s10052-019-7479-8}\BibitemShut
  {NoStop}%
\bibitem [{\citenamefont {Dutta}\ \emph {et~al.}(2016)\citenamefont {Dutta},
  \citenamefont {Gandhi}, \citenamefont {Kayser}, \citenamefont {Masud},\ and\
  \citenamefont {Prakash}}]{Dutta:2016glq}%
  \BibitemOpen
  \bibfield  {author} {\bibinfo {author} {\bibfnamefont {D.}~\bibnamefont
  {Dutta}}, \bibinfo {author} {\bibfnamefont {R.}~\bibnamefont {Gandhi}},
  \bibinfo {author} {\bibfnamefont {B.}~\bibnamefont {Kayser}}, \bibinfo
  {author} {\bibfnamefont {M.}~\bibnamefont {Masud}}, \ and\ \bibinfo {author}
  {\bibfnamefont {S.}~\bibnamefont {Prakash}},\ }\href {\doibase
  10.1007/JHEP11(2016)122} {\bibfield  {journal} {\bibinfo  {journal} {JHEP}\
  }\textbf {\bibinfo {volume} {11}},\ \bibinfo {pages} {122} (\bibinfo {year}
  {2016})},\ \Eprint {http://arxiv.org/abs/1607.02152} {arXiv:1607.02152
  [hep-ph]} \BibitemShut {NoStop}%
\bibitem [{\citenamefont {Singha}\ \emph {et~al.}(2022)\citenamefont {Singha},
  \citenamefont {Ghosh}, \citenamefont {Majhi},\ and\ \citenamefont
  {Mohanta}}]{Singha:2022btw}%
  \BibitemOpen
  \bibfield  {author} {\bibinfo {author} {\bibfnamefont {D.~K.}\ \bibnamefont
  {Singha}}, \bibinfo {author} {\bibfnamefont {M.}~\bibnamefont {Ghosh}},
  \bibinfo {author} {\bibfnamefont {R.}~\bibnamefont {Majhi}}, \ and\ \bibinfo
  {author} {\bibfnamefont {R.}~\bibnamefont {Mohanta}},\ }\href@noop {} {\
  (\bibinfo {year} {2022})},\ \Eprint {http://arxiv.org/abs/2211.01816}
  {arXiv:2211.01816 [hep-ph]} \BibitemShut {NoStop}%
\bibitem [{\citenamefont {Berryman}\ \emph {et~al.}(2015)\citenamefont
  {Berryman}, \citenamefont {de~Gouv\^ea}, \citenamefont {Kelly},\ and\
  \citenamefont {Kobach}}]{Berryman:2015nua}%
  \BibitemOpen
  \bibfield  {author} {\bibinfo {author} {\bibfnamefont {J.~M.}\ \bibnamefont
  {Berryman}}, \bibinfo {author} {\bibfnamefont {A.}~\bibnamefont
  {de~Gouv\^ea}}, \bibinfo {author} {\bibfnamefont {K.~J.}\ \bibnamefont
  {Kelly}}, \ and\ \bibinfo {author} {\bibfnamefont {A.}~\bibnamefont
  {Kobach}},\ }\href {\doibase 10.1103/PhysRevD.92.073012} {\bibfield
  {journal} {\bibinfo  {journal} {Phys. Rev. D}\ }\textbf {\bibinfo {volume}
  {92}},\ \bibinfo {pages} {073012} (\bibinfo {year} {2015})},\ \Eprint
  {http://arxiv.org/abs/1507.03986} {arXiv:1507.03986 [hep-ph]} \BibitemShut
  {NoStop}%
\bibitem [{\citenamefont {Gandhi}\ \emph {et~al.}(2015)\citenamefont {Gandhi},
  \citenamefont {Kayser}, \citenamefont {Masud},\ and\ \citenamefont
  {Prakash}}]{Gandhi:2015xza}%
  \BibitemOpen
  \bibfield  {author} {\bibinfo {author} {\bibfnamefont {R.}~\bibnamefont
  {Gandhi}}, \bibinfo {author} {\bibfnamefont {B.}~\bibnamefont {Kayser}},
  \bibinfo {author} {\bibfnamefont {M.}~\bibnamefont {Masud}}, \ and\ \bibinfo
  {author} {\bibfnamefont {S.}~\bibnamefont {Prakash}},\ }\href {\doibase
  10.1007/JHEP11(2015)039} {\bibfield  {journal} {\bibinfo  {journal} {JHEP}\
  }\textbf {\bibinfo {volume} {11}},\ \bibinfo {pages} {039} (\bibinfo {year}
  {2015})},\ \Eprint {http://arxiv.org/abs/1508.06275} {arXiv:1508.06275
  [hep-ph]} \BibitemShut {NoStop}%
\bibitem [{\citenamefont {Agarwalla}\ \emph {et~al.}(2016)\citenamefont
  {Agarwalla}, \citenamefont {Chatterjee},\ and\ \citenamefont
  {Palazzo}}]{Agarwalla:2016xxa}%
  \BibitemOpen
  \bibfield  {author} {\bibinfo {author} {\bibfnamefont {S.~K.}\ \bibnamefont
  {Agarwalla}}, \bibinfo {author} {\bibfnamefont {S.~S.}\ \bibnamefont
  {Chatterjee}}, \ and\ \bibinfo {author} {\bibfnamefont {A.}~\bibnamefont
  {Palazzo}},\ }\href {\doibase 10.1007/JHEP09(2016)016} {\bibfield  {journal}
  {\bibinfo  {journal} {JHEP}\ }\textbf {\bibinfo {volume} {09}},\ \bibinfo
  {pages} {016} (\bibinfo {year} {2016})},\ \Eprint
  {http://arxiv.org/abs/1603.03759} {arXiv:1603.03759 [hep-ph]} \BibitemShut
  {NoStop}%
\bibitem [{\citenamefont {Reyimuaji}\ and\ \citenamefont
  {Liu}(2020)}]{Reyimuaji:2019wbn}%
  \BibitemOpen
  \bibfield  {author} {\bibinfo {author} {\bibfnamefont {Y.}~\bibnamefont
  {Reyimuaji}}\ and\ \bibinfo {author} {\bibfnamefont {C.}~\bibnamefont
  {Liu}},\ }\href {\doibase 10.1007/JHEP06(2020)094} {\bibfield  {journal}
  {\bibinfo  {journal} {JHEP}\ }\textbf {\bibinfo {volume} {06}},\ \bibinfo
  {pages} {094} (\bibinfo {year} {2020})},\ \Eprint
  {http://arxiv.org/abs/1911.12524} {arXiv:1911.12524 [hep-ph]} \BibitemShut
  {NoStop}%
\bibitem [{\citenamefont {Denton}\ \emph {et~al.}(2022)\citenamefont {Denton},
  \citenamefont {Giarnetti},\ and\ \citenamefont {Meloni}}]{Denton:2022pxt}%
  \BibitemOpen
  \bibfield  {author} {\bibinfo {author} {\bibfnamefont {P.~B.}\ \bibnamefont
  {Denton}}, \bibinfo {author} {\bibfnamefont {A.}~\bibnamefont {Giarnetti}}, \
  and\ \bibinfo {author} {\bibfnamefont {D.}~\bibnamefont {Meloni}},\
  }\href@noop {} {\  (\bibinfo {year} {2022})},\ \Eprint
  {http://arxiv.org/abs/2210.00109} {arXiv:2210.00109 [hep-ph]} \BibitemShut
  {NoStop}%
\bibitem [{\citenamefont {Choubey}\ \emph {et~al.}(2017)\citenamefont
  {Choubey}, \citenamefont {Dutta},\ and\ \citenamefont
  {Pramanik}}]{Choubey:2017cba}%
  \BibitemOpen
  \bibfield  {author} {\bibinfo {author} {\bibfnamefont {S.}~\bibnamefont
  {Choubey}}, \bibinfo {author} {\bibfnamefont {D.}~\bibnamefont {Dutta}}, \
  and\ \bibinfo {author} {\bibfnamefont {D.}~\bibnamefont {Pramanik}},\ }\href
  {\doibase 10.1103/PhysRevD.96.056026} {\bibfield  {journal} {\bibinfo
  {journal} {Phys. Rev. D}\ }\textbf {\bibinfo {volume} {96}},\ \bibinfo
  {pages} {056026} (\bibinfo {year} {2017})},\ \Eprint
  {http://arxiv.org/abs/1704.07269} {arXiv:1704.07269 [hep-ph]} \BibitemShut
  {NoStop}%
\bibitem [{\citenamefont {Rubbia}(1977)}]{Rubbia:1977zz}%
  \BibitemOpen
  \bibfield  {author} {\bibinfo {author} {\bibfnamefont {C.}~\bibnamefont
  {Rubbia}},\ }\href@noop {} {\  (\bibinfo {year} {1977})}\BibitemShut
  {NoStop}%
\bibitem [{\citenamefont {Gandhi}\ \emph {et~al.}(2008)\citenamefont {Gandhi},
  \citenamefont {Ghoshal}, \citenamefont {Goswami},\ and\ \citenamefont
  {Sankar}}]{Gandhi:2008zs}%
  \BibitemOpen
  \bibfield  {author} {\bibinfo {author} {\bibfnamefont {R.}~\bibnamefont
  {Gandhi}}, \bibinfo {author} {\bibfnamefont {P.}~\bibnamefont {Ghoshal}},
  \bibinfo {author} {\bibfnamefont {S.}~\bibnamefont {Goswami}}, \ and\
  \bibinfo {author} {\bibfnamefont {S.~U.}\ \bibnamefont {Sankar}},\ }\href
  {\doibase 10.1103/PhysRevD.78.073001} {\bibfield  {journal} {\bibinfo
  {journal} {Phys. Rev. D}\ }\textbf {\bibinfo {volume} {78}},\ \bibinfo
  {pages} {073001} (\bibinfo {year} {2008})},\ \Eprint
  {http://arxiv.org/abs/0807.2759} {arXiv:0807.2759 [hep-ph]} \BibitemShut
  {NoStop}%
\bibitem [{\citenamefont {Parke}\ and\ \citenamefont
  {Zhang}(2020{\natexlab{a}})}]{PhysRevD.101.056005}%
  \BibitemOpen
  \bibfield  {author} {\bibinfo {author} {\bibfnamefont {S.~J.}\ \bibnamefont
  {Parke}}\ and\ \bibinfo {author} {\bibfnamefont {X.}~\bibnamefont {Zhang}},\
  }\href {\doibase 10.1103/PhysRevD.101.056005} {\bibfield  {journal} {\bibinfo
   {journal} {Phys. Rev. D}\ }\textbf {\bibinfo {volume} {101}},\ \bibinfo
  {pages} {056005} (\bibinfo {year} {2020}{\natexlab{a}})}\BibitemShut
  {NoStop}%
\bibitem [{\citenamefont {Li}\ \emph {et~al.}(2018)\citenamefont {Li},
  \citenamefont {Ling}, \citenamefont {Xu},\ and\ \citenamefont
  {Yue}}]{Li_2018}%
  \BibitemOpen
  \bibfield  {author} {\bibinfo {author} {\bibfnamefont {W.}~\bibnamefont
  {Li}}, \bibinfo {author} {\bibfnamefont {J.}~\bibnamefont {Ling}}, \bibinfo
  {author} {\bibfnamefont {F.}~\bibnamefont {Xu}}, \ and\ \bibinfo {author}
  {\bibfnamefont {B.}~\bibnamefont {Yue}},\ }\href {\doibase
  10.1007/jhep10(2018)021} {\bibfield  {journal} {\bibinfo  {journal} {Journal
  of High Energy Physics}\ }\textbf {\bibinfo {volume} {2018}} (\bibinfo {year}
  {2018}),\ 10.1007/jhep10(2018)021}\BibitemShut {NoStop}%
\bibitem [{\citenamefont {Behera}\ \emph {et~al.}(2017)\citenamefont {Behera},
  \citenamefont {Ghosh}, \citenamefont {Choubey}, \citenamefont {Datar},
  \citenamefont {Mishra},\ and\ \citenamefont {Mohanty}}]{Behera:2016kwr}%
  \BibitemOpen
  \bibfield  {author} {\bibinfo {author} {\bibfnamefont {S.~P.}\ \bibnamefont
  {Behera}}, \bibinfo {author} {\bibfnamefont {A.}~\bibnamefont {Ghosh}},
  \bibinfo {author} {\bibfnamefont {S.}~\bibnamefont {Choubey}}, \bibinfo
  {author} {\bibfnamefont {V.~M.}\ \bibnamefont {Datar}}, \bibinfo {author}
  {\bibfnamefont {D.~K.}\ \bibnamefont {Mishra}}, \ and\ \bibinfo {author}
  {\bibfnamefont {A.~K.}\ \bibnamefont {Mohanty}},\ }\href {\doibase
  10.1140/epjc/s10052-017-4851-4} {\bibfield  {journal} {\bibinfo  {journal}
  {Eur. Phys. J. C}\ }\textbf {\bibinfo {volume} {77}},\ \bibinfo {pages} {307}
  (\bibinfo {year} {2017})},\ \Eprint {http://arxiv.org/abs/1605.08607}
  {arXiv:1605.08607 [hep-ph]} \BibitemShut {NoStop}%
\bibitem [{\citenamefont {Gandhi}\ and\ \citenamefont
  {Ghoshal}(2012)}]{Gandhi:2011jg}%
  \BibitemOpen
  \bibfield  {author} {\bibinfo {author} {\bibfnamefont {R.}~\bibnamefont
  {Gandhi}}\ and\ \bibinfo {author} {\bibfnamefont {P.}~\bibnamefont
  {Ghoshal}},\ }\href {\doibase 10.1103/PhysRevD.86.037301} {\bibfield
  {journal} {\bibinfo  {journal} {Phys. Rev. D}\ }\textbf {\bibinfo {volume}
  {86}},\ \bibinfo {pages} {037301} (\bibinfo {year} {2012})},\ \Eprint
  {http://arxiv.org/abs/1108.4360} {arXiv:1108.4360 [hep-ph]} \BibitemShut
  {NoStop}%
\bibitem [{\citenamefont {Thakore}\ \emph {et~al.}(2018)\citenamefont
  {Thakore}, \citenamefont {Devi}, \citenamefont {Kumar~Agarwalla},\ and\
  \citenamefont {Dighe}}]{Thakore:2018lgn}%
  \BibitemOpen
  \bibfield  {author} {\bibinfo {author} {\bibfnamefont {T.}~\bibnamefont
  {Thakore}}, \bibinfo {author} {\bibfnamefont {M.~M.}\ \bibnamefont {Devi}},
  \bibinfo {author} {\bibfnamefont {S.}~\bibnamefont {Kumar~Agarwalla}}, \ and\
  \bibinfo {author} {\bibfnamefont {A.}~\bibnamefont {Dighe}},\ }\href
  {\doibase 10.1007/JHEP08(2018)022} {\bibfield  {journal} {\bibinfo  {journal}
  {JHEP}\ }\textbf {\bibinfo {volume} {08}},\ \bibinfo {pages} {022} (\bibinfo
  {year} {2018})},\ \Eprint {http://arxiv.org/abs/1804.09613} {arXiv:1804.09613
  [hep-ph]} \BibitemShut {NoStop}%
\bibitem [{\citenamefont {Acero}\ \emph {et~al.}(2022)\citenamefont {Acero}
  \emph {et~al.}}]{Acero:2022wqg}%
  \BibitemOpen
  \bibfield  {author} {\bibinfo {author} {\bibfnamefont {M.~A.}\ \bibnamefont
  {Acero}} \emph {et~al.},\ }\href@noop {} {\  (\bibinfo {year} {2022})},\
  \Eprint {http://arxiv.org/abs/2203.07323} {arXiv:2203.07323 [hep-ex]}
  \BibitemShut {NoStop}%
\bibitem [{\citenamefont {Parke}\ and\ \citenamefont
  {Zhang}(2020{\natexlab{b}})}]{Parke:2019jyu}%
  \BibitemOpen
  \bibfield  {author} {\bibinfo {author} {\bibfnamefont {S.~J.}\ \bibnamefont
  {Parke}}\ and\ \bibinfo {author} {\bibfnamefont {X.}~\bibnamefont {Zhang}},\
  }\href {\doibase 10.1103/PhysRevD.101.056005} {\bibfield  {journal} {\bibinfo
   {journal} {Phys. Rev. D}\ }\textbf {\bibinfo {volume} {101}},\ \bibinfo
  {pages} {056005} (\bibinfo {year} {2020}{\natexlab{b}})},\ \Eprint
  {http://arxiv.org/abs/1905.01356} {arXiv:1905.01356 [hep-ph]} \BibitemShut
  {NoStop}%
\bibitem [{\citenamefont {Chattopadhyay}\ \emph {et~al.}(2022)\citenamefont
  {Chattopadhyay}, \citenamefont {Chakraborty}, \citenamefont {Dighe},\ and\
  \citenamefont {Goswami}}]{Chattopadhyay:2022ftv}%
  \BibitemOpen
  \bibfield  {author} {\bibinfo {author} {\bibfnamefont {D.~S.}\ \bibnamefont
  {Chattopadhyay}}, \bibinfo {author} {\bibfnamefont {K.}~\bibnamefont
  {Chakraborty}}, \bibinfo {author} {\bibfnamefont {A.}~\bibnamefont {Dighe}},
  \ and\ \bibinfo {author} {\bibfnamefont {S.}~\bibnamefont {Goswami}},\
  }\href@noop {} {\  (\bibinfo {year} {2022})},\ \Eprint
  {http://arxiv.org/abs/2204.05803} {arXiv:2204.05803 [hep-ph]} \BibitemShut
  {NoStop}%
\bibitem [{\citenamefont {Banuls}\ \emph {et~al.}(2001)\citenamefont {Banuls},
  \citenamefont {Barenboim},\ and\ \citenamefont {Bernabeu}}]{Banuls:2001zn}%
  \BibitemOpen
  \bibfield  {author} {\bibinfo {author} {\bibfnamefont {M.~C.}\ \bibnamefont
  {Banuls}}, \bibinfo {author} {\bibfnamefont {G.}~\bibnamefont {Barenboim}}, \
  and\ \bibinfo {author} {\bibfnamefont {J.}~\bibnamefont {Bernabeu}},\ }\href
  {\doibase 10.1016/S0370-2693(01)00723-7} {\bibfield  {journal} {\bibinfo
  {journal} {Phys. Lett. B}\ }\textbf {\bibinfo {volume} {513}},\ \bibinfo
  {pages} {391} (\bibinfo {year} {2001})},\ \Eprint
  {http://arxiv.org/abs/hep-ph/0102184} {arXiv:hep-ph/0102184} \BibitemShut
  {NoStop}%
\bibitem [{\citenamefont {Huber}\ \emph {et~al.}(2005)\citenamefont {Huber},
  \citenamefont {Lindner},\ and\ \citenamefont {Winter}}]{Huber:2004ka}%
  \BibitemOpen
  \bibfield  {author} {\bibinfo {author} {\bibfnamefont {P.}~\bibnamefont
  {Huber}}, \bibinfo {author} {\bibfnamefont {M.}~\bibnamefont {Lindner}}, \
  and\ \bibinfo {author} {\bibfnamefont {W.}~\bibnamefont {Winter}},\ }\href
  {\doibase 10.1016/j.cpc.2005.01.003} {\bibfield  {journal} {\bibinfo
  {journal} {Comput. Phys. Commun.}\ }\textbf {\bibinfo {volume} {167}},\
  \bibinfo {pages} {195} (\bibinfo {year} {2005})},\ \Eprint
  {http://arxiv.org/abs/hep-ph/0407333} {arXiv:hep-ph/0407333 [hep-ph]}
  \BibitemShut {NoStop}%
\bibitem [{\citenamefont {Gandhi}\ \emph {et~al.}(2007)\citenamefont {Gandhi},
  \citenamefont {Ghoshal}, \citenamefont {Goswami}, \citenamefont {Mehta},
  \citenamefont {Sankar},\ and\ \citenamefont {Shalgar}}]{Gandhi:2007td}%
  \BibitemOpen
  \bibfield  {author} {\bibinfo {author} {\bibfnamefont {R.}~\bibnamefont
  {Gandhi}}, \bibinfo {author} {\bibfnamefont {P.}~\bibnamefont {Ghoshal}},
  \bibinfo {author} {\bibfnamefont {S.}~\bibnamefont {Goswami}}, \bibinfo
  {author} {\bibfnamefont {P.}~\bibnamefont {Mehta}}, \bibinfo {author}
  {\bibfnamefont {S.~U.}\ \bibnamefont {Sankar}}, \ and\ \bibinfo {author}
  {\bibfnamefont {S.}~\bibnamefont {Shalgar}},\ }\href {\doibase
  10.1103/PhysRevD.76.073012} {\bibfield  {journal} {\bibinfo  {journal} {Phys.
  Rev. D}\ }\textbf {\bibinfo {volume} {76}},\ \bibinfo {pages} {073012}
  (\bibinfo {year} {2007})},\ \Eprint {http://arxiv.org/abs/0707.1723}
  {arXiv:0707.1723 [hep-ph]} \BibitemShut {NoStop}%
\bibitem [{\citenamefont {Choubey}\ and\ \citenamefont
  {Roy}(2006)}]{Choubey:2005zy}%
  \BibitemOpen
  \bibfield  {author} {\bibinfo {author} {\bibfnamefont {S.}~\bibnamefont
  {Choubey}}\ and\ \bibinfo {author} {\bibfnamefont {P.}~\bibnamefont {Roy}},\
  }\href {\doibase 10.1103/PhysRevD.73.013006} {\bibfield  {journal} {\bibinfo
  {journal} {Phys. Rev. D}\ }\textbf {\bibinfo {volume} {73}},\ \bibinfo
  {pages} {013006} (\bibinfo {year} {2006})},\ \Eprint
  {http://arxiv.org/abs/hep-ph/0509197} {arXiv:hep-ph/0509197} \BibitemShut
  {NoStop}%
\bibitem [{Note1()}]{Note1}%
  \BibitemOpen
  \bibinfo {note} {In the appendix we have shown that with non-zero $\Delta
  _{21}$ in the Cayley Hamilton method we get better fit at these regions as
  well as at very low energies.}\BibitemShut {Stop}%
\bibitem [{\citenamefont {Fogli}\ \emph {et~al.}(1994)\citenamefont {Fogli},
  \citenamefont {Lisi},\ and\ \citenamefont {Montanino}}]{Fogli:1993ck}%
  \BibitemOpen
  \bibfield  {author} {\bibinfo {author} {\bibfnamefont {G.~L.}\ \bibnamefont
  {Fogli}}, \bibinfo {author} {\bibfnamefont {E.}~\bibnamefont {Lisi}}, \ and\
  \bibinfo {author} {\bibfnamefont {D.}~\bibnamefont {Montanino}},\ }\href
  {\doibase 10.1103/PhysRevD.49.3626} {\bibfield  {journal} {\bibinfo
  {journal} {Phys. Rev. D}\ }\textbf {\bibinfo {volume} {49}},\ \bibinfo
  {pages} {3626} (\bibinfo {year} {1994})}\BibitemShut {NoStop}%
\bibitem [{\citenamefont {Abi}\ \emph {et~al.}(2020{\natexlab{b}})\citenamefont
  {Abi} \emph {et~al.}}]{DUNE:2020lwj}%
  \BibitemOpen
  \bibfield  {author} {\bibinfo {author} {\bibfnamefont {B.}~\bibnamefont
  {Abi}} \emph {et~al.} (\bibinfo {collaboration} {DUNE}),\ }\href {\doibase
  10.1088/1748-0221/15/08/T08008} {\bibfield  {journal} {\bibinfo  {journal}
  {JINST}\ }\textbf {\bibinfo {volume} {15}},\ \bibinfo {pages} {T08008}
  (\bibinfo {year} {2020}{\natexlab{b}})},\ \Eprint
  {http://arxiv.org/abs/2002.02967} {arXiv:2002.02967 [physics.ins-det]}
  \BibitemShut {NoStop}%
\bibitem [{\citenamefont {Acciarri}\ \emph {et~al.}(2016)\citenamefont
  {Acciarri} \emph {et~al.}}]{DUNE:2016hlj}%
  \BibitemOpen
  \bibfield  {author} {\bibinfo {author} {\bibfnamefont {R.}~\bibnamefont
  {Acciarri}} \emph {et~al.} (\bibinfo {collaboration} {DUNE}),\ }\href@noop {}
  {\  (\bibinfo {year} {2016})},\ \Eprint {http://arxiv.org/abs/1601.05471}
  {arXiv:1601.05471 [physics.ins-det]} \BibitemShut {NoStop}%
\bibitem [{\citenamefont {Alion}\ \emph {et~al.}(2016)\citenamefont {Alion}
  \emph {et~al.}}]{DUNE:2016ymp}%
  \BibitemOpen
  \bibfield  {author} {\bibinfo {author} {\bibfnamefont {T.}~\bibnamefont
  {Alion}} \emph {et~al.} (\bibinfo {collaboration} {DUNE}),\ }\href@noop {} {\
   (\bibinfo {year} {2016})},\ \Eprint {http://arxiv.org/abs/1606.09550}
  {arXiv:1606.09550 [physics.ins-det]} \BibitemShut {NoStop}%
\bibitem [{\citenamefont {Honda}\ \emph {et~al.}(2015)\citenamefont {Honda},
  \citenamefont {Athar}, \citenamefont {Kajita}, \citenamefont {Kasahara},\
  and\ \citenamefont {Midorikawa}}]{PhysRevD.92.023004}%
  \BibitemOpen
  \bibfield  {author} {\bibinfo {author} {\bibfnamefont {M.}~\bibnamefont
  {Honda}}, \bibinfo {author} {\bibfnamefont {M.~S.}\ \bibnamefont {Athar}},
  \bibinfo {author} {\bibfnamefont {T.}~\bibnamefont {Kajita}}, \bibinfo
  {author} {\bibfnamefont {K.}~\bibnamefont {Kasahara}}, \ and\ \bibinfo
  {author} {\bibfnamefont {S.}~\bibnamefont {Midorikawa}},\ }\href {\doibase
  10.1103/PhysRevD.92.023004} {\bibfield  {journal} {\bibinfo  {journal} {Phys.
  Rev. D}\ }\textbf {\bibinfo {volume} {92}},\ \bibinfo {pages} {023004}
  (\bibinfo {year} {2015})}\BibitemShut {NoStop}%
\bibitem [{\citenamefont {Barger}\ \emph {et~al.}(2016)\citenamefont {Barger},
  \citenamefont {Bhattacharya}, \citenamefont {Chatterjee}, \citenamefont
  {Gandhi}, \citenamefont {Marfatia},\ and\ \citenamefont
  {Masud}}]{Barger:2014dfa}%
  \BibitemOpen
  \bibfield  {author} {\bibinfo {author} {\bibfnamefont {V.}~\bibnamefont
  {Barger}}, \bibinfo {author} {\bibfnamefont {A.}~\bibnamefont
  {Bhattacharya}}, \bibinfo {author} {\bibfnamefont {A.}~\bibnamefont
  {Chatterjee}}, \bibinfo {author} {\bibfnamefont {R.}~\bibnamefont {Gandhi}},
  \bibinfo {author} {\bibfnamefont {D.}~\bibnamefont {Marfatia}}, \ and\
  \bibinfo {author} {\bibfnamefont {M.}~\bibnamefont {Masud}},\ }\href
  {\doibase 10.1142/S0217751X16500202} {\bibfield  {journal} {\bibinfo
  {journal} {Int. J. Mod. Phys. A}\ }\textbf {\bibinfo {volume} {31}},\
  \bibinfo {pages} {1650020} (\bibinfo {year} {2016})},\ \Eprint
  {http://arxiv.org/abs/1405.1054} {arXiv:1405.1054 [hep-ph]} \BibitemShut
  {NoStop}%
\bibitem [{\citenamefont {Ternes}\ \emph {et~al.}(2019)\citenamefont {Ternes},
  \citenamefont {Gariazzo}, \citenamefont {Hajjar}, \citenamefont {Mena},
  \citenamefont {Sorel},\ and\ \citenamefont
  {T\'ortola}}]{PhysRevD.100.093004}%
  \BibitemOpen
  \bibfield  {author} {\bibinfo {author} {\bibfnamefont {C.~A.}\ \bibnamefont
  {Ternes}}, \bibinfo {author} {\bibfnamefont {S.}~\bibnamefont {Gariazzo}},
  \bibinfo {author} {\bibfnamefont {R.}~\bibnamefont {Hajjar}}, \bibinfo
  {author} {\bibfnamefont {O.}~\bibnamefont {Mena}}, \bibinfo {author}
  {\bibfnamefont {M.}~\bibnamefont {Sorel}}, \ and\ \bibinfo {author}
  {\bibfnamefont {M.}~\bibnamefont {T\'ortola}},\ }\href {\doibase
  10.1103/PhysRevD.100.093004} {\bibfield  {journal} {\bibinfo  {journal}
  {Phys. Rev. D}\ }\textbf {\bibinfo {volume} {100}},\ \bibinfo {pages}
  {093004} (\bibinfo {year} {2019})}\BibitemShut {NoStop}%
\bibitem [{\citenamefont {Suzuki}\ \emph {et~al.}(1987)\citenamefont {Suzuki},
  \citenamefont {Measday},\ and\ \citenamefont {Roalsvig}}]{PhysRevC.35.2212}%
  \BibitemOpen
  \bibfield  {author} {\bibinfo {author} {\bibfnamefont {T.}~\bibnamefont
  {Suzuki}}, \bibinfo {author} {\bibfnamefont {D.~F.}\ \bibnamefont {Measday}},
  \ and\ \bibinfo {author} {\bibfnamefont {J.~P.}\ \bibnamefont {Roalsvig}},\
  }\href {\doibase 10.1103/PhysRevC.35.2212} {\bibfield  {journal} {\bibinfo
  {journal} {Phys. Rev. C}\ }\textbf {\bibinfo {volume} {35}},\ \bibinfo
  {pages} {2212} (\bibinfo {year} {1987})}\BibitemShut {NoStop}%
\bibitem [{\citenamefont {Klinskikh}\ \emph {et~al.}(2008)\citenamefont
  {Klinskikh}, \citenamefont {Brianson}, \citenamefont {Brudanin} \emph
  {et~al.}}]{Themuonlifetime}%
  \BibitemOpen
  \bibfield  {author} {\bibinfo {author} {\bibfnamefont {A.}~\bibnamefont
  {Klinskikh}}, \bibinfo {author} {\bibfnamefont {S.}~\bibnamefont {Brianson}},
  \bibinfo {author} {\bibfnamefont {V.}~\bibnamefont {Brudanin}},  \emph
  {et~al.},\ }\href {\doibase https://doi.org/10.3103} {\bibfield  {journal}
  {\bibinfo  {journal} {Bull. Russ. Acad. Sci. Phys.}\ }\textbf {\bibinfo
  {volume} {72}} (\bibinfo {year} {2008}),\
  https://doi.org/10.3103}\BibitemShut {NoStop}%
\bibitem [{\citenamefont {Ohlsson}\ and\ \citenamefont
  {Snellman}(2000)}]{Ohlsson:1999xb}%
  \BibitemOpen
  \bibfield  {author} {\bibinfo {author} {\bibfnamefont {T.}~\bibnamefont
  {Ohlsson}}\ and\ \bibinfo {author} {\bibfnamefont {H.}~\bibnamefont
  {Snellman}},\ }\href {\doibase 10.1063/1.533270} {\bibfield  {journal}
  {\bibinfo  {journal} {J. Math. Phys.}\ }\textbf {\bibinfo {volume} {41}},\
  \bibinfo {pages} {2768} (\bibinfo {year} {2000})},\ \bibinfo {note}
  {[Erratum: J.Math.Phys. 42, 2345 (2001)]},\ \Eprint
  {http://arxiv.org/abs/hep-ph/9910546} {arXiv:hep-ph/9910546} \BibitemShut
  {NoStop}%
\bibitem [{\citenamefont {Kamo}\ \emph {et~al.}(2003)\citenamefont {Kamo},
  \citenamefont {Yajima}, \citenamefont {Higasida}, \citenamefont {Kubota},
  \citenamefont {Tokuo},\ and\ \citenamefont {Ichihara}}]{Kamo:2002sj}%
  \BibitemOpen
  \bibfield  {author} {\bibinfo {author} {\bibfnamefont {Y.}~\bibnamefont
  {Kamo}}, \bibinfo {author} {\bibfnamefont {S.}~\bibnamefont {Yajima}},
  \bibinfo {author} {\bibfnamefont {Y.}~\bibnamefont {Higasida}}, \bibinfo
  {author} {\bibfnamefont {S.-I.}\ \bibnamefont {Kubota}}, \bibinfo {author}
  {\bibfnamefont {S.}~\bibnamefont {Tokuo}}, \ and\ \bibinfo {author}
  {\bibfnamefont {J.-I.}\ \bibnamefont {Ichihara}},\ }\href {\doibase
  10.1140/epjc/s2003-01138-0} {\bibfield  {journal} {\bibinfo  {journal} {Eur.
  Phys. J. C}\ }\textbf {\bibinfo {volume} {28}},\ \bibinfo {pages} {211}
  (\bibinfo {year} {2003})},\ \Eprint {http://arxiv.org/abs/hep-ph/0209097}
  {arXiv:hep-ph/0209097} \BibitemShut {NoStop}%
\bibitem [{\citenamefont {Akhmedov}\ \emph {et~al.}(2004)\citenamefont
  {Akhmedov}, \citenamefont {Johansson}, \citenamefont {Lindner}, \citenamefont
  {Ohlsson},\ and\ \citenamefont {Schwetz}}]{Akhmedov:2004ny}%
  \BibitemOpen
  \bibfield  {author} {\bibinfo {author} {\bibfnamefont {E.~K.}\ \bibnamefont
  {Akhmedov}}, \bibinfo {author} {\bibfnamefont {R.}~\bibnamefont {Johansson}},
  \bibinfo {author} {\bibfnamefont {M.}~\bibnamefont {Lindner}}, \bibinfo
  {author} {\bibfnamefont {T.}~\bibnamefont {Ohlsson}}, \ and\ \bibinfo
  {author} {\bibfnamefont {T.}~\bibnamefont {Schwetz}},\ }\href {\doibase
  10.1088/1126-6708/2004/04/078} {\bibfield  {journal} {\bibinfo  {journal}
  {JHEP}\ }\textbf {\bibinfo {volume} {04}},\ \bibinfo {pages} {078} (\bibinfo
  {year} {2004})},\ \Eprint {http://arxiv.org/abs/hep-ph/0402175}
  {arXiv:hep-ph/0402175} \BibitemShut {NoStop}%
\end{thebibliography}%

\end{document}